\documentclass[article,onecolumn,superscriptaddress,nofootinbib,floatfix,onecolumn,prd,aps]{revtex4-2}

\usepackage{graphicx} % Required for inserting images
\usepackage{amsmath}
\usepackage{xcolor}

\usepackage{amssymb}
\usepackage[colorlinks,linkcolor={red!50!black},citecolor={blue!80!black},urlcolor={blue!80!black}]{hyperref}
\usepackage{comment}
\usepackage{gensymb}
\usepackage{makecell}
\usepackage[normalem]{ulem}

\let\savedegree\degree
\let\degree\relax
\let\savecorresponds\corresponds
\let\corresponds\relax
\usepackage{mathabx}
\let\degree\savedegree
\let\corresponds\savecorresponds

\usepackage{amsmath}
\usepackage{mathrsfs}
\usepackage{amsfonts}
\usepackage{geometry}
\usepackage[none]{hyphenat}
\usepackage{slashed}
\usepackage{amssymb}
\usepackage{amscd}
\usepackage{enumerate}
\usepackage{graphicx}
\usepackage{float}
\usepackage[lofdepth,lotdepth]{subfig}
\usepackage[countmax]{subfloat}
\usepackage{color}
\usepackage{etoolbox}
\usepackage{multirow}
\usepackage{caption}
\newcommand{\pp}{p-p}

\newcommand{\epem}{\mathrm{e}^+\mathrm{e}^-}

\newcommand{\mumu}{\mu^+\mu^-}
\newcommand{\tautau}{\tau^+\tau^-}
\newcommand{\gaga}{\gamma\gamma}
\newcommand{\gag}{g_{\mathrm{a}\gamma}}
\newcommand{\ma}{m_\mathrm{a}}
\newcommand{\mV}{m_\mathrm{V}}
\newcommand{\gaG}{g_{\mathrm{G}\gamma}}
\newcommand{\mG}{m_\mathrm{G}}
\newcommand{\mDM}{m_\mathrm{DM}}

\newcommand{\MP}{M_\mathrm{P}}
\newcommand{\alphaG}{\alpha_\mathrm{G}}
\newcommand{\alphaUniv}{\alpha_\mathrm{G}/\MP}

\newcommand{\sqrts}{\sqrt{s}}

\newcommand{\sqrtsnn}{\sqrt{s_{_\text{NN}}}}

\newcommand{\jpsi}{J/\psi}
\newcommand{\ETmiss}{\not{\hbox{\kern-4pt E}}_\mathrm{T}}
\providecommand{\ccbar}{\mathrm{c}\overline{\mathrm{c}}}
\providecommand{\bbbar}{\mathrm{b}\overline{\mathrm{b}}}
\providecommand{\ttbar}{\mathrm{t}\overline{\mathrm{t}}}

\def\ttt#1{\texttt{\small #1}}
\def\order#1{\mathcal{O}{(#1)}}

\newcommand{\Lumi}{\mathcal{L}}

\newcommand{\madgraph}{\textsc{MadGraph5\_aMC@NLO}}

\newcommand{\gammaUPC}{\ttt{gamma-UPC}}

\newcommand{\ufo}{\textsc{ufo}}

\def\be{\begin{equation*}}
\def\ee{\end{equation*}}
\def\bsp#1\esp{\begin{split}#1\end{split}} 
\def\bpm{\begin{pmatrix}}
\def\epm{\end{pmatrix}}

\usepackage{xspace}
\newcommand*{\eg}{e.g.,\@\xspace}
\newcommand*{\ie}{i.e.,\@\xspace} 
\newcommand*{\cm}{c.m.\@\xspace}

\newcommand{\orcid}[1]{\href{https://orcid.org/#1}{\hspace*{0.1em}\raisebox{-0.45ex}{\includegraphics[width=1em]{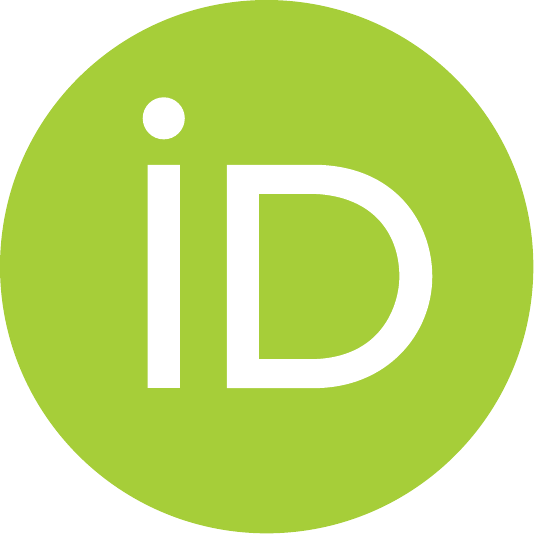}}}}

\begin{document}

%%%%%%%%%%%%%%%%%%%%%%%%%%%%%%%%%%%%%%%%%%%%%%%%%%%%%%%%%%%%%%%%%%%%%%%%%%%%%%%%%%%%
\title{Bounds on massive graviton-like particles from searches for\\ axion-like particles coupling to photons}

\author{Jordan Gué\orcid{0009-0000-4383-285X}}\email{jgue@ifae.es}
\affiliation{Institut de F\'{ı}sica d’Altes Energies (IFAE), The Barcelona Institute of Science and Technology, Campus UAB, 08193 Bellaterra (Barcelona), Catalonia, Spain}
\author{David~d'Enterria\orcid{0000-0002-5754-4303}}\email{david.d'enterria@cern.ch}
\affiliation{CERN, EP Department, CH-1211 Geneva, Switzerland}

\begin{abstract}
\noindent Limits on spin-0 axion-like-particles (ALPs) coupling to photons are reinterpreted as constraints on massive spin-2 graviton-like-particles (GLPs) with universal coupling $\alphaUniv$ (where $\MP$ is the reduced Planck mass) to the Standard Model fields. A minimally model-dependent recasting is performed, exploiting the formally analogous production and detection mechanisms for both particle types, based on the Primakoff and Gertsenshtein effects, i.e., photon-axion/graviton conversion.
Constraints originally derived in the ALP mass vs.\ photon-coupling plane $(\ma,\gag)$ are translated into the corresponding bounds in the GLP $(\mG,\alphaUniv)$ parameter space over the full mass range, $m_\mathrm{a,G} \approx 10^{-20}$--$10^{14}$~eV probed in current and future experimental setups including cavity-based detectors (haloscopes and resonant upconversion devices), helioscopes, magnetometers, optical interferometers, beam dumps, fixed-target, and collider experiments, as well as astrophysical and cosmological constraints. 
Generic scenarios are considered in which GLPs are a dark matter candidate and not. Whereas current ALP searches do not set stronger bounds on massive spin-2 particles than fifth-force tests, future magnetometers, two-beam interferometers, and upconversion experiments have the potential to provide very strong sensitivity, down to $\alphaUniv\approx 10^{-32}$~GeV$^{-1}$, for light graviton-like particles with $\mG\lesssim 10^{-8}$~eV. These future detectors exhibit comparatively greater sensitivity to massive gravitons than to axions. 
For massive gravitons at the TeV scale, exclusive diphoton decay searches, employed in ALP studies, offer a complementary approach to standard searches for spin-2 resonances in other inclusive final states.

\end{abstract}

\maketitle

\vspace{-0.5cm}

\tableofcontents

%%%%%%%%%%%%%%%%%%%%%%%%%%%%%%%%%%%%%%%%%%
\section{Introduction}

Axion-like particles (ALPs) are hypothetical spin-0 pseudoscalar bosons proposed as an extension of the Standard Model (SM) field content to address a variety of unsolved issues in particle physics and cosmology. In its original incarnation, the axion is a pseudo-Nambu--Goldstone boson (pNGB) with mass and SM-coupling inversely proportional to a high energy scale related to the breaking of a new (Peccei--Quinn, PQ) global U$(1)$ symmetry, introduced to explain the absence of violation of the charge-conjugation and parity (CP) symmetry in the strong interaction as described by quantum chromodynamics (QCD)~\cite{Peccei:1977hh,Weinberg:1977ma,Wilczek:1977pj}. More generally, ALPs also arise as light pseudoscalars that can serve as dark matter (DM) candidates~\cite{Duffy:2009ig,Marsh:2015xka}, or mediators between DM and SM fields~\cite{Nomura:2008ru,Dolan:2014ska,Kozaczuk:2015bea}. They also appear as pNGBs from spontaneously broken U(1) global symmetries, in composite or extended Higgs sectors~\cite{Cacciapaglia:2019bqz,Branco:2011iw}, in string-theory-inspired models~\cite{Ringwald:2012cu}, or as ``relaxions'' in theories addressing the electroweak (EW) hierarchy problem~\cite{Graham:2015cka,Choi:2020rgn}. Unlike the QCD axion, whose mass and couplings are tied to a single underlying scale, the ALP mass and SM~couplings are independent parameters and can be varied separately.\\

Although experimentally and theoretically less studied than spin-0 ALPs, hypothetical spin-2 massive graviton-like particles (called hereafter GLPs), are also predicted in different models of physics beyond the Standard Model (BSM). At variance with the standard graviton of General Relativity (GR) ---which is massless as guaranteed by diffeomorphism invariance~\cite{Hinterbichler:2011tt}, mediates long-range gravity, has two polarizations (two tensor modes) in four dimensions (4-D), and couples universally to the energy-momentum tensor--- a generic massive spin-2 particle has five physical polarizations in 4-D (two tensor modes, two vectors, and one scalar), mediates a short-range (Yukawa-type) force, may or may not couple to the metric of GR, and does not necessarily couple universally to matter. Such massive spin-2 degrees of freedom arise in  a variety of GR extensions. First, since the existence of DM is inferred solely from its gravitational effects, extensions of the gravitational sector have been considered, \eg\ bimetric theories~\cite{Hassan:2011vm}, that introduce an additional massive spin-2 field as a viable dark matter candidate\footnote{In this case, the massive field is not responsible for ordinary gravitation, and these theories may therefore evade the very stringent constraints arising from positivity requirements~\cite{Bellazzini:2023nqj,Blas:2024kps}.}~\cite{Babichev:2016bxi,Marzola:2017lbt,Armaleo:2020,Cembranos:2017vgi,Blas:2024kps}.  
Extradimensional theories of gravity, such as the Arkani-Hamed--Dimopoulos--Dvali (ADD)~\cite{Arkani-Hamed:1998sfv} and Randall--Sundrum (RS)~\cite{Randall:1999ee} models, predict massive tensor states appearing as Kaluza--Klein (KK) excitations of the extra spatial dimensions forming a spectrum of either quasi-continuum masses or well-separated resonances, respectively. Models with extra dimensions at the micron scale also predict KK modes, sometimes known as ``dark gravitons'', which provide a natural DM candidate~\cite{Babichev:2016bxi,Aoki:2016zgp,Gonzalo:2022jac}. Last but not least, the ghost-free nonlinear theory of a massive spin-2 field developed by de Rham--Gabadadze--Tolley (dRGT)~\cite{deRham:2016nuf} offers an alternative explanation for cosmic acceleration by modifying gravity on large scales that obviates the need for a cosmological constant or dark energy.\\

\begin{figure}[htpb!]
\centering
\includegraphics[width=0.9\textwidth]{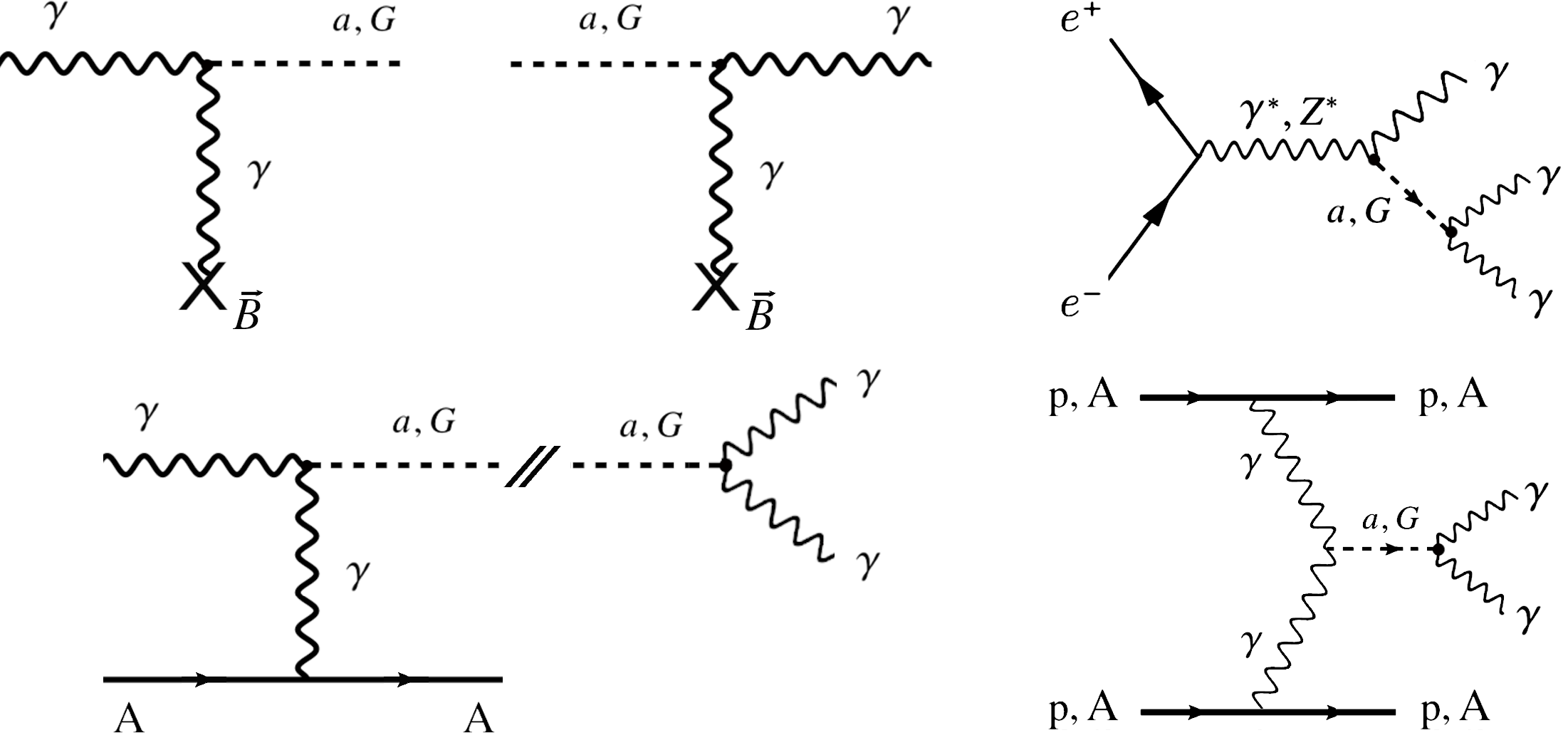}
\caption{\small Representative diagrams of ALP/GLP coupled to photons produced and/or decaying in Primakoff/Gertsenshtein ($\gamma \Leftrightarrow a,G$) processes in a strong magnetic field $\vec B$ (top left and center), beam dump experiments (bottom left and center, the `A' symbol indicates a nucleus, and `//' a long decay volume), triphoton production at $\epem$ colliders (top right), and photon-photon fusion at hadron colliders (bottom right, the `p' symbol stands for protons).}
\label{fig:diags}
\end{figure}

The experimental searches for
ALPs~\cite{Graham:2015ouw,Jaeckel:2015jla,Dobrich:2015jyk,Bauer:2017ris,Knapen:2016moh, Irastorza:2021tdu,Agrawal:2021dbo,dEnterria:2021ljz} have mostly exploited their coupling to photons via the different processes shown in Fig.~\ref{fig:diags}. All those ALP production processes are closely related to the ``Primakoff effect'' describing the conversion of photons into ALPs (and vice versa) in the presence of an external electromagnetic (EM) field, typically a strong static magnetic field $\vec B$. Interestingly, the underlying principle of the ALP searches ---the conversion of a spin-0 boson into a photon in the presence of an EM field--- has a formal analog in the GLP case. Indeed, spin-2 tensor states can also interact with photons through the so-called ``Gertsenshtein effect''~\cite{Gertsenshtein:1962kfm}, whereby a GLP converts into a photon in a strong external (gravitational or electromagnetic) field. Since the structure of this interaction resembles that of the Primakoff effect, one can reinterpret ALP-photon conversion limits as bounds on spin-2 particles, with appropriate modifications to account for their differences in spin and other properties. This involves appropriately adjusting the production and decay mechanisms to account for the tensor nature of the graviton-like coupling, the different polarization dependencies, and the possible presence of additional degrees of freedom, among others. Such a reinterpretation provides an efficient means of extending existing experimental results to new classes of BSM scenarios, through a well-controlled modification of the theoretical framework, and constitutes the primary motivation for this work.

In this study, we review searches for ALPs coupled to photons at all existing (and future) experimental setups through the production/decay mechanisms schematically shown in Fig.~\ref{fig:diags}, and translate all existing (expected) limits into the corresponding GLP parameter space. We explicitly consider searches performed with haloscopes (sensitive to DM ALPs) and helioscopes (for ALPs produced inside the Sun)~\cite{Sikivie:2020zpn}, constraints from astrophysical observations~\cite{Caputo:2024oqc}, and at accelerators including beam dumps~\cite{Blumlein:1990ay,Dobrich:2015jyk,Dobrich:2019dxc,Jerhot:2022chi}, fixed-target~\cite{Harland-Lang:2019zur}, and collider experiments~\cite{Mimasu:2014nea,Jaeckel:2015jla,Brivio:2017ije,Bauer:2017ris,Agrawal:2021dbo, dEnterria:2021ljz,Biekotter:2025fll}. We do not explore ``light-shining-through-wall'' (LSW) experiments~\cite{Redondo:2010dp,OSQAR:2015qdv,ALPSII:2025eri} %such as ALPS II, 
---whereby laser photons convert into ALPs in a magnetic field, pass through a wall, and reconvert to photons behind it--- which currently provide less sensitive limits than the other methods.\\

The paper is organized as follows. In Section~\ref{sec:th}, we remind the basic theoretical description of massive gravitons based on the Fierz--Pauli Lagrangian with a universal coupling to SM fields, and ALPs based on an effective field theory (EFT) approach. Section~\ref{sec:DM} presents the reinterpretation of light ALP searches into the corresponding GLP parameter space under the assumption that both particles constitute dark matter candidates. This includes searches with haloscopes, magnetometers, setups involving a spacetime-dependent EM field (birefringence, upconversion, and conversion in Earth's $B$ field), as well as astrophysical DM constraints. Section~\ref{sec:notDM} derives the ALP-GLP mapping in the absence of the DM assumption, focusing on detection in helioscopes and limits based on astrophysical observations. Section~\ref{sec:accel} exploits searches for heavier ALPs ($\ma\gtrsim 1$~MeV) at beam dumps, fixed-target, and collider experiments to set the corresponding bounds on the GLP mass and couplings. In Section~\ref{sec:results}, we provide the ALP-GLP ``dictionary'' and couplings mapping for all 16 search methods considered, and collect all derived upper-limit curves in the $(\mG,\alphaUniv)$ plane, and compare those with massive graviton bounds obtained from alternative methods, such as fifth-force probes, GW detectors, and astrophysical (stellar energy losses) observations. Finally, Section~\ref{sec:summ} summarizes the main conclusions of the study. Throughout the paper, we use the  $(-+++)$ metric convention and natural units $\hbar = c = 1$.

%%%%%%%%%%%%%%%%%%%%%%%%%%%%%%%%%%%%%%%%%%
\section{General theory}
\label{sec:th}

This chapter briefly recalls the basic theoretical description of GLP and ALP dynamics, presenting key expressions (Lagrangian densities, decay widths, etc.) used in the remainder of the paper.

\subsection{Massive gravitons}

\begin{figure}[htpb!]
\centering
\includegraphics[width=0.8\textwidth]{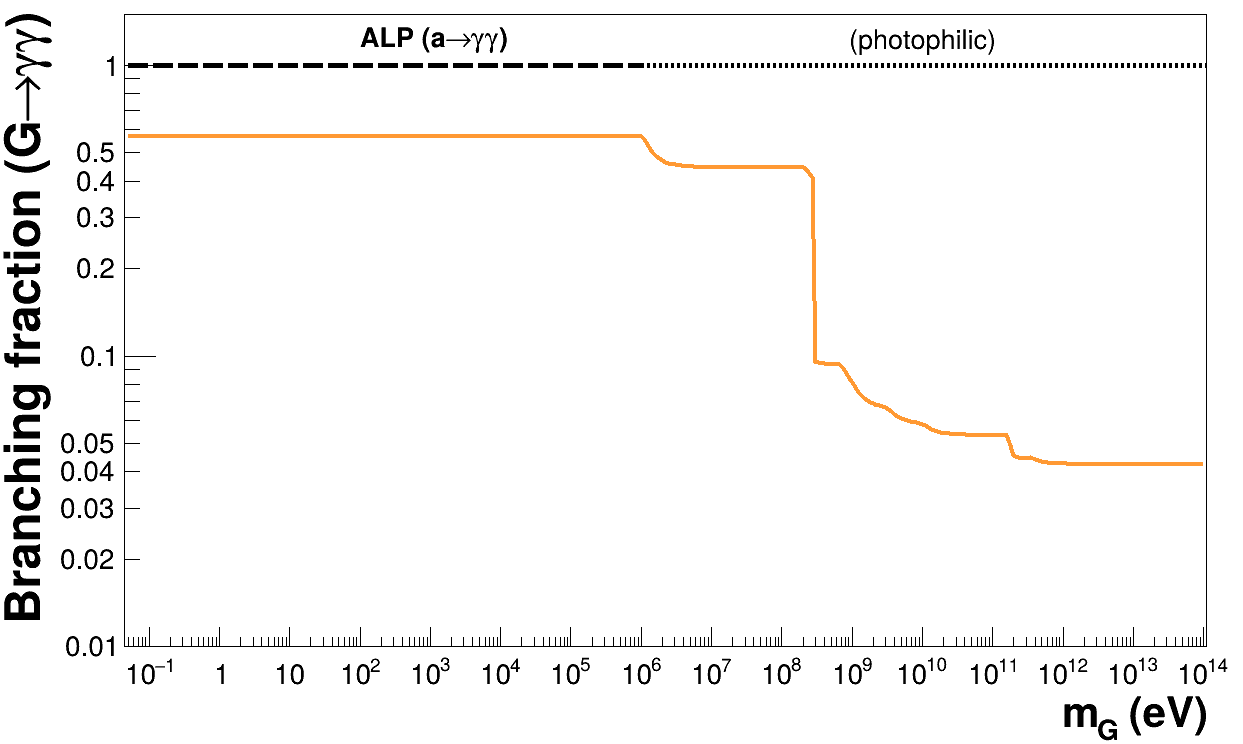}
\caption{\small Diphoton decay branching fraction of GLPs with universal coupling to SM fields versus GLP mass, as obtained from Eqs.~(\ref{eq:Gamma_G}). The different thresholds are explained in the text. The upper dashed curve shows the $\mathcal{B}_{\mathrm{a}\to\gaga}=1$ value (which at high masses is also typically considered in ``photophilic'' ALP scenarios).}
\label{fig:GLP_BRgamgam}
\end{figure}

The motion of a free spin-2 field $G^{\mu\nu}$ of mass $\mG$ propagating on a spacetime with metric $g_{\mu\nu}$ can be described by the Lorentz-invariant Fierz--Pauli Lagrangian~\cite{Fierz:1939ix} (see \eg~\cite{Marzola:2017lbt, Armaleo:2020})
\begin{equation}\label{eq:Lgraviton}
\mathcal{L}_\mathrm{FP} = -\frac{1}{2} (\partial_\rho G_{\mu\nu})^2 + \partial_\mu G_{\nu\rho} \partial^\nu G^{\mu\rho}
-\partial_\mu G^{\mu \nu} \partial_\nu G + \frac{1}{2} (\partial_\rho G)^2 
-\frac{1}{2}\mG^2 \left( (G_{\mu\nu})^2-G^2\right),
\end{equation}
defined with such a structure so as to satisfy the consistency of the theory, avoiding ghosts and keeping exactly five physical polarizations. Indeed, solving the Euler--Lagrange equation, Eq.~(\ref{eq:Lgraviton}) leads to a null trace of the massive graviton field: $G\equiv G^\mu_{\:\mu} =\partial_\mu G^{\mu\nu}=0$, \ie\ the traceless and transverse (TT) conditions are fulfilled at the equation-of-motion level and are not a gauge choice. This leaves five degrees of freedom for the $G^{\mu\nu}$ field: two tensor modes ($+,\times$), two vector modes ($R,L$), and one scalar mode ($S$). In a $(\vec p, \vec q, \vec k)$ frame where $\vec k$ denotes the direction of propagation of the GLPs, and $\hat p, \hat q$ are two unit vectors in the transverse plane, the polarization tensor reads~\cite{Armaleo:2019gil}
\begin{align}\label{eq:G_tensor}
    G &= \frac{\mathcal{G}_0}{\sqrt{2}}\begin{pmatrix}
        G_+ - G_S/\sqrt{3} & G_\times & G_R \, \\
        G_\times & -G_+-G_S/\sqrt{3} & G_L \, \\
        G_R & G_L & 2G_S/\sqrt{3} \, \\
    \end{pmatrix} \equiv 
        \mathcal{G}_0\cdot\varepsilon\, ,
\end{align}
where each component of the tensor has been normalized such that $G^2_+ + G^2_\times + G^2_R + G^2_L + G^2_S =1$.
The dimension-5 interacting Lagrangian of such a massive graviton $G^{\mu\nu}$ coupling to the SM stress energy tensor $T_{\mu\nu}$ with dimensionless coupling strength $\alphaG$ reads
\begin{align}\label{eq:int_lagrangian}
    \mathcal{L}_\text{int,GLP} &= -\frac{\alphaG}{\MP}\,T_{\mu\nu}G^{\mu\nu} \, ,
\end{align}
where $\MP = 1/\sqrt{8\pi\,G_\mathrm{N}} = 2.4\cdot 10^{18}$~GeV is the reduced Planck mass (and $G_\mathrm{N} = 6.709\cdot 10^{-39}$~GeV$^{-2}$, the Newton constant), and $g \equiv \det(g_{\mu\nu})$. 
In its simplest and most natural realization, the coupling $\alphaG$ is universal for all SM fields (non-universality tends to introduce ghosts, or violate perturbative unitarity in the $\mG \to 0$ regime in scatterings with SM particles~\cite{Das:2016pbk}), though such a constraint might be relaxed in some scenarios~\cite{Das:2016pbk,Blas:2024kps}.
In the SM, the energy-momentum tensors of the gauge fields $V$ (the $\mathrm{SU(2)}_L\times \mathrm{U(1)}_Y$ EW gauge bosons $W,B$, and the $\mathrm{SU(3)}_c$ gluon $g$) and matter fields $f$ (including quarks $q$, charged leptons $\ell^\pm$, and left-handed neutrinos $\nu_f$) read~\cite{Das:2016pbk}, respectively,
%\begin{align}
\begin{equation}\bsp
 T^V_{\mu\nu} & =
      -g_{\mu\nu}\left[-\frac{1}{4}F^{\rho\sigma}F_{\rho\sigma}+\delta_{\mV,0}\left(\left(\partial^{\rho}\partial^{\sigma}V_{\sigma}\right)V_{\rho}+\frac{1}{2}\left(\partial^{\rho}V_{\rho}\right)^2\right)\right] \\
  &\quad-F_{\mu}^{\: \rho}F_{\nu\rho}+\delta_{\mV,0}\left[\left(\partial_{\mu}\partial^{\rho}V_{\rho}\right)V_{\nu}
+\left(\partial_{\nu}\partial^{\rho}V_{\rho}\right)V_{\mu}\right]\,,\\
  T^f_{\mu\nu} & =
    -g_{\mu\nu}\left[\bar{\psi}_f\left(i\gamma^{\rho}D_{\rho}-m_f\right)\psi_f-\frac{1}{2}\partial^{\rho}\left(\bar{\psi}_fi\gamma_{\rho}\psi_f\right)\right]\\
  &\quad
   +\left[\frac{1}{2}\bar{\psi}_fi\gamma_{\mu}D_{\nu}\psi_f-\frac{1}{4}\partial_{\mu}\left(\bar{\psi}_fi\gamma_{\nu}\psi_f\right)+\left(\mu\leftrightarrow \nu\right)\right]\,,
\esp \end{equation}
where $F_{\mu\nu}$ denotes the field-strength tensor of the gauge field $V$ (hereafter, unless stated otherwise, $F_{\mu\nu}$ will be used to refer to the photon field-strength tensor), and the gauge-fixing term in $T^V_{\mu\nu}$, proportional to the Kronecker delta $\delta_{\mV,0}$, contributes only in the massless case $\mV=0$ (\ie\ for $\rm V=g,\gamma$), and with indices corresponding to other quantum numbers (such as colour) implicitly understood. After spontaneous EW symmetry breaking, one gets the mass eigenstates of the EW bosons ($\rm Z, W^{\pm},\gamma$) and the SM Higgs boson H (the expression of the energy-momentum tensor for the Higgs doublet coupling to the massive graviton can be found, \eg\ in Ref.~\cite{Das:2016pbk}).
From the Lagrangian density above, one can derive the corresponding leading-order (LO) partial decay widths of a spin-2 particle $\mathrm{G}$ into various SM particles\footnote{The expressions for decays of a heavy graviton into (pairs of) W, Z, and Higgs bosons, omitted here, can be found \eg\ in Ref.~\cite{Das:2016pbk}.}, 
\begin{equation}\label{eq:Gamma_G}
\bsp
& \Gamma(\mathrm{G}\to \gaga)=\left(\frac{\alphaG}{\MP}\right)^2\,\frac{\mG^3}{80\pi},\\
& \Gamma(\mathrm{G}\to \nu_f\bar{\nu}_f)=\left(\frac{\alphaG}{\MP}\right)^2\,\frac{\mG^3}{320\pi} \quad \text{(per $\nu_f$ flavour)\,,} \\
& \Gamma(\mathrm{G}\to f\bar{f})=\left(\frac{\alphaG}{\MP}\right)^2\,\frac{N_c^f \mG^3}{160\pi}(1-4r_f)^{3/2}\left(1+\frac{8}{3} r_f\right)  \quad \text{(per } f \in \{ \ell^\pm, q \})\,,\\ % \quad f\neq \nu\\
& \Gamma(\mathrm{G}\to gg)=\left(\frac{\alphaG}{\MP}\right)^2\,\frac{\mG^3}{10\pi},\quad \text{(valid for $\mG>2m_\pi\approx 270$~MeV)}\\
\esp
\end{equation} 
where $r_f={m_f^2}/{\mG^2}$ is a dimensionless quantity that sets the kinematic threshold of the difermion decay (for heavy quarks, $m_f$ corresponds to the charm and bottom masses, whereas it corresponds to the pion (kaon) mass for ud (s) light quarks), and $N_c^f$ is the number of colours of the fermion $f$ (\ie\ $N_c^f=1$ for charged leptons and $N_c^f=3$ for quarks). 
All decay widths show a typical cubic dependence with $\mG$, which implies that for decreasing masses, the decay rate will eventually become so small that the GLP can be considered stable. Figure~\ref{fig:GLP_BRgamgam} shows the diphoton branching fraction as a function of GLP mass. For masses below $\mG \approx 1$~MeV, where the dielectron decay is not yet kinematically accessible, the diphoton final state dominates the GLP decays, $\mathcal{B}_{\mathrm{G}\to\gaga}\approx 60\%$, with the dineutrino final state\footnote{It is worth noting that massive graviton studies have in the past often omitted the $\mathrm{G}\to\nu\bar \nu$ decay in their estimates, leading to factor of two stronger limits for light GLPs than those derived here.} accounting for the other 40\%. Between $\mG\approx 1$~MeV (dielectron threshold) and $\mG \approx 270$~MeV (once dimuon and dihadron decays are accessible), GLPs still dominantly decay into diphotons, with a branching fraction above 40\%. This fraction decreases to about 10--5\% across the $\mG\approx 0.27$--1000~GeV mass range, with different thresholds shown in Fig.~\ref{fig:GLP_BRgamgam} indicating heavier SM final states ($\tautau$, $\ccbar$, $\bbbar$, VV, HH, $\ttbar$,...) becoming kinematically accessible. For $\mG\gtrsim 270$~MeV, the hadronic final states effectively account for about 80\% of the GLP decays~\cite{Das:2016pbk}. 

In this study, we focus on the most generic and natural scenario of GLPs with a single universal coupling to all SM fields, in which case one can reabsorb the interaction term (\ref{eq:int_lagrangian}) into a redefinition of the metric $g_{\mu\nu}\to g_{\mu\nu}+(2\alphaUniv)\,G_{\mu\nu} $, and the effect of the spin-2 field becomes analogous to a gravitational wave (GW), \ie\ a perturbation of the Minkowski metric, 
\begin{align}\label{eq:h_G_equality}
    h_{\mu\nu} &= \frac{2\alphaG}{\MP}G_{\mu\nu} \, .
\end{align}
For searches for heavy GLPs coupling to photons, one can often find in the literature the Lagrangian densities~(\ref{eq:int_lagrangian}) or (\ref{eq:h_G_equality}) written with the coupling substitution $\gaG \equiv \alphaUniv$~\cite{dEnterria:2023}.

\subsection{Axion-like particles}

The most general effective Lagrangian (up to dimension-5 operators) describing a spin-0 CP-odd ALP field `$a$' interacting with SM fields, written in the unbroken phase of the EW symmetry, reads~\cite{Georgi:1986df,Bauer:2017ris,Biekotter:2025fll} 
\begin{equation}\label{eq:Leff_ALP}
\begin{aligned}
   \mathcal{L}_\text{eff,ALP}
   &= \frac12 \left( \partial_\mu a\right)\!\left( \partial^\mu a\right)
    - \frac{\ma^2}{2}\,a^2 
    + \frac{\partial^\mu a}{\Lambda} \sum_f\,\bar\psi_f\,\mathcal{X}_f\,\gamma_\mu\,\psi_f \\[-1mm]
   &\quad\mbox{}+ g_s^2\,\frac{C_{gg}}{\Lambda}\,a\,\mathcal{G}_{\mu\nu}^A\,\tilde{\mathcal{G}}^{\mu\nu,A}
    + g^2\,\frac{C_{WW}}{\Lambda}\,a\,W_{\mu\nu}^A\,\tilde W^{\mu\nu,A}
    + g^{\prime\,2}\,\frac{C_{BB}}{\Lambda}\,a\,B_{\mu\nu}\,\tilde B^{\mu\nu} \,,
\end{aligned}
\end{equation}
where $\mathcal{G}_{\mu\nu}^A$, $W_{\mu\nu}^A$, and $B_{\mu\nu}$ are the field-strength tensors, respectively, of SU(3)$_c$, SU(2)$_L$, and U(1)$_Y$ (with dual field-strength tensors defined as $\tilde B^{\mu\nu}=\frac12\epsilon^{\mu\nu\alpha\beta} B_{\alpha\beta}$, etc.\ with $\epsilon^{0123}=1$), and $g_s$, $g$, and $g'$ are their corresponding gauge coupling constants. The $\Lambda$ and $C_{VV}$ parameters of (\ref{eq:Leff_ALP}) are, respectively, the high energy scale of new physics and the effective Wilson coefficients of the EFT. The value of $\Lambda$ controls how suppressed the ALP interactions are ---related \eg\ to the scale of the anomalous U(1) global symmetry breaking in concrete ultraviolet (UV) completions--- and the dimensionless Wilson coefficients, which are typically $\mathcal{O}(1)$, encode the detailed structure of the UV physics (charges, anomalies, loops) that parameterize the ALP couplings to gauge bosons. The parameters $\Lambda$ and $C_{VV}$ for ALPs play thus the same roles as $\MP$ and $\alphaG$ for GLPs, Eq.~(\ref{eq:int_lagrangian}). In the case of the QCD axion, the CP-violating $\theta$ term of the QCD Lagrangian is eliminated by absorbing it into the axion field, $a = \bar a - \bar\theta f_a$, with the ``axion decay constant'' $f_a$, defined such that $\Lambda/|C_{gg}|=32\pi^2 f_a$, playing the role of the high-energy scale. At the same time, the particle excitation corresponding to the QCD axion field acquires a pNGB mass that is inversely proportional to the decay constant $\ma \approx (m_{\pi}f_{\pi}/f_a)\, \sqrt{m_u m_d}/(m_u+m_d)$~\cite{Weinberg:1977ma}. Such a relationship between $\ma$ and $f_a$ is, however, missing in the case of generic ALPs. The sum in the first line of Eq.~(\ref{eq:Leff_ALP}) extends over the SM chiral fermion multiplets $f$, and $\mathcal{X}_f$ are hermitian matrices in generation (flavour) space. The fact that the gauge couplings are pulled out of the ALP interaction terms with the SM gauge fields implicitly implies that, in the most generic case, the bosonic ALP couplings are loop suppressed, while the fermionic ALP couplings arise at tree level. After EW symmetry breaking, the effective Lagrangian (\ref{eq:Leff_ALP}) contains the explicit couplings of the ALP to $\gaga$, $\gamma$Z, and ZZ,
\begin{equation}\label{eq:Leff_ALP2}
\begin{aligned}
   \mathcal{L}_\text{eff,ALP} %^{D\le 5}
   & \supset \frac12 \left( \partial_\mu a\right)\!\left( \partial^\mu a\right)
    - \frac{\ma^2}{2}\,a^2 \\[-1mm]
    & + e^2\,\,\frac{C_{\gaga}}{\Lambda}\,a\,F_{\mu\nu}\,\tilde F^{\mu\nu}
    + \frac{2e^2}{s_w c_w}\,\frac{C_{\gamma Z}}{\Lambda}\,a\,F_{\mu\nu}\,\tilde Z^{\mu\nu}
    + \frac{e^2}{s_w^2 c_w^2}\,\frac{C_{ZZ}}{\Lambda}\,a\,Z_{\mu\nu}\,\tilde Z^{\mu\nu} \,,
\end{aligned}
\end{equation}
where $e$ is the EM coupling constant and $s_w=\sin\theta_w$ ($c_w=\cos\theta_w$) denotes the (co)sine of the Weinberg angle, and the effective Wilson coefficients read
\begin{equation}\label{Cgagadef}
   C_{\gaga} = C_{WW} + C_{BB} \,, \qquad
   C_{\gamma Z} = c_w^2\,C_{WW} - s_w^2\,C_{BB} \qquad
   C_{ZZ} = c_w^4\,C_{WW} + s_w^4\,C_{BB}\,.
\end{equation}
At low masses ($\ma\lesssim 1$~MeV), the ALP can only effectively convert or decay into photons, and the $\gamma$-ALP coupling %$\gag\equiv e^2\, C_{\gaga}/\Lambda$
is the only object of phenomenological study, with Lagrangian density
\begin{equation}
    \mathcal{L}_{\gamma\text{-ALP}} \supset  -\frac{\gag}{4}a F^{\mu \nu}\tilde{F}_{\mu\nu} = \gag \,a\, \mathbf{E} \cdot \mathbf{B},\,\mbox{ and }\gag\equiv e^2\, C_{\gaga}/\Lambda,
\end{equation}
where the right equality expresses the interaction in terms of the electric $\bf E$ and magnetic $\bf B$ vector fields. In the QCD axion case, the adimensional coupling $C_{\gaga}=E/N-1.92(4)$ is defined in terms of the electromagnetic ($E$) and color anomaly ($N$) coefficients, which depend on the particular PQ charges assigned to the particles in the concrete axion model realization~\cite{DiLuzio:2020wdo}, and the numerical constant depends on ratios of the $m_\mathrm{u,d}$ quark masses~\cite{Irastorza:2021tdu}. For higher masses, one can set limits on the ALP-EW couplings $g_\mathrm{aV} \equiv f(e,c_w,s_w)\,C_{VV}/\Lambda$, but most often limits on $\gag$ are set assuming the photon-dominance or \textit{photophilic} scenario\footnote{The opposite physical limit is the \textit{photophobic} scenario, whereby $C_{BB} = -C_{WW}$ at the UV scale, and the ALP does not directly couple to photons~\cite{Craig:2018kne}. It is worth noting that even in this photophobic case, ALPs can decay into photons through EW loops, and the expression of the $\mathrm{a}\to\gaga$ decay width, Eq.~(\ref{eq:Gamma_ALP}), is identical to the photophilic one, with the $\gag$ coupling now generated effectively at the loop level.}~\cite{Agrawal:2021dbo}, given by $C_{WW} \approx C_{BB} $.

From the general Lagrangian (\ref{eq:Leff_ALP}), one can obtain the partial widths kinematically open for ALP decays into SM particles~\cite{Bauer:2017ris,Biekotter:2025fll}:
\begin{align}
\begin{split}\label{eq:Gamma_ALP}
\Gamma(\mathrm{a} \to \gaga) &= \frac{\gag^2}{64 \pi}\ma^3,  \\
\Gamma(\mathrm{a} \to f\bar{f})  &= \frac{g_\mathrm{af}^2}{8 \pi} m_f^2 \, N_c^f\, \ma \sqrt{1 - \frac{4 m_f^2}{\ma^2}} %\, ,\\ % \quad f \in \{ e, \mu, \tau, c, b, t \} ,   
 \quad \text{(per } f \in \{ \ell^\pm, q \})\,,\\
% \Gamma(a \to \text{light hadrons}) &=
\Gamma(\mathrm{a} \to gg) &= \frac{g_\mathrm{ag}^2}{8 \pi^3} \ma^3 \left[ 1+ \frac{83}{4} \frac{\alpha_s}{\pi} \right] , \quad \text{(valid for $\ma\gtrsim 3m_\pi\approx 420$~MeV)} \, . 
\end{split}
\end{align}
Gauge boson decay widths display the characteristic cubic $\ma$ dependence, while fermion decay modes depend linearly on $\ma$ and, like the Higgs boson, are proportional to the square of the fermion mass (at variance with the GLP case, where the fermionic and bosonic widths depend on $m_G^3$ for both types of decays). For light quarks, the (effective) threshold fermion mass $m_f$ corresponds to\footnote{\label{note1}The radiative hadronic decay $\mathrm{a}\to 2\pi\gamma$ is  accessible for somewhat lower ALP masses, but is electromagnetically suppressed compared with the 3-hadron strong decays.} $\ma\gtrsim 3m_\pi\approx 420$~MeV (see \eg\ Ref.~\cite{Blinov:2021say,Alda:2025nsz} for detailed descriptions of different hadronic ALP decay modes). The difermion decay width in Eq.~(\ref{eq:Gamma_ALP}) should be also valid for (Dirac) neutrinos~\cite{Darme:2020sjf} and thus, at variance with the massive graviton case, Eq.~(\ref{eq:Gamma_G}), the $\mathrm{a}\to\nu\bar\nu$ channel (being proportional to $m_\nu^2$) is completely negligible (unless the $g_{a\nu}$ coupling is enormously large). The decay rate into gluons (which, as for the quarks case, is only valid for ALPs with masses above the 3-pion threshold) includes one-loop QCD corrections, identical to the Higgs boson case~\cite{Spira:1995rr}. The expressions for decays into massive EW gauge bosons, $\Gamma(\mathrm{a} \to \rm Z\gamma,\,ZZ,\,WW)$ (enhanced by inverse powers of $s_w$ with respect to the photon decay rate) can be found \eg\ in Refs.~\cite{Bauer:2017ris,Biekotter:2025fll}, but are not listed here as they are only relevant for heavy ALPs and most experimental searches have focused on the photophilic case with\footnote{The assumption of a purely diphoton decay is not justified for $\ma>m_\mathrm{Z}$, where an ALP coupling to photons is also expected to decay via $\rm a \to Z \gamma$.} $\Gamma(\rm a\to\gaga)\approx 1$.
For large ALP masses, and equal effective couplings $\gag\approx g_\mathrm{ag} \approx g_\mathrm{aq}$, decays to gluons and the more massive quarks dominate compared with the photonic case. However, any hadronic ALP decays will ultimately produce also final-state photons (mostly through $\pi^0$ decays), and some photon-excess searches from astrophysical ALP sources also exploit such a decay signature.

It is worth noting that accelerator searches for heavy ALPs have mostly assumed the photophilic scenario, in which all decay branching fractions except the diphoton one are taken to be negligible, \ie\ $\gag\gg g_\mathrm{aX}$ and $\mathcal{B}_{\mathrm{a}\to\gaga}\approx 1$. This assumption contrasts with the case of GLPs with universal coupling, for which this fraction is $\mathcal{B}_{\mathrm{G}\to\gaga}\approx 0.1$--0.05 for $\ma \gtrsim 270$~MeV (Fig.~\ref{fig:GLP_BRgamgam}). As discussed in Section~\ref{sec:accel}, this implies that limits on heavy GLPs are expected to be weaker by a factor of about $[\mathcal{B}_{\mathrm{a}\to\gaga}/\mathcal{B}_{\mathrm{G}\to\gaga}]^{1/2}\approx 3$--5 compared with those set on ALPs, if all other aspects were identical for the two particles.

%%%%%%%%%%%%%%%%%%%%%%%%%%%%%%%%%%%%%%%%%%
\section{Massive gravitons as dark matter}
\label{sec:DM}

In this work, we reinterpret ALP searches spanning many orders-of-magnitude in mass, in terms of the corresponding GLP parameter space. For (very) low masses and couplings, both ALPs and GLPs can serve as cosmologically stable dark matter candidates, with a lower mass bound typically set by the requirement that DM must allow the formation of large-scale structure, namely, that the de~Broglie wavelength of the DM particle must not exceed the characteristic size of dwarf galaxies\footnote{For GLPs, a much weaker theoretical bound exists on the minimum mass allowed for spin-2 particles (Higuchi bound), following from the requirement to avoid ghost-like (negative norm) helicity-0 modes in the background spacetime~\cite{Higuchi:1986py}, that translates into $\mG\gtrsim \sqrt{2}H_0 \approx 2.1\cdot 10^{-33}$~eV for the present-day universe (approximately de~Sitter-like due to dark energy) with Hubble parameter $H_0 = 1.5\cdot 10^{-33}$~eV.}. In standard models of galaxy formation, the galactic frame is the DM rest frame, such that in the heliocentric frame, DM is nonrelativistic with average velocity $v_\mathrm{DM} \sim 10^{-3}$, namely the kinetic energy of DM particles is tiny compared with their rest mass: $E_\mathrm{kin} \approx 1/2\,\mDM v_\mathrm{DM}^2 \approx 10^{-6} \mDM$. Assuming the typical size of dwarf galaxies of some thousands light-years, this implies that the minimum mass allowed for DM particles is $\order{10^{-22}~\text{eV}}$. More recent studies of individual galactic dynamics suggest a somehow larger $\order{10^{-21}~\text{eV}}$ lower bound~\cite{Zimmermann:2024xvd}. 

\begin{figure}[htpb!]
\centering
\includegraphics[width=0.9\textwidth]{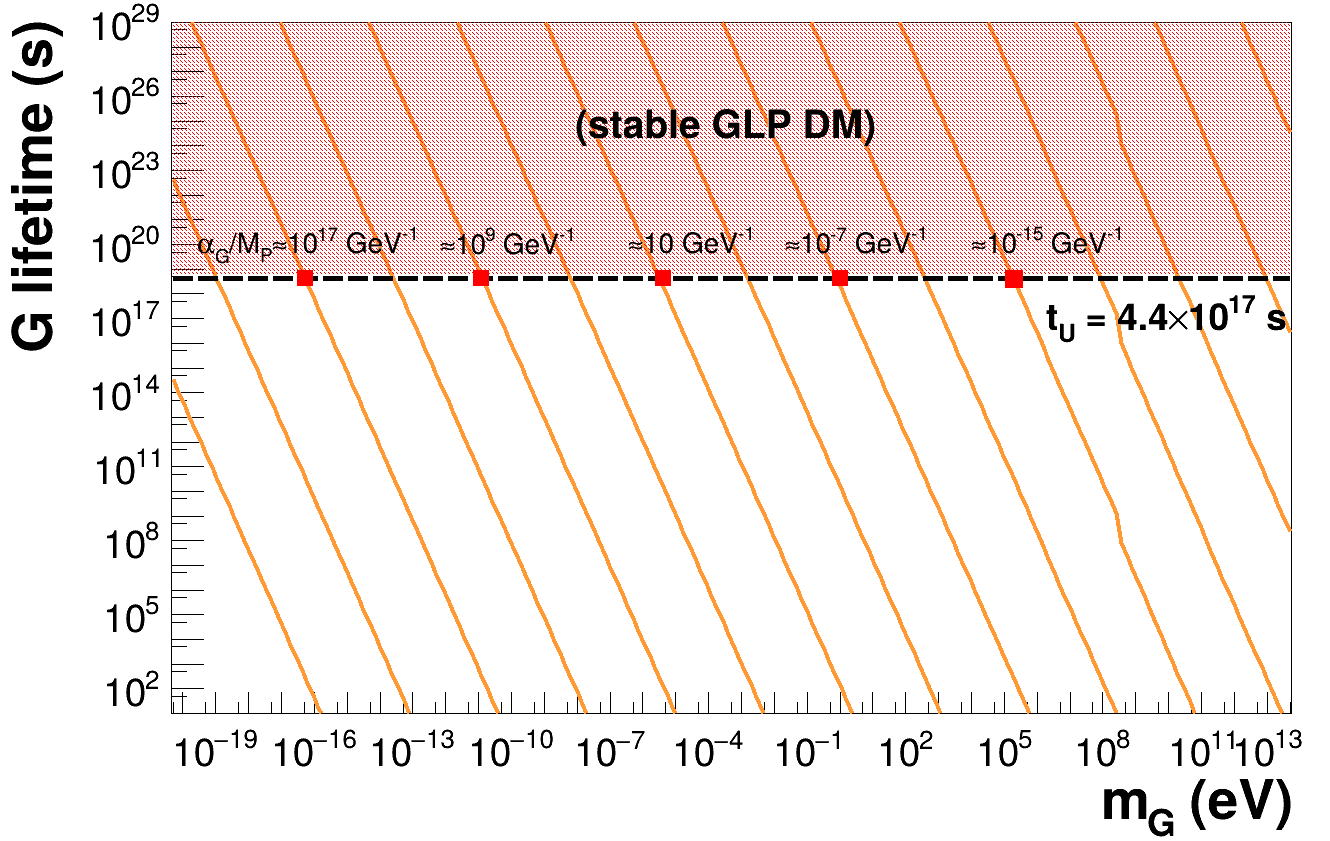}
\caption{\small GLP lifetime versus its mass $\mG$ for varying values of the universal coupling $\alphaUniv$ (diagonal lines). The horizontal dashed line marks the current age of the universe $t_\mathrm{U}$. For a given $\mG$, the red boxes highlight a few representative values of $\alphaUniv$ for which the GLP lifetime matches the age of the universe. GLPs with lifetimes above it can be considered cosmologically stable (red dotted area). 
\label{fig:lifetime}}
\end{figure}

The lifetime of the ALP and GLP is an important quantity for its consideration as a potential DM candidate and for setting limits for its existence based on the non-observation of excesses of SM particles (most commonly photons) produced in its decay. Given the current age of the universe $t_\mathrm{U} \approx 13.8~\text{Gyr}\approx 4.35\cdot 10^{17}$~s, any GLP with a total decay width below $\Gamma_\mathrm{G} = t_\mathrm{U}^{-1} = 1.5\cdot 10^{-33}$~eV is effectively stable cosmologically. The total decay width of a GLP exhibits a strong dependence on both its mass and coupling, scaling approximately as $\propto (\mG)^3(\alphaUniv)^2$ as per Eqs.~(\ref{eq:Gamma_G}). Figure~\ref{fig:lifetime} displays the GLP lifetime vs.\ mass for different $\alphaUniv$ couplings (diagonal lines), compared with the current age of the universe $t_\mathrm{U}$ (horizontal dashed line). In essence, light GLPs ($\mG<1$~eV) are effectively stable unless endowed with unrealistically large universal couplings, whereas heavier GLPs decay rapidly unless their coupling to SM fields is very suppressed. As an example, GLPs with mass below $\mG\approx 1$~eV can be considered stable on cosmological scales (and, therefore, viable DM candidates) for any universal coupling $\alphaUniv\gtrsim 10^{-7}$~GeV$^{-1}$, whereas above this mass and for the same lower-limit coupling, the GLPs will decay and may appear as excesses of SM particles (mostly photons) with respect to astrophysical backgrounds. Quantitative bounds in the $(\mG,\alphaUniv)$ plane, based on astrophysical DM decay limits, are provided later in Fig.~\ref{fig:himass_bounds_current}.\\

Hereafter, we consider the so-called ultralight DM regime for GLP candidates covering the $\mDM \approx 10^{-20}$--1~eV mass range. In this regime, the typical occupation number of galactic DM is extremely large, $n_\mathrm{DM}= \rho_\mathrm{DM}/\mDM \approx 3\cdot10^{28}$--$10^{8}$~cm$^{-3}$ for the local DM energy density $\rho_\mathrm{DM}\approx0.4$~GeV\,cm$^{-3}$~\cite{Cirelli:2024ssz}, such that the field can be treated as a coherent collection of many particles, with negligible quantum fluctuations, in the same way as classical electromagnetism. In this case, a massive spin-2 DM field behaves as a classical oscillating field of the form,
\begin{align}\label{eq:Gmunu}
    G_{\mu\nu} &= \frac{\sqrt{2\rho_\mathrm{DM}}}{\omega_\mathrm{G}}\varepsilon_{\mu\nu}\cos(\omega_\mathrm{G} t-\vec k_\mathrm{G} \cdot \vec x+\phi_\mathrm{G})\, ,
\end{align}
obtained by setting $\mathcal{G}_0 = \sqrt{2\rho_\mathrm{DM}}/\omega_\mathrm{G}\cos(\omega_\mathrm{G} t-\vec k_\mathrm{G} \cdot \vec x+\phi_\mathrm{G})\equiv G_0\cos(\omega_\mathrm{G} t-\vec k_\mathrm{G} \cdot \vec x+\phi_\mathrm{G})$ in Eq.~\eqref{eq:G_tensor}, where $\omega_\mathrm{G}=\mG$ is the Compton frequency of the massive graviton field, $\vec k_\mathrm{G}= \omega_\mathrm{G} v_\mathrm{DM}\hat k_\mathrm{G}$ its (nonrelativistic) momentum\footnote{Throughout this paper, the $\hat{}$ symbol indicates a unit vector.}, and $\phi_\mathrm{G}$ an arbitrary phase. Note that, in the equation above, we have assumed a common phase  $\phi_\mathrm{G}$ for all polarizations, which need not be the case in general.  
Note that galactic DM must be virialized~\cite{Freese:2012xd,Kuhlen:2013tra} such that its velocity distribution acquires a characteristic width $\sigma_v \sim v_\mathrm{DM}$, leading to a typical finite coherence time and length of the field, above which it behaves as a incoherent sum of classical waves of different amplitudes and phases.
Neglecting corrections of order $v_\mathrm{DM}$, the DM wave therefore oscillates at frequency $\mG$. From $\partial_\mu G^{\mu\nu}=0$, one can show that the $G_{00}$ and $G_{0i}$ components of Eq.~(\ref{eq:G_tensor}) are suppressed, respectively, by a linear and quadratic $v_\mathrm{DM}$ factor compared to $G_{ij}$.\\

Similarly, if the axion is identified as DM, its wave equation can be expressed as 
\begin{align}
a = \frac{\sqrt{2\rho_\mathrm{DM}}}{\omega_\mathrm{a}}\cos(\omega_\mathrm{a} t-\vec k_\mathrm{a}\cdot \vec x +\phi_\mathrm{a}) \equiv a_0 \cos(\omega_\mathrm{a} t-\vec k_\mathrm{a}\cdot \vec x +\phi_\mathrm{a})\,,    
\end{align}
for the same corresponding variables, \ie\ only the polarization differs with respect to the massive graviton case, Eq.~(\ref{eq:Gmunu}), as expected.

Below we provide details on multiple experimental approaches designed to detect axion dark matter. As will be shown, these setups are also sensitive to spin-2 DM and can therefore place stringent constraints on the corresponding GLP parameter space.

\subsection{Laboratory experiments involving a magnetostatic background}

We start by considering laboratory experiments that use magnetostatic fields, and rely on the axion-photon conversion occurring in such a background. More specifically, we are interested in a DM background sourcing an EM field inside the experimental apparatus, which can then be measured. Such a system is controlled by Maxwell's equations that, in the presence of background spin-0 or spin-2 DM fields, are modified. In particular, the axion-photon coupling leads to an effective EM current that appears in Ampere's law and reads~\cite{Domcke:2022rgu,Sikivie:2020zpn}
\begin{align}\label{eq:j_alps}
    j^\mu_a &= \gag\partial_\nu(a\tilde F^{\nu\mu})\, \longrightarrow\, \vec j_a = \omega_\mathrm{a}\gag a_0\vec B\sin(\omega_\mathrm{a} t+\phi_\mathrm{a})\, ,
\end{align}
where the right equality assumes that the background $\vec B$ is purely magnetic and that the ALP's de~Broglie wavelength is much larger than the typical experiment size such that the field can be considered as homogeneous. 

Similarly, the massive graviton can also source an EM field through an effective current analogous to that of a GW~\cite{Berlin:2022,Domcke:2022rgu},
\begin{align}\label{eq:j_glps}
    j^\mu_\mathrm{G} &= \frac{2\alphaG}{\MP}\left(-G^\nu_{\:\alpha}\partial_\nu F^{\alpha\mu}+F^{\alpha\nu}\partial_\nu G^\mu_{\:\alpha}\right)\, \longrightarrow\, j^i_\text{G} = \frac{2\alphaG}{\MP}\left(-\epsilon_{lki}\varepsilon_{jk}G_0\partial_j B_l+\epsilon_{ljk} \varepsilon_{ij}B_l\partial_k G_0\right) \, ,
\end{align}
whose exact form depends on the structure of the background magnetic field (in particular if it is uniform or not, see below). Note that we have set terms of the form $G^{\mu}_{\beta}\partial_\alpha F^{\alpha\beta}=0$, as interactions in regions outside of the source of the background EM fields will always be assumed.  
Three concrete detection setups based on DM ALP-photon conversion are considered below: haloscopes, magnetometers, and Earth's magnetic field.

\subsubsection{Haloscopes}

Haloscopes search for galactic DM ALPs through their Primakoff conversion into photons in magnetic cavities (Fig.~\ref{fig:diags}, top left), which can become resonant if the axion frequency matches one eigenfrequency of the cavity. Examples of haloscope experiments include ADMX~\cite{ADMX:2020ote}, CAPP~\cite{Lee:2020cfj}, HAYSTAC~\cite{Brubaker:2016ktl}, ORGAN~\cite{Quiskamp:2022pks}, RADES~\cite{Ahyoune:2024klt}, and QUAX~\cite{Alesini:2019ajt} in their operating phase, and GrAHal~\cite{Grenet:2021vbb} as a future facility. Though the emphasis of these experiments is on the QCD axion, they are of course sensitive to any ALP or GLP signal.
Since DM is nonrelativistic, the converted photon energy is dominated by the ALP/GLP rest-mass contribution, $\omega_\mathrm{a,G}\simeq m_\mathrm{a,G}$. While haloscopes can also search for relic axions (from the cosmic axion background generated in the early universe in the same way as the photon CMB), the associated bounds are much weaker because their density should be much smaller than for the DM case, and we focus on the limits set for ALPs/GLPs as DM candidates. 
In haloscopes, the background magnetic field is uniform such that the GLP effective current can be simplified 
\begin{align}\label{eq:eff_j_magnetostatic_uniform}
    \vec j_\mathrm{G} = \frac{2\omega_\mathrm{G} v_\mathrm{DM}\alphaG}{\MP} G_0\, \vec \varepsilon \cdot (\hat k_\mathrm{G} \times \vec B) \sin(\omega_\mathrm{G}t+\phi_\mathrm{G}) \, ,
\end{align}
where $(\vec \varepsilon \cdot (\hat k_\mathrm{G} \times \vec B))_i = \epsilon_{jkl}\varepsilon_{ij}\hat k_\mathrm{G}^k B_l$. Comparing Eqs.~\eqref{eq:eff_j_magnetostatic_uniform} and~\eqref{eq:j_alps}, one can already notice that the GLP current will be suppressed by a factor of $v_\mathrm{DM}\approx 10^{-3}$ compared with the ALP one. The reason for this difference comes from parity:  the effective current must be even under parity operation while the background magnetic field is a pseudovector. Since the ALP is also parity-odd, no spatial gradient (a parity-odd operator) is needed to make the ALP effective current even; while for the GLP which is parity-even, one must include a spatial gradient for the effective current to have the correct parity. Because the background magnetic field is uniform, this spatial gradient necessarily acts on the GLP field, which is suppressed by the nonrelativistic velocity of the field.
Combining Ampere's law (which contains the DM effective current) and Faraday's law, one obtains
\begin{align}
    \nabla \times \nabla \times \delta \vec E + \partial^2_t \delta \vec E &= -\partial_t \vec j_\mathrm{a,G}\, ,
\end{align}
where $\delta \vec E =  \mathcal{O}(\gag; \alphaUniv)$ is the EM field generated by the ALP or GLP.
Further decomposing such a field as a sum over eigenmodes of the cavity, \ie\ $\delta \vec E = \sum_n e_n(t) \vec E_n(\vec x)$, where $e_n$ is the time-dependent amplitude of the eigenmode $\vec E_n$, this equation is equivalent to~\cite{Berlin:2022} 
\begin{align}\label{eq:eom_mode}
    \left(\partial^2_t e_n +\frac{\omega_n}{Q_n}\partial_t e_n +\omega^2_n e_n\right) = -\frac{\int \mathrm{d}\mathcal{V} \partial_t \vec j_\mathrm{a,G} \cdot \vec E_n}{\int \mathrm{d}\mathcal{V} |\vec E_n|^2}\, ,
\end{align}
where $\mathcal{V}$ is the cavity volume, and $Q_n$ is the quality factor of the eigenmode accounting for losses at the cavity boundaries. For this latter equation, only the right-hand side differs between ALP and GLP signals, which read, respectively,
\begin{subequations}
    \begin{align}
        &- \omega^2_\mathrm{a} \gag a\frac{\int \mathrm{d}\mathcal{V} \vec B \cdot \vec E_n}{\int \mathrm{d}\mathcal{V} |\vec E_n|^2} \quad \text{for the ALP,} \\
        &- \frac{2\omega^2_\mathrm{G} v_\mathrm{DM}\alphaG}{\MP}\epsilon_{ljk} G_{ij}\hat k^k_\text{G}\frac{\int \mathrm{d}\mathcal{V} B_l E^i_{n}}{\int \mathrm{d}\mathcal{V} |\vec E_n|^2} \quad \text{for the GLP.}
    \end{align}
\end{subequations} 
From this result, in order to convert the coupling $\gag$ sensitivity of haloscopes into the corresponding universal $\alphaUniv$ coupling, we just need to equate the (absolute value of) both expressions above
\begin{align}
    \frac{2v_\mathrm{DM}\alphaG}{\MP} \left|\epsilon_{ljk} \varepsilon_{ij}\hat k^k_\text{G} \int \mathrm{d}\mathcal{V} \hat B_l \hat E^i_{n}\right|  &= \gag \left|\int \mathrm{d}\mathcal{V} \hat B \cdot \hat E_n\right| \, \label{eq:halo_coupling_convers}
\end{align}
where $a_0=G_0$ and $\omega_\mathrm{a} = \omega_\mathrm{G}$ are used since we are converting sensitivities at a fixed mass for the same amplitudes of the spin-2 or spin-0 DM fields. In practice, most haloscopes are cylindrical cavities with the background magnetic field aligned with the symmetry axis $\hat z$. In addition, for maximum sensitivity to ALPs, the signal mode used is the Transverse Magnetic (TM) mode with radial, azimuthal, and longitudinal mode numbers $\{nml\}=\{010\}$, which is purely along the $\hat z$ direction, such that this left-hand side can be simplified to $\frac{2v_\mathrm{DM}\alphaG}{\MP} \left|\epsilon_{Bjk} \varepsilon_{Bj}\hat k^k_\text{G} \int \mathrm{d}\mathcal{V} \hat B \cdot \hat E_{n}\right|$, where the subscript $B$ refers to the direction of the $B$ field in the massive spin-2 frame. The left-hand side of Eq.~(\ref{eq:halo_coupling_convers}) is nontrivial as it depends on the orientation of the cavity with respect to the GLP polarization and propagation.
For simplicity, we consider that $\left|2\varepsilon_{Bj}\epsilon_{Bjk}\hat k^\mathrm{G}_k\right| = \mathcal{O}(1)$ such that the ALP-GLP coupling conversion factor becomes
\begin{align}\label{eq:conversion_haloscopes}
    \frac{\alphaG}{\MP}\bigg|_\text{GLP,haloscope} & \sim \frac{\gag}{v_\mathrm{DM}} \bigg|_\text{ALP,haloscope}\, ,
\end{align}
\ie\ haloscopes are less sensitive to GLPs than to ALPs of the same mass by a factor of $v_\mathrm{DM}\approx 10^{-3}$. Note that massive gravitons, through their coupling with matter, would also induce a force and a stress on the cavity walls. This would perturb the EM boundary conditions, similarly to GWs, see e.g.~\cite{Gue:2026kga}, and lead to an additional signal. However, compared with the EM signal above, such a contribution is suppressed because of the nonrelativistic nature of the GW. 

\subsubsection{Magnetometers}

Beyond haloscopes, several existing experiments also exploit static magnetic fields to convert DM  ALPs into photons. However, rather than directly detecting the resulting photon in a resonant cavity mode, these searches probe the small oscillating magnetic field generated by the effective current induced by the ALP field. As for haloscopes, these experiments probe small ALP masses such that their associated DM wave can be approximated as homogeneous over the scale of the experiment. Therefore, the difference in the received magnetic flux between GLPs and ALPs arises solely from differences in the effective EM current. One can separate the magnetometers in two categories: those using a background toroidal magnetic field, like ABRACADABRA~\cite{Kahn:2016aff, Ouellet:2019tlz}, SHAFT~\cite{Gramolin:2020ict}, and the future  DMRadio-50L~\cite{DMRadio:2022pkf} and DMRadio-GUT~\cite{DMRadio:2022jfv} magnetometers; and those using a solenoidal magnetic field, like the future WISPLC~\cite{Zhang:2021bpa} and DM-radio-m$^3$~\cite{DMRadio:2022pkf}.

In the toroidal setups, the magnetic field is not uniform. Because the DM wavelength is much larger than the scale of the experiment, the first term of Eq.~\eqref{eq:j_glps} dominates. Ultimately, considering a toroidal magnetic field with symmetry axis along $\hat z$, the observable is the magnetic flux that flows through a pickup coil oriented in the $x-y$ plane, sourced by such current, see e.g.~\cite{Domcke:2022rgu}
\begin{align}
    \Phi^z_\text{a,G} = \int_\text{loop} d^2\vec{r}{\:'} B^z_\text{a,G}(\vec r{\:'}) &= \int_\text{loop} d^2\vec{r}{\:'}\int_\text{toroid} d\mathcal{V} \frac{ (j^\rho_\text{a,G} \hat e_\phi -j^\phi_\text{a,G} \hat e_\rho)\cdot (\vec r{\:'}-\vec r)}{4\pi|\vec r{\:'}-\vec r|} \, .
\end{align}
In general, the pickup loop used for the detection is perfectly circular. In this case, one can show that the dominant contribution to $j_\text{G}$ does not lead to any signal. The same result arises for GW, as discussed in~\cite{Domcke:2022rgu}. Then, the only remaining term is the second one in Eq.~\eqref{eq:j_glps}, and again neglecting $\mathcal{O}(1)$ geometric factors that are related to the relative orientation between the magnetic field and the GLP polarization and propagation directions, we find the same ALP-GLP coupling conversion factor as with haloscopes, Eq.~\eqref{eq:conversion_haloscopes}, namely
\begin{align}\label{eq:conversion_magnetometer_toroidal_circ_loop}
    \frac{\alphaG}{\MP}\bigg|_\text{GLP, tor.\ magnetom., circ.\ loop} & \sim \frac{\gag}{v_\mathrm{DM}} \bigg|_\text{ALP, tor.\ magnetom., circ.\ loop}\,.
\end{align}
\ie\ these magnetometers are about three orders-of-magnitude less sensitive to GLP than to ALP signals of the same mass.

The conversion factor (\ref{eq:conversion_magnetometer_toroidal_circ_loop}) will be applied for already existing magnetometers that employ toroidal magnetic fields, such as ABRACADABRA and SHAFT. 
%As pointed out in Ref.~\cite{Domcke:2022rgu}, by slightly modifying the shape of the pickup loop, in particular considering the so-called ``figure-8'' configuration, made of two oppositely oriented semicircles, the leading order term in Eq.~\eqref{eq:j_glps} remains.
As pointed out in Ref.~\cite{Domcke:2022rgu}, the leading-order term in Eq.~\eqref{eq:j_glps} can be preserved by slightly modifying the geometry of the pickup loop, in particular by adopting the so-called ``figure-8'' configuration, consisting of two oppositely oriented semicircles. In that case, comparing it with Eq.~\eqref{eq:j_alps} for ALPs, the conversion factor reads
\begin{align}\label{eq:conversion_magnetometer_toroidal_fig8}
    \frac{\alphaG}{\MP}\bigg|_\text{GLP, tor.\ magnetom., fig.\ 8} & \sim \gag \cdot(\omega_\text{G}d_B)\bigg|_\text{ALP, tor.\ magnetom., fig.\ 8}\,,
\end{align}
where $d_B$ is the typical transverse size of the magnet. 
Future experiments employing toroidal magnets, such as DMRadio-50L and DMRadio-GUT, will respectively use magnets with an approximate radius $d_B \approx 0.1\mbox{--}1$~m, equivalent to $d_B \approx 5\cdot 10^5\mbox{--}5\cdot 10^6$~eV$^{-1}$ in natural units, and probe masses $\ma=\mG \approx 10^{-11}\mbox{--}10^{-6}$~eV. Therefore, those experiments have an enhanced sensitivity to GLPs compared with ALPs of order $\lambda_G/d_B \sim 1\mbox{--}10^6$, where $\lambda_\text{G}=2\pi/\mG$ is the Compton wavelength of the field. Therefore, we will assume these experiments will perform an optimized data-taking strategy for GW detection, as proposed in~\cite{Domcke:2022rgu}, employing the ``figure-8'' configuration, such that the conversion factor in Eq.~\eqref{eq:conversion_magnetometer_toroidal_fig8} can be applied.

In the solenoidal case, the magnetic field is uniform over the region of detection and therefore, one can use the same ALP-GLP coupling conversion factor, as with toroidal magnetometers employing circular pickup loops, namely
\begin{align}\label{eq:conversion_magnetometer_solenoidal}
    \frac{\alphaG}{\MP}\bigg|_\text{GLP, sol.\ magnetom.} & \sim \frac{\gag}{v_\mathrm{DM}} \bigg|_\text{ALP, sol.\ magnetom.}\,.
\end{align}

\subsubsection{Conversion in Earth's magnetic field}

If galactic DM is composed of either ALPs or GLPs, their conversion to photons in the Earth's static magnetic field can also be expected. This kind of axion searches have been performed by several experiments: Snipe Hunt~\cite{Sulai:2023zqw}, Eskdalemuir~\cite{Nishizawa:2025xka}, and SuperMAG~\cite{Arza:2021ekq,Friel:2024shg}. The main difference with the previous experiments considered in this section is the spatial form of the magnetic field. In particular, the geomagnetic field $\vec B_E= -\nabla V_0$, where $V_0$ is a scalar potential that is a sum of spherical harmonics $Y^m_\ell(\theta,\phi)$~\cite{Arza:2021ekq}, such that the induced EM currents are 
\begin{subequations}
    \begin{align}
        \vec j_a &= \omega_\mathrm{a} \gag a \vec B_E \, \\
        j^i_\mathrm{G} &= \frac{2\alphaG }{\MP}\left(\varepsilon_{jk}\epsilon_{lki}G_0\partial_j \partial_l V_0-\epsilon_{ljk} \varepsilon_{ij}(\partial_l V_0)(\partial_k G_0)\right) \label{eq:current_g_Earth}\, ,
    \end{align}
\end{subequations}
for the ALP and GLP cases, respectively. Due to the form of $V_0$, we have $\partial_i V_0 \sim V_0/r$, while $\partial_k G_0\sim k^k_\text{G} G_0$, where $k_\text{G}$ is the GLP wavenumber. Because these experiments probe low-mass DM particles, where the field is homogeneous over the Earth radius $R_{\Earth}$, one has $k_G R_{\Earth} \ll 1$, and the first term of Eq.~\eqref{eq:current_g_Earth} dominates. Neglecting $\mathcal{O}(1)$ geometric factors, the comparison of axion and massive graviton currents leads to the following coupling conversion factor between both types of particles,
\begin{align}
    \frac{\alphaG}{\MP}\bigg|_\text{GLP, geomag.} &\sim \gag\cdot(\omega_\mathrm{G} R_{\Earth})\bigg|_\text{ALP, geomag.} \, .
\end{align}
In natural units, $R_{\Earth} \approx 3\cdot10^{-13}$~eV$^{-1}$, such that $(\omega_\mathrm{G} R_{\Earth}) \sim 1$ and searches for DM with masses $\mG = \omega_\mathrm{G}\approx 10^{-14}$~eV, based on conversion in the Earth magnetic field, are roughly equally sensitive to ALPs and GLPs.

%%%%%%%%%%%%%%%%%%%%%%%%%%%%%%%%%%%%%%%%%%%%%%%%%%%%%%%%%%%%%%%%%%%%%%%%%%%%%%%%%%%%
\subsection{Laboratory experiments involving an alternating electromagnetic background}

Setups that aim at detecting DM ALPs by exploiting its impact on a spacetime-dependent EM background generated by a laser (via two-beam interferometry), or an eigenmode of an electromagnetic cavity (via upconversion), are discussed next.

\subsubsection{Two-beams interferometry}

Different experiments aim to detect axion-photon conversion via light birefringence, \ie\ by the observation of the different phase velocity of light depending on its polarization, induced by a DM axion background. Specifically, starting from the modified Maxwell's equations in the presence of an axion, the equation of motion for the EM vector potential, $A_i$, is~\cite{Obata:2018vvr} 
\begin{align}\label{eq:EOM_EM_a}
    \partial^2_t A_i -\nabla^2 A_i + \gag(\partial_t a)\epsilon_{ijk}\partial_j A_k &=0 \, ,
\end{align}
as written in temporal and Coulomb gauges, and assuming that the axion field is homogeneous over the scale of the detector. This leads to a modification of the phase velocity of circularly polarized light of the form~\cite{Obata:2018vvr},
\begin{align}\label{eq:wp_a}
    \omega_\gamma &=  k_\gamma\sqrt{1\pm \frac{\gag\partial_t a}{k_\gamma}} \approx k_\gamma \mp  \frac{ \gag\omega_\mathrm{a} a_0 }{2}\sin(\omega_\mathrm{a} t +\phi_\text{a})\, ,
\end{align}
where $\omega_\gamma$ and $k_\gamma$ refer, respectively, to the frequency and the momentum of light, the $\pm$ sign depends on the (left or right) helicity of the background polarization, and the last approximation assumes $\gag\partial_t a/k_\gamma \ll 1$. This effect can be probed in cavities by measuring the phase velocity of light of different circular polarizations, see e.g.~\cite{Obata:2018vvr, Pandey:2024dcd, Heinze:2023nfb}, and has also been proposed as a potential target for GW detectors~\cite{Nagano:2019rbw,Gue25,Yao25, YaoR25}. Ultimately, one accesses this signal via interferometric measurement between two optical lasers. The phase shift between two optical lasers of respectively left and right helicities after traveling the same distance $L$ reads
\begin{align}\label{eq:delta_phi_a}
    \Delta \varphi_\text{a}(t) &= \int^t_{t-L} \mathrm{d}t'\Delta \omega_\gamma(t') = 2\gag a_0\sin\left(\frac{\omega_\text{a} L}{2}\right)\sin\left(\omega_\text{a} \left(t-\frac{L}{2}\right)+\phi_\text{a}\right)\,.
\end{align}

For the case of massive gravitons, a similar equation to (\ref{eq:wp_a}) can be derived from the interaction Lagrangian of the massive gravitons with the EM energy momentum tensor, Eq.~\eqref{eq:int_lagrangian}, 
\begin{align}\label{eq:Lint_glps}
    \mathcal{L}_\mathrm{int} &= \frac{\alphaG}{\MP}G^{\mu\nu}(-F_{\mu\rho}F^{\: \rho}_{\nu}+\frac{1}{4}g_{\mu\nu}F_{\alpha\beta}F^{\alpha\beta}) \, .
\end{align} 
At first order in perturbation theory, the second term disappears due to the traceless condition of $G$.
The full equation of motion of light in vacuum (including the free Lagrangian) reads
\begin{align}\label{eq:EOM_EM}
    \partial^\alpha\partial_\alpha A^\beta - \partial^\beta \partial_\alpha A^\alpha+\frac{2\alphaG}{\MP}\left[G^{\alpha\nu}(\partial_\alpha (\partial_\nu A^\beta -\partial^\beta A_\nu)-(\partial_\alpha G^{\beta\nu})(\partial_\nu A^\alpha - \partial^\alpha A_\nu)\right] &=0 \, ,
\end{align}
In the temporal and Coulomb gauges, and neglecting $G_{0\mu}$ components, the spatial part of this equation takes the form
\begin{align}
 \partial^\alpha\partial_\alpha A_i+\frac{2\alphaG}{\MP}\big[G_{jk}\partial_j\partial_k A_i - G_{kj}\partial_k\partial_i A_j - \partial_j G_{ik}\partial_k A_j + \partial_\alpha G_{ij}\partial^\alpha A_j\big] &= 0\, ,
\end{align}
while the corresponding physical results remain gauge independent.
In general, in experiments aiming at detecting vacuum birefringence, the EM field is in the form of an optical laser whose frequency is several orders-of-magnitude larger than the DM frequency of interest. Therefore, the third and fourth terms inside the brackets are highly suppressed. Taking the plane-wave ansatz for the vector potential, $\vec A = \vec A_0 e^{-i(\omega_\gamma t-\vec k_\gamma \cdot \vec x)}$, the second term in the bracket also disappears due to the orthogonality condition $\vec k_\gamma \cdot \vec A=0$. Therefore, the equation of motion simply becomes, 
\begin{align}
    \partial^2_t A_i - \nabla^2 A_i +\frac{2\alphaG k^2_\gamma}{\MP}G_{jk}\hat k^j_\gamma\hat k^k_\gamma A_i &= 0 \, .
\end{align}
The solution of this equation leads to a change of the dispersion relation of light in the presence of a GLP background, of the form,
\begin{align}\label{eq:wp_g}
    \omega_\gamma &= k_\gamma\sqrt{1+\frac{2\alphaG}{\MP}G_{jk}\hat k^j_\gamma \hat k^k_\gamma} \approx k_\gamma + \frac{\alphaG k_\gamma G_0}{\MP}\varepsilon_{jk}\hat k^j_\gamma \hat k^k_\gamma \cos(\omega_\mathrm{G} t+\phi_\mathrm{G}) \, ,
\end{align}
which is the equivalent of Eq.~(\ref{eq:wp_a}) for ALPs. Finally, the phase shift induced by the GLP background is 
\begin{align}\label{eq:delta_phi_G}
    \Delta\varphi_\text{G}(t) &= \int_{t-L}^t \mathrm{d}t' \: \Delta\omega_\gamma (t')= \frac{2\alphaG k_\gamma}{\MP\omega_\mathrm{G}}\varepsilon_{jk}(\hat k^j_{\gamma_1}\hat k^k_{\gamma_1}-\hat k^j_{\gamma_2}\hat k^k_{\gamma_2})G_0\sin\left(\frac{\omega_\mathrm{G} L}{2}\right)\cos\left(\omega_\mathrm{G} \left(t-\frac{L}{2}\right)+\phi_\mathrm{G}\right) \, ,
\end{align}
where $\gamma_1,\gamma_2$ refer to the photons from the beam 1 and 2, respectively. There are several important implications of this result compared with the axion case described by Eq.~(\ref{eq:delta_phi_a}). For the ALP, the modification of dispersion relation of light appears at linear order only for circularly polarized light\footnote{Vacuum birefringence induced by ALPs can also be probed using a linearly polarized light beam, by detecting photons whose polarization has rotated into the orthogonal linear state (see, e.g.~\cite{Nagano:2019rbw}); however, this approach requires dedicated configurations of optical benches.}. 
For GLPs, no assumptions on the background light polarization state have been made to obtain Eq.~\eqref{eq:wp_g}, \ie\ every polarization feels the same effect, and therefore the phase shift Eq.~\eqref{eq:delta_phi_G} does not depend on the polarization of the background. This is consistent with the fact that, in standard GR, GWs (and, by extension, massive spin-2 fields) do not induce light birefringence, but only a phase shift of light. This means that optical interferometers employing linear polarization are naturally sensitive to GLPs. However, the downside is that if the two arms of the interferometer are parallel, the effect cancels as $\varepsilon_{jk}(\hat k^j_{\gamma_1}\hat k^k_{\gamma_1}-\hat k^j_{\gamma_2}\hat k^k_{\gamma_2})=0$. 
If the two arms of the interferometer are close to be orthogonal, which happens for instance with Michelson-like interferometers, the factor $\varepsilon_{jk}(\hat k^j_{\gamma_1}\hat k^k_{\gamma_1}-\hat k^j_{\gamma_2}\hat k^k_{\gamma_2})$ is a priori not small compared to unity. This means that, by comparing the amplitude of the phase shift given by Eqs.~\eqref{eq:delta_phi_a} and \eqref{eq:delta_phi_G}, one can derive the coupling conversion between the two types of signals as 
\begin{align}\label{eq:birefringence_corresp_a_g}
    \frac{\alphaG}{\MP}\bigg|_\text{GLP, ortho. Michelson-like interf.} \sim \gag\cdot\left(\frac{\omega_\mathrm{G}}{k_\gamma}\right)\bigg|_\text{ALP, ortho. Michelson-like interf.}\,,
\end{align}
where we have considered $\varepsilon_{jk}(\hat k^j_{\gamma_1} \hat k^k_{\gamma_1} - \hat k^j_{\gamma_2} \hat k^k_{\gamma_2}) = \mathcal{O}(1)$. This latter geometric factor, which depends on the relative arm angle and on the direction of propagation of light in the spin-2 frame (and is therefore highly dependent on the spacetime position where the experiment is performed), should not be much smaller than unity in most actual realizations. In conclusion, we find that the birefringence signal induced by GLPs is enhanced by a factor of $k_\gamma/\omega_\mathrm{G} \approx 10^{20} \cdot (10^{-20}\mathrm{eV}/\mG)$ compared with the ALPs case, assuming a background optical laser.

Other experiments aiming at detecting ALPs through birefringence do not use orthogonal arms optical interferometers. One can separate them into two categories: detectors such as DANCE in its original proposal~\cite{Obata:2018vvr}, which use a longitudinally stretched bow-tie cavity and measure the phase-shift induced by axion DM between left and right circularly polarized beams with different optical paths; and detectors such as LIDA~\cite{Heinze:2023nfb}, DarkGEO~\cite{Heinze:2024bdc}, ADBC~\cite{Pandey:2024dcd}, and the first DANCE data-taking~\cite{Oshima:2023csb},  which use linearly polarized light and measure the sidebands of orthogonal polarization sourced by the axion field through polarimetric heterodyne readout. In the former case, the conversion factor depends on the geometric factor $\hat k^j_{\gamma_1}\hat k^k_{\gamma_1}-\hat k^j_{\gamma_2}\hat k^k_{\gamma_2}$, \ie\ whether the photons in each beam travel along the same direction or not. Bow-tie cavities, as proposed in~\cite{Obata:2018vvr,Oshima:2023csb} are stretched in the longitudinal direction, such that defining the length of the bow-tie to be $L$ and the width to be $\ell$, with $L \gg \ell$, we have $\varepsilon_{jk}(\hat k^j_{\gamma_1}\hat k^k_{\gamma_1}-\hat k^j_{\gamma_2}\hat k^k_{\gamma_2}) \;\propto\; \ell/L +\mathcal{O}(\ell/L)^3$. Then, the conversion factor becomes
\begin{align}\label{eq:Circ_pol_corresp_a_g}
    \frac{\alphaG}{\MP}\bigg|_\text{GLP, long.\ bow-tie, 2-beams interf.} \sim \gag\cdot\left(\frac{L}{\ell}\right)\cdot\left(\frac{\omega_\mathrm{G}}{k_\gamma}\right)\bigg|_\text{ALP, long.\ bow-tie, 2-beams interf.}\,.
\end{align}
For our estimates, we will assume $L/\ell = 100$. Since such an experiment works at relatively low DM masses, it features a significantly enhanced sensitivity to massive gravitons, despite the additional correction factor $\ell/L\ll 1$.

For the latter case of polarimetric heterodyne readout, in the current version of the detectors, one rotates the polarization of the pump field using a half-wave plate for the interference to work in the axion case. However, since spin-2 fields do not induce birefringence, the pump and signal fields have identical polarization, preventing their separation, and thus spoiling the spin-2 measurement. Therefore, currently, experiments such as LIDA, DarkGEO, and ADBC are insensitive to massive gravitons (and by extension to GWs). Such detectors could become sensitive to spin-2 fields by, \eg\ removing the half-wave plate. However, this would increase the expected noise level, since it would also compromise the common-mode suppression inherent to the polarimetric heterodyne scheme. Therefore, we will assume that at that level, no simple conversion factor can be estimated, and we leave for future work a more thorough analysis of such an experimental setup.

\subsubsection{Upconversion}

Experiments that search for axions via upconversion in electromagnetic cavities rely on a concept closely related to that of haloscopes but, instead of a magnetostatic field, the background EM field is provided by a pump mode $\vec{E}_0$ oscillating at frequency $\omega_0$. In this case, the axion field with frequency $\omega_\mathrm{a}$ can, if coupled to the EM field, populate a signal mode $\vec E_1$ at frequency $\omega_1=\omega_0 \pm \omega_\mathrm{a}$. This is the idea of \eg \textsc{upload}~\cite{Thomson:2023moc}, but other experiments of this kind exist such as the anyon cavity conversion~\cite{Bourhill:2022alm}, and the superconducting radio frequency (SRF) cavity proposal from~\cite{Li:2025pyi}. In addition, this setup is used to probe axions with much lower frequencies than with haloscopes, \ie\ one typically has $\omega_\mathrm{a} \ll \omega_0,\omega_1$.

In the case of a spacetime-dependent background, the effective current induced by axions remains unchanged at leading order and is given by Eq.~\eqref{eq:j_alps}, whereas the EM current induced by massive gravitons becomes
\begin{align}\label{eq:eff_j_heterodyne}
    j^i_\mathrm{G} &= -\frac{2\alphaG}{\MP}\left(\epsilon_{ilk}\partial_jB^l_0G_{jk}+\omega_\mathrm{G}G_{ij}E^j_0\right) \, ,
\end{align}
where we neglected terms suppressed by powers of $v_\text{DM}$, which we assume to be negligible, as described below. Then, the right-hand side of Eq.~\eqref{eq:eom_mode} reads
\begin{subequations}
    \begin{align}
        \omega_0 \omega_\mathrm{a}\gag a \frac{\int \mathrm{d}\mathcal{V} \vec B_0 \cdot \vec E_1}{\int \mathrm{d}\mathcal{V} |\vec E_1|^2} &\quad \text{for the ALP,} \\
        -\frac{2\alphaG}{\MP}\omega_0\left(\epsilon_{ilk}G_{jk}\frac{\int \mathrm{d}\mathcal{V} E^i_1\partial_j B^l_0}{\int \mathrm{d}\mathcal{V} |\vec E_1|^2}+\omega_\mathrm{G} G_{ij}\frac{\int \mathrm{d}\mathcal{V} E^i_{0}  E^j_{1}}{\int \mathrm{d}\mathcal{V} |\vec E_1|^2}\right) &\quad \text{for the GLP.} \label{eq:spin2_heterodyne_source_EM}
    \end{align}
\end{subequations}
The overlap integrals above depend on the specific choice of pump and signal modes. In the axion case, maximal sensitivity is achieved when the polarizations of the pump magnetic field and of the signal electric field are aligned.
On the other hand, since the second term of Eq.~\eqref{eq:spin2_heterodyne_source_EM} is suppressed by $\omega_\mathrm{G}/\omega_0 \ll 1$, maximum sensitivity to massive gravitons is reached when the pump magnetic and signal electric modes are orthogonal. In addition, in the case of massive gravitons, the sensitivity also depends on the projection of the spin-2 polarization tensor and wavevector on the experiment. For our estimates, we consider two kinds of experiments. For already existing experiments, such as \textsc{upload}, which have used collinear modes for maximum sensitivity to ALPs, we only consider the second term of Eq.~\eqref{eq:spin2_heterodyne_source_EM}, as the first one vanishes. In this case, making the same approximation as for haloscopes, \ie\ $|\int \mathrm{d}\mathcal{V} \vec B_0 \cdot \vec E_1|\sim 2| \varepsilon_{ij}\int \mathrm{d}\mathcal{V}  E^i_0 E^j_1|$, the conversion factor between GLP and ALP couplings is\footnote{Note that $\omega_\mathrm{G}/\omega_0 \gg v_\text{DM}$ for \textsc{upload}, thereby justifying the neglect of terms linear in $v_\text{DM}$ in Eq.~\eqref{eq:eff_j_heterodyne}.}
\begin{align}
    \frac{\alphaG}{\MP}\bigg|_\text{GLP,\textsc{upload}}&\sim \gag\bigg|_\text{ALP,\textsc{upload}} \, . 
\end{align}
Therefore, in this specific upconversion setup, one obtains the same sensitivity to ALPs and GLPs signals.

For planned experiments such as the anyon resonator~\cite{Bourhill:2022alm,Li:2025pyi}, as well as for future \textsc{upload} operation, we assume that an optimized data-taking can be performed for massive gravitons search. This requires the use of different pump and signal modes compared with the axion case, as the same set of modes cannot simultaneously maximize both signals. This could also modify the frequency band to which the experiment is sensitive. However, those proposed experiments are broadband, suggesting that the frequency of the pump and signal modes are already assumed to be shifted continuously during operation. In this optimized case, the coupling conversion factor is improved by
\begin{align}
    \frac{\alphaG}{\MP}\bigg|_\text{GLP, heterodyne, $B_0\perp E_1$}&\sim \gag\cdot\frac{\omega_\mathrm{G}}{\omega_0}\bigg|_\text{ALP, heterodyne, $B_0\perp E_1$} \, .
\end{align}
As in the vacuum birefringence case, the signal induced by the massive graviton is enhanced by a factor of $\omega_0/\omega_\mathrm{G}$, which can be large given the range of DM frequencies (kHz--MHz) targeted by such an experiment. 

The \textsc{upload} setup as well as the heterodyne detector proposal from~\cite{Li:2025pyi} operate at frequencies well above the fundamental mechanical resonances of a metallic resonator, while the anyon cavity probes very small DM masses, \ie\ much smaller than the typical fundamental acoustic mode of a metal. In both cases, because the contribution from the mechanical vibration is non resonant, we always neglect this contribution\footnote{Using~\cite{Gue:2026kga}, one can show that the ratio between the off-resonant mechanical contribution to the signal and the EM contribution scales like $v^2_\text{DM}$ when $\omega_\text{G} \gg \omega_m$ (relevant for~\cite{Li:2025pyi} and \textsc{upload}), where $\omega_m$ is the fundamental acoustic resonant frequency. This is analogous to the so-called freely falling limit in GW context, see \cite{Gue:2026kga}. When $\omega_\text{G}\ll \omega_m$ (relevant for the anyon cavity), this ratio is $\mathcal{O}(1)$, therefore the inclusion of the mechanical response will not change significantly our estimates.}. 
A more detailed description of the signal expected from massive gravitons could be considered following, \eg~\cite{Gue:2026kga}, and this will be the subject of a future work. 

\subsection{Astrophysical constraints}
\label{sec:DM_decay}

Dark matter can also be searched for by exploiting astrophysical or cosmological observations. In particular, in a scenario where a DM particle eventually decays into SM particles, most often photons because those are the only particles light enough to be kinematically accessible for low-mass DM candidates, distortions of their observed spectrum on Earth will occur with respect to the expected backgrounds. Several studies have constrained the axion parameter space from analyses of various astrophysical or cosmological signals~\cite{Caputo:2024oqc,Bolliet:2020ofj,Thorpe-Morgan:2020rwc,Foster:2021ngm,Wadekar:2021qae,Roach:2022lgo,Calore:2022pks,Carenza:2023qxh, Wang:2023imi,Todarello:2023hdk,Pinetti:2025owq,Sun:2023acy,Saha:2025any}.

The photon flux from DM decays depends linearly on its decay rate, which is the only parameter that will differ between different DM candidates. For ALPs, only the decay to two photons has been considered following the decay rate dependence on $\ma$ and $\gag$ given by Eq.~\eqref{eq:Gamma_ALP}. On the other hand, because massive gravitons couple universally to all types of matter and energy, one needs to consider all the different decay channels with their individual decay widths from Eq.~(\ref{eq:Gamma_G}). However, only final-state decay photons can be measured by the telescopes. Therefore, in principle, depending on the mass of the GLP, one should account for multiple decay channels of the form $\mathrm{G} \to \rm XX \to \gamma Y$, where $\rm X,Y$ are different types of SM particles (hadronic decays, in particular, will dominantly produce neutral pions that themselves will decay into a pair of photons). For our estimates, we only consider the $\mathrm{G} \to \gaga$ decay and neglect potential additional channels that produce photons in the final state. Therefore, our decay rate could be underestimated especially for very massive gravitons (Fig.~\ref{fig:GLP_BRgamgam}). From the two decay rates to photons in Eqs.~\eqref{eq:Gamma_G} and \eqref{eq:Gamma_ALP}, we find $\sqrt{\Gamma_{\mathrm{a}\to \gaga}/\Gamma_{\mathrm{G}\to \gaga}}=\sqrt{5}/2$, \ie\ the ALP-GLP coupling conversion factor is found to be close to unity,
\begin{align}
    \frac{\alphaG}{\MP}\bigg|_\text{GLP,DM decay} &= \frac{\sqrt{5}}{2}\gag\bigg|_\text{ALP,DM decay} \, ,
\end{align}
which is exact for $m_\mathrm{DM}\lesssim 10^{8}$~eV and, above this mass, not a bad approximation if one takes into account the expected dominant hadronic decays of ALPs and GLPs that, ultimately, also produce photons via neutral pion decays. Note that, as stated previously, these bounds are meaningful only if the total decay width of the given DM particle is $\Gamma_\text{DM} \ll 1/t_\text{U}$, such that DM is cosmologically stable and has not been completely depleted today (Fig.~\ref{fig:lifetime}).

%%%%%%%%%%%%%%%%%%%%%%%%%%%%%%%%%%%%%%%%%%%%%%%%%%%%%%%%%%%%%%%%%%%%%%%%%%%%%%%%%%%%
\section{Non-dark-matter massive gravitons}
\label{sec:notDM}

In this section, the requirement of ALPs/GLPs to constitute a DM candidate is not imposed, and the production of both BSM weakly interacting particles proceeds via scattering with SM particles in astrophysical environments, such as the Sun, globular cluster (GC) stars, or supernovae (in particular, SN1987A).

%%%%%%%%%%%%%%%%%%%%%%%%%%%%%%%%%%%%%%%%%%
\subsection{Helioscopes}

Helioscopes such as CAST~\cite{CAST:2017uph} and the future IAXO~\cite{IAXO:2019mpb} employ powerful magnets aligned with the Sun to search for axions produced in the solar core, which are subsequently reconverted into photons via the inverse Primakoff process within the magnetic field of the experiment (Fig.~\ref{fig:diags}, top center). Unlike haloscope searches, helioscope experiments do not rely on axions constituting the local DM density. The photon flux produced in the detector is parameterized as
\begin{align}
    \Phi_\gamma &= \Phi_\mathrm{a,G} \times P_{\mathrm{a,G}\to \gamma} \, ,
\end{align}
where $\Phi_\mathrm{a,G}$ is the ALP or GLP flux from the Sun and $P_{\mathrm{a,G} \to \gamma}$ is the conversion probability inside the detector.
In the most general case of a homogeneous magnetic field of strength $B_0$ over a distance $L$, the axion conversion probability into photons is given by~\cite{Sikivie:2020zpn}
\begin{align}
    P_{\mathrm{a}\to \gamma}=\frac{\gag^2 B^2_0}{v_\mathrm{a}}\sin^2\left(\frac{qL}{2}\right)\frac{1}{q^2} \, ,
\end{align}
where $v_\mathrm{a}$ represents the speed of the incident axion, and
 $q=|\vec k_\gamma -\vec k_\mathrm{a}|$ is the absolute difference between the photon and axion momentum vectors, which can be written as
$q=\omega_{\gamma}-\sqrt{\omega_{\gamma}^2-\ma^2}\equiv \omega_{\gamma}\left(1-\sqrt{1-\ma^2/\omega_{\gamma}^2}\right)$ when the conversion is collinear (valid in the relevant $\omega_{\gamma} L \gg 1$ regime). The conversion happens when the energies of both particles are the same (because the process takes place within a static magnetic field), \ie\ when the ALP and photon frequencies are identical. The helioscope conversion is resonant as long as $q\ll 1/L$. For instance, CAST~\cite{CAST:2017uph} is sensitive to X-ray photons with energies $\omega_\gamma\approx 4$~keV~\cite{Sikivie:2020zpn}, corresponding to the plasma energy inside the solar core~\cite{Guarini:2020hps}, for a magnetic field spanning a distance of $L \approx 9.3$~m. In this case, the resonant condition occurs for $\ma \lesssim 10^{-2}$~eV, with a decreased experimental sensitivity above this signal mass. This implies $\ma \ll \omega_{\gamma}$, \ie\ $q \approx \ma^2/(2\omega_{\gamma})$, \ie\ the axions are relativistic, $v_\mathrm{a} \sim 1$. The conversion probability then becomes
\begin{align}\label{eq:P_agamma_CAST}
    P^\mathrm{CAST}_{\mathrm{a}\to \gamma}\approx\frac{\gag^2B^2_0 L^2}{4}\, .
\end{align}

For the massive graviton case, still in the $q \ll 1/L $ and $\omega_{\gamma} L \gg 1$ regime, the direction of propagation of the incoming GLP and the produced photon are the same, and the GLP-to-photon conversion probability is given by~\cite{Garcia-Cely:2025ula}
\begin{subequations}
\begin{align}
    P^\mathrm{CAST}_{\mathrm{G}\to \gamma} &\approx \frac{\alphaG^2B^2_0 L^2}{2\MP^2} \, ,
\end{align}
\end{subequations}
for each of the tensor modes. The vector modes decouple completely because they can only produce photons with a propagation direction different from that of the incoming massive graviton; and the scalar mode will lead to subdominant signals, as explained below.

The flux of massive gravitons from the solar core is computed by considering both bremsstrahlung and Primakoff processes for scalar and tensor modes. As shown in Ref.~\cite{Garcia-Cely:2025ula}, the flux from the tensor modes dominates over the scalar flux. At the photon energy sensitivity corresponding to the solar core, $\omega_\gamma \approx 4$~keV, the conversion factor between the solar fluxes of axion and massive gravitons is $\alphaUniv \sim \gag$. Thus, accounting for the photon conversion probability and solar fluxes, the ALP-GLP coupling conversion factor for helioscopes reads 
\begin{align}
    \frac{\alphaG}{\MP}\bigg|_\text{GLP, helioscope} \sim \frac{\gag}{\sqrt{2}}\bigg|_\text{ALP, helioscope} \,,
\end{align}
namely, helioscopes are roughly equally sensitive to ALP and GLP signals.

\subsection{Astrophysical constraints}

Astrophysical environments, such as stars, supernovae, and galactic centers, offer alternative laboratories for probing the production of BSM particles. When kinematically accessible, ALPs and/or GLPs can be copiously produced in the hot and dense plasma present within or around such environments, often through the Primakoff/Gertsenshtein process. Depending on their masses and coupling to photons and other SM particles, these particles can either result in modifications in the photon spectrum and/or flux observed on Earth conflicting with standard sources, or carry away a significant amount of energy and thereby alter stellar evolution.

\subsubsection{Modulation of photon spectra}

In astrophysical environments with external magnetic fields present, such as galactic centers or blazars, photons can quantum-mechanically interconvert into ALPs/GLPs via the Primakoff effect, inducing irregularities in their power-law spectrum observed on Earth. The relevant observable is the conversion probability $P_{\gamma\to \mathrm{a,G}}$ and, in analogy with neutrino oscillations, its energy dependence would imprint characteristic modulations on the observed photon spectrum. Spectral distortions induced by axion-photon mixing have been extensively searched for, \eg\ in X- and gamma-ray spectra~\cite{Wouters:2013hua, Marsh:2017yvc, Reynolds:2019uqt, Reynes:2021bpe, Fermi-LAT:2016nkz, Davies:2022wvj, MAGIC:2024arq, Li:2020pcn, Li:2021gxs, Li:2024zst} setting bounds on axions of $\ma\lesssim 10^{-7}$~eV. The conversion probability of photons to ALPs in a plasma background patch of size $L$ with constant transverse magnetic field $B_0$ is only possible for photons whose polarization is along the $B_0$ direction, and reads~\cite{Raffelt:1987im}
\begin{align}\label{eq:prob_conv_spin0_plasma}
    P_{\gamma_\parallel\to \mathrm{a}} &=  \frac{4\omega_{\gamma}^2 B^2_0\gag^2}{(\omega^2_{\gamma\,\text{pl},\parallel}-\ma^2)^2}\sin^2\left(\frac{(\omega^2_{\gamma\,\text{pl},\parallel}-\ma^2)L}{4\omega_{\gamma}}\right)\, ,
\end{align}
where $\omega_{\gamma\,\text{pl},\parallel}$ is the plasma frequency in the direction parallel to the magnetic field, which plays the role of an effective mass for the photon.
When the argument inside the sine is small, \ie for $\ma \approx \omega_{\gamma\,\text{pl},\parallel} \approx 10^{-11}$~eV~\cite{Wouters:2013hua}, the conversion probability becomes equivalent to
its resonant expression in vacuum, %that in vacuum on resonance, 
Eq.~\eqref{eq:P_agamma_CAST}.
Note that for very light ALPs, with $\ma \ll \omega_{\gamma\,\text{pl}}$, the conversion probability also becomes independent of their mass.

In the case of massive gravitons, their different polarizations do not couple identically to photons~\cite{Biggio:2006im, Garcia-Cely:2025ula}. In particular, the vector modes decouple as a consequence of angular momentum conservation. In the limit of small couplings, $(\alphaUniv)B_0 \ll (\mG^2-\omega^2_{\gamma\,\text{pl}})/2\omega_{\gamma}$, which is fulfilled assuming typical magnetic field-strength $B_0 \sim \mu G$ in galaxy centers~\cite{Marsh:2017yvc} or around single magnetic white dwarfs with $B_0 \sim 10^8~G$~\cite{Dessert:2022yqq}, the conversion probabilities are~\cite{Biggio:2006im, Garcia-Cely:2025ula}
\begin{subequations}\label{eq:prob_conv_spin2_plasma}
    \begin{align}
        P_{\gamma_{\parallel} \to G_{\times2}} &= \frac{8\omega_{\gamma}^2B^2_0\alphaG^2}{\MP^2(\omega^2_{\gamma\,\text{pl},\parallel}-\mG^2)^2}\sin^2\left(\frac{(\omega^2_{\gamma\,\text{pl},\parallel}-\mG^2)L}{4\omega_{\gamma}}\right)  \, \\
        P_{\gamma_{\perp} \to G_{+2}} &=  \frac{8\omega_{\gamma}^2B^2_0\alphaG^2}{\MP^2(\omega^2_{\gamma\,\text{pl},\perp}-\mG^2)^2}\sin^2\left(\frac{(\omega^2_{\gamma\,\text{pl},\perp}-\mG^2)L}{4\omega_{\gamma}}\right) \, \\
        P_{\gamma_{\perp} \to G_{0}} &=\frac{8\omega_{\gamma}^2B^2_0\alphaG^2}{3\MP^2(\omega^2_{\gamma\,\text{pl},\perp}-\mG^2)^2}\sin^2\left(\frac{(\omega^2_{\gamma\,\text{pl},\perp}-\mG^2)L}{4\omega_{\gamma}}\right)
    \end{align}
where $\gamma_{\parallel, \perp}$ denote, respectively, photons with polarization parallel and perpendicular to the magnetic field, and $\omega_{\gamma\,\text{pl},\parallel},\omega_{\gamma\,\text{pl},\perp}$ indicate, respectively, the photon plasma frequency in the direction of the magnetic field and orthogonal to it. In addition, $G_{\times 2,+2,0}$ denote, respectively, the two tensor polarizations and the scalar one. 
\end{subequations}

Because of the smallness of the magnetic field-strength, $|B_0|e/(m_\mathrm{e}\omega_{\gamma}) \ll1$, \ie\ anisotropies in the refractive index can be neglected, and therefore $\omega_{\gamma\,\text{pl},\parallel} \sim \omega_{\gamma\,\text{pl},\perp}$. 
In this case, taking into account the scalar, vector, and tensor modes of massive gravitons, the ALP-GLP coupling conversion factor is 
\begin{align}
 \frac{\alphaG}{\MP} \bigg|_{\text{GLP,}\gamma\,\text{oscil.}} \sim \sqrt{\frac{3}{14}} \gag\bigg|_{\text{ALP,}\gamma\,\text{oscil.}}\,,
\end{align}
assuming that each spin-2 tensor component has equal energy density. Here, we neglect the potential signal contribution from the massive gravitons production through, \eg\ Bremsstrahlung in the core plasma of the cluster, which will then lead to oscillation to photons. In principle, this contribution could also produce modulations of the received spectrum, but the observable would not only be the survival probability of photons but also their absolute number in the spectrum. We leave a potential dedicated signal modeling and data analysis for future work.

While the magnetic field has in general a nontrivial spatial distribution in astrophysical environments, all conversion probabilities scale identically with it, such that its exact distribution is not needed when computing a conversion factor here. Note also that we assume that the distribution of the polarization of the incoming photons is isotropic, \ie\ that the number of photons with $\hat k_\parallel$ and $\hat k_\perp$ is the same. Indeed, the photons can be produced by an unpolarized process, not tied to the magnetic field, or they can be produced through a process directly related to the magnetic field. In the former case, it is expected that the distribution of polarization is isotropic in the transverse plane. In the latter case, the symmetry would be broken by the magnetic field, leading to one of the polarization being dominant over the other one. However, across the cluster, the magnetic field is not homogeneous and has a nontrivial profile with many different orientations along the line of sight. In this case, averaging over all the different patches with homogeneous magnetic field, no polarization state will convert preferably to massive gravitons.

\subsubsection{Anomalous X- and gamma-ray fluxes}

The production and emission rates of light weakly interacting particles in stellar interiors, through Primakoff and bremsstrahlung processes, leads to an additional flux of highly energetic photons (typically in the keV energy range).
At low masses, the production of ALPs is dominated by the Primakoff process and the ALP is relativistic, \ie\ with negligible mass with respect to the star temperature $T$. 
For all experiments under consideration, the energy of the measured photons is higher than the core temperature of the star where the ALPs/GLPs are produced. As shown in~\cite{Garcia-Cely:2025ula}, in this regime, the Primakoff production rate of ALPs is approximately equivalent to that of GLPs, taking into account all polarizations.   
Using that the conversion probability Eqs.~\eqref{eq:prob_conv_spin0_plasma} and \eqref{eq:prob_conv_spin2_plasma} are equivalent, up to $\mathcal{O}(1)$ factors, the coupling conversion factor becomes 
\begin{align}
 \frac{\alphaG}{\MP} \bigg|_\text{GLP, anomalous flux} \sim \gag\bigg|_\text{ALP, anomalous flux}\,,
\end{align}
which can be used to reinterpret the constraints derived in~\cite{Xiao:2020pra, Hoof:2022xbe,Calore:2021hhn,Dessert:2020lil,Ning:2024eky,Meyer:2020vzy,Manzari:2024jns,Fiorillo:2025gnd}.

\subsubsection{Other astrophysical  constraints}

The production of new weakly interacting particles in dense astrophysical environments ---such as GCs, supernovae, or stars--- followed by their free escape, provides additional energy-loss channels beyond standard cooling mechanisms driven by the emissions of photons (from the surface) and neutrinos (from the core). Even if this particle flux cannot be measured, the backreaction on stars can lead to observable consequences because such energy losses modify stellar evolution, which is tightly constrained by observations.
The strongest limits for generic photon-coupled ALPs come from observations of horizontal branch stars in GCs~\cite{Dolan:2022kul}, as well as from the absence of a $\gamma$-ray burst (from the photon-ALP conversion via the Primakoff effect in the star core, followed by photon conversion in the galactic magnetic field) in coincidence with a neutrino burst during the SN1987A explosion~\cite{Fiorillo:2025yzf}.

In particular, for $m_\text{a,G} \gtrsim 1$~keV, which correspond roughly to the temperature of the core plasma, the ALP/GLP mass is not negligible anymore, and the computations of stellar GLPs fluxes in~\cite{Garcia-Cely:2025ula} do not apply. Therefore, in order to convert the constraints derived, \eg\ in Refs.~\cite{Hoof:2022xbe,Caputo:2022mah, Muller:2023vjm,Lucente:2020whw,Beaufort:2023zuj,Fiorillo:2025yzf,Candon:2025ypl,Diamond:2023cto}, dedicated numerical calculations of the GLP production rates through various processes such as photon-photon scattering, bremsstrahlung, Compton, and electron-positron annihilation need to be performed, as done in~\cite{Cembranos:2017vgi}. Because analytical conversion factor cannot be obtained in this case, we do not treat them explicitly and leave them for future work. By comparing constraints on GLPs from~\cite{Cembranos:2017vgi} and on ALPs from e.g.~\cite{Vinyoles:2015aba, Dolan:2022kul,Caputo:2022mah}, we expect the conversion factor to be $\gag \sim N (\alphaUniv)$, with $N \approx 10\mbox{--}100$ depending on the system producing the ALPs/GLPs. Therefore, stellar production can lead to more stringent constraints for GLPs, as expected from their universal coupling to SM particles.

Limits on ALPs can also be obtained by considering their irreducible background production from the thermal plasma in the early universe~\cite{Langhoff:2022bij}. 
Such ALPs, which do not constitute the DM of our universe, can decay into photons and be constrained by DM decay searches, see Sec.~\ref{sec:DM_decay}. Importantly, this bound escapes the assumption of cosmologically stable ALPs, because there is no assumptions about DM here, and their irreducible ``freeze-in'' density is not constrained by observations. In order to compute the equivalent constraint for GLPs, one would need to compute their production rate in the thermal plasma considering all the processes mentioned above, which goes beyond the scope of the present study.

%%%%%%%%%%%%%%%%%%%%%%%%%%%%%%%%%%%%%%%%%%%%%%%%%%%%%%%%%%%%%%%%%%%%%%%%%%%%%%%%%%%%
\section{Heavy graviton searches at accelerators}
\label{sec:accel}

Heavy spin-0 pseudoscalar particles ($\ma\gtrsim 1$~MeV) coupling to photons have been extensively searched-for at accelerators, either in fixed-target or collider experiments~\cite{Blumlein:1990ay,Dobrich:2015jyk,Dobrich:2019dxc,Jerhot:2022chi,Harland-Lang:2019zur,Mimasu:2014nea,Jaeckel:2015jla,Brivio:2017ije,Bauer:2017ris,Agrawal:2021dbo, dEnterria:2021ljz,Biekotter:2025fll}. Similar massive spin-2 states decaying into photons have received comparatively less attention, aside from dedicated searches for TeV-scale ADD and RS gravitons at the LHC~\cite{Landsberg:2015pka,ATLAS:2024fdw,CMS:2024nht}, and only a few exploratory studies exist at colliders~\cite{Atwood:1999zg,Ahern:2000jn,Fichet:2014uka,Inan:2018jza,Cembranos:2021vdv, Shao:2022cly,dEnterria:2023} and at fixed-target facilities via the Primakoff process~\cite{Voronchikhin:2022rwc,Jodlowski:2023yne}. In this section, we extend the work of Ref.~\cite{dEnterria:2023} to include limits from beam-dump and fixed-target setups, and extract new bounds on the GLP $(\mG,\alphaUniv)$ parameter space by properly recasting an up-to-date set of searches for heavier ALPs coupling to photons performed recently at colliders.

Searches for ALPs at beam dumps exploit high-intensity proton or electron beams impinging on dense targets, where ALPs and/or GLPs are generated via Primakoff-like processes. 
These bosons decay into two photons, $\mathrm{G,a} \to \gaga$, either within the detector volume or travel a macroscopic distance before decaying, creating a distinctive displaced vertex signal (Fig.~\ref{fig:diags}, bottom left). At higher energies, $\epem$ collisions can produce heavier ALPs and/or GLPs via $\epem\to (\mathrm{G,a})\,\gamma \to 3\gamma$ processes (Fig.~\ref{fig:diags}, top~right). At hadron colliders (including protons and heavy-ions beams), ALPs/GLPs can originate in photon-fusion processes in the so-called ultraperipheral collisions (UPCs)~\cite{Baltz:2007kq}, $\gaga\to (\mathrm{G,a})\to\gaga$ (Fig.~\ref{fig:diags}, bottom~right) on top of the SM light-by-light scattering continuum~\cite{dEnterria:2013zqi}, extending the accessible mass reach up to $\ma\approx 2$~TeV~\cite{Bruce:2018yzs,dEnterria:2022sut}.

In the accelerator setups, the typical ALP signature is a narrow diphoton resonance, produced either promptly at the interaction point or, longer-lived, displaced in space. Search strategies are optimized depending on the ALP's mass and associated decay length, which scales inversely with the cube of its mass and the square of the photon coupling, following Eq.~(\ref{eq:Gamma_ALP}), $L_\text{decay} = \Gamma_\mathrm{a}^{-1} \,\propto\, m_\mathrm{a}^{-3}\gag^{-2}$. 
The ALPs limits set at beam dumps are for masses $\ma\lesssim 400$~MeV, in a range where their dominant decay channel is the diphoton one because most of all other decay modes, except the dielectron and dimuon ones, are not kinematically accessible, see Eq.~(\ref{eq:Gamma_ALP}), and the lightest quark/gluon decay, $\mathrm{a}\to 3\pi$, only opens up for $\ma\gtrsim 420$~MeV. For the higher ALP masses probed at colliders, the limits on $\gag$ have been set most often assuming the photon-dominance scenario, in which all decay branching fractions except the diphoton one are neglected, \ie\ $\mathcal{B}_{\mathrm{a}\to\gaga}\approx 1$. For the generic GLPs considered here, with universal couplings to SM particles, the diphoton decay dominates only below the two-pion threshold at $\mG\approx 270$~MeV corresponding to the $\mathrm{G}\to\pi\pi$ decay, the lightest hadronic channel allowed by conservation laws (and forbidden in the ALP case). Above this threshold, hadronic modes open up and quickly dominate, reducing the diphoton branching ratio to $\mathcal{B}_{\mathrm{G} \to \gaga}\approx 0.05$ (Fig.~\ref{fig:GLP_BRgamgam}).

%%%%%%%%%%%%%%%%%%%%%%%%%%%%%%%%%%%%%%%%%%
\subsection{Beam-dumps and fixed-target collisions}

Searches for ALPs in fixed-target collisions performed at beam-dump or collider facilities, include data from the following setups: SLAC E141~\cite{Riordan:1987aw} and E137~\cite{Batell:2014mga}; Fermilab MiniBooNE~\cite{MiniBooNEDM:2018cxm,MiniBooNE:2020pnu,Capozzi:2023ffu}; U70's NuCal~\cite{Dobrich:2019dxc}; CERN CHARM~\cite{CHARM:1985anb}, NA64~\cite{NA64:2020qwq}, FASER~\cite{FASER:2024bbl}, and SHiP~\cite{SHiP:2021nfo} (future); and Fermilab SeaQuest~\cite{Blinov:2021say,Blinov:2021say} (future). 
In beam-dump experiments, a high-intensity proton or electron beam strikes a thick target (``dump'') where the quasireal photon field of the charged beam particles, or the real photons from secondary photons from decays of neutral mesons produced in hadronic interactions of the beam with the target material, interact with the Coulomb field of the target nuclei A of charge $Z$ and can produce ALPs via the Primakoff effect, $\rm \gamma A\to a A$ (Fig.~\ref{fig:diags}, bottom left and center). Collisions of the quasireal or real photon beams with the Coulomb field of the target electrons are suppressed by a $Z^2$ factor, and ignored hereafter. While the large flux of secondary photons typically dominates the production of ALPs, quasireal photons from the incoming beam (whose flux is described within the Equivalent Photon Approximation, EPA~\cite{Brodsky:1971ud,Budnev:1975poe}), can produce ALPs with a slightly larger boost, providing sensitivity to shorter lifetimes (\ie\ to heavier and/or more weakly coupled ALPs). 
The dump is followed by tens to hundreds of meters of shielding where the SM particles are absorbed, but a weakly coupled ALP travels essentially unattenuated.
Further downstream, a long vacuum provides a volume for the ALP to decay, followed by detector systems where the $\rm a\to\gaga$ decay photons are searched for. Beam-dump experiments are thus more competitive than colliders at probing long-lived ALPs, \ie\ with small mass and coupling values.

The Primakoff cross section, $\rm \gamma A\to X\, A$, for the production of a particle X of even-spin ($J=0,2,...$) and diphoton width $\Gamma(\mathrm{X}\to\gaga)$ through the collision of an incoming quasireal $\gamma$ (EPA from the beam) or a real $\gamma$ (from meson decays) with the Coulomb field of the target nuclei, of nuclear mass $A$ and charge $Z$, can be written in the \cm\ frame as
\begin{eqnarray}
\sigma(\gamma\mathrm{A}\to\mathrm{X\,A}) = (2J+1)\frac{4 e^2Z^2}{m_\mathrm{X}^2}\, \Gamma(\mathrm{X}\to\gaga) \int_{t_{\min}}^{t_{\max}} \mathrm{d}t\,
\frac{F^2(t)}{t^2} \left(t - t_{\min}\right)
\label{eq:sigma_Primakoff}
\end{eqnarray}
where $F(t)$ is the nuclear/atomic EM form factor as a function of the momentum transfer $t=(P'-P)^2<0$, whose minimal kinematically allowed value is $t_{\min}\approx m_\mathrm{X}^4/(4E_\gamma^2)$. The EM form-factor encodes the charge distribution actually seen by the incoming photon, which can be (partially) screened by the electron shells at increasingly large impact parameters. The nucleus Coulomb field is often parametrized with the Helm form-factor: $F_A(t)\approx \exp(-t R_A^2/6)$ with nuclear radius $R_A=1.2A^{1/3}$, for a nucleus of mass number $A$. The maximum momentum transfer is bounded by the inverse of $R_A$ required to preserve the coherence of the nuclear Coulomb field, implying $t_{\max}\approx R_A^{-2}$. At very small $t$, atomic electrons screen the charge distribution, and a common parametrization is $F(t)= b^2t/(1+b^2t) \cdot F_A(t)$, with $b\approx 111/(Z^{1/3}m_e)$~\cite{Bjorken:2009mm}. 

From the generic expression (\ref{eq:sigma_Primakoff}), and using the (\ref{eq:Gamma_G}) and~(\ref{eq:Gamma_ALP}) diphoton widths, the Primakoff production cross sections of ALPs and GLPs, decaying back to diphotons, can be schematically written as
\begin{eqnarray}
\sigma(\gamma\mathrm{A}\to\mathrm{a(\gaga)\,A}) = 
 \frac{\pi}{16} \gag^2\ma \,\mathcal{B}_{\mathrm{a}\to \gaga}\;F\!F(t\approx\ma^2)\,,\;\text{for ALPs,}
\label{eq:sigma_beamdump_ALP}
\end{eqnarray}
\begin{eqnarray}
\sigma(\gamma\mathrm{A}\to\mathrm{G(\gaga)\,A}) = 
 \frac{\pi}{4} \left(\frac{\alphaG}{\MP}\right)^2\mG \,\mathcal{B}_{\mathrm{G}\to \gaga}\,
   F\!F(t\approx\mG^2)\,,\;\text{for GLPs,}
\label{eq:sigma_beamdump_GLP}
\end{eqnarray}
where $F\!F(t)$ encodes the integral of the atomic/nuclear form factor, which should be identical for states with the same mass. The additional GLP helicities enhance the Primakoff cross section by roughly a factor of four compared with the ALP case. However, since production is dominated by the helicity-2 configurations of the $\gaga$ system, a naive fully averaged spin counting $(2J+1)=5$, as given by Eq.~(\ref{eq:sigma_beamdump_ALP}), slightly overestimates the GLP yields; this approximation nevertheless remains adequate within our level of accuracy.
By equating both expressions, the ALP-GLP coupling conversion factor in beam-dump collisions at a given mass $m_\mathrm{a,G}$, reads 
\begin{align}\label{eq:beamdumps}
 \frac{\alphaG}{\MP} \bigg|_{\text{GLP, beam\,dump}} \sim \frac{1}{2}\cdot\left(\frac{\mathcal{B}_{\mathrm{a}\to \gaga}}{\mathcal{B}_{\mathrm{G}\to \gaga}}\right)^{1/2} \gag\bigg|_{\text{ALP, beam\,dump}}\,.
\end{align}
The ALP limits placed at beam dumps cover the $\ma\approx 0.1$--400~MeV range, where their dominant decay channel is the diphoton one, \ie\ $\mathcal{B}_{\mathrm{a}\to \gaga}\approx 1$, modulo potential small reductions from the $\mathrm{a}\to\epem,\mumu$ decays above the respective $\ma\approx 1,\,200$~MeV thresholds.
In the same mass range, the GLP diphoton decay width is $\mathcal{B}_{\mathrm{G}\to \gaga}\approx 0.4$, and above the two-pion threshold at $\mG\approx 270$~MeV, it becomes further reduced to $\mathcal{B}_{\mathrm{G}\to \gaga}\approx 0.1$ (Fig.~\ref{fig:GLP_BRgamgam}). 
These different branching fractions also modify the average distance $L$ traveled by the GLP through the experiment-dependent beam-dump and decay volumes relative to the ALP case, thereby affecting the extraction of the corresponding GLP exclusion limits for realistic experimental configurations. A proper treatment of these decay effects requires dedicated studies on case-by-case basis, for each experimental configuration considered here~\cite{Riordan:1987aw,Batell:2014mga,MiniBooNEDM:2018cxm,MiniBooNE:2020pnu,Capozzi:2023ffu,Dobrich:2019dxc,CHARM:1985anb,NA64:2020qwq,FASER:2024bbl,SHiP:2021nfo,Blinov:2021say,Blinov:2021say}, that go beyond the scope of this work; however, within our approximations, the effective conversion factor between ALP and GLP bounds for the same mass point, can be taken to be reasonably close to $\alphaUniv \approx \gag$.

%%%%%%%%%%%%%%%%%%%%%%%%%%%%%%%%%%%%%%%%%%
\subsection{Particle colliders}

Heavy photon-coupled ALPs are being sought-for both at $\epem$ and hadron colliders through, respectively, triphoton final states $\epem\to \mathrm{a}(\gaga)\gamma$ (Fig.~\ref{fig:diags}, top~right), and photon-photon fusion processes $\gaga \to \mathrm{a}\to\gaga$ (Fig.~\ref{fig:diags}, bottom~right). At $\epem$ colliders, limits have been set from LEP~\cite{L3:1994shn,Baillargeon:1995dg,OPAL:2002vhf}, Belle~II~\cite{Dolan:2017osp,Belle-II:2020jti,Belle-II:2026znr}, and BES-III~\cite{BESIII:2022rzz,BESIII:2024hdv} data, in nonresonant triphoton final-states as well as searching for exotic decays of resonances such as $\mathrm{Z},\,\Upsilon,\,\jpsi \to \mathrm{a}(\gaga)\gamma$. Additional bounds have also been placed for light ALPs from searches for monophoton final states with missing energy from long-lived ``invisible'' ALPs, via $\epem\to\gamma\mathrm{a}$ as well as from radiative $\Upsilon \to \gamma\mathrm{a}$ decays, at CLEO~\cite{CLEO:1994hzy} and BaBar~\cite{CLEO:1994hzy}, but those are usually less competitive than the beam-dump results and not discussed hereafter. Estimated ALP limits at future $\epem$ and hadron-hadron colliders, such as FCC-ee~\cite{RebelloTeles:2023uig,Polesello:2025gwj} and FCC-hh~\cite{FCC:2025lpp,deBlas:2025gyz,RebelloDdE2026}, have also been recently derived.

The most stringent ALP limits over the $\ma\approx 5$--100~GeV mass range have been set by searches in ultraperipheral Pb-Pb collisions by CMS~\cite{CMS:2018erd,CMS:2024bnt} and ATLAS~\cite{ATLAS:2020hii}. The final state of interest corresponds to a resonant diphoton excess exclusively produced~\cite{Knapen:2016moh}, \ie\ with no other hadronic activity in the event, on top of the light-by-light continuum~\cite{dEnterria:2013zqi}. In \pp\ collisions, the presence of a very large number of pileup events hinders the observation of such a process unless one can tag one or both protons in very forward spectrometers to remove overwhelming hadronic backgrounds~\cite{TOTEM:2021zxa, CMS:2022zfd, ATLAS:2023zfc}. 
Additional constraints on ALP production have been also derived in \pp\ collisions at the LHC by exploiting \textit{inclusive} diphoton or triphoton final states through reinterpretations~\cite{Jaeckel:2015jla,Knapen:2016moh,Bauer:2018uxu,Bauer:2017ris} of generic searches for resonant bumps targeting spin-0 (scalar and pseudoscalar) resonances as well as spin-2 ADD and RS states, or for $\rm Z\to a(\gaga)\gamma$ decays, and cover the invariant mass range $m_{\gaga}\approx 10$~GeV--2.6~TeV~\cite{dEnterria:2021ljz}.
However, such ALP limits are model-dependent and more complicated to recast into the corresponding GLP bounds, and will not be considered hereafter.

A fraction of the existing ALP limits set in $\epem$ collisions~\cite{Belle-II:2020jti,BESIII:2022rzz,Baillargeon:1995dg} and in UPCs at the LHC~\cite{dEnterria:2021ljz,CMS:2018erd,ATLAS:2023zfc,TOTEM:2021zxa,CMS:2022zfd,ATLAS:2023zfc} have been transformed into the corresponding GLP bounds in Ref.~\cite{dEnterria:2023}. In this Section, we extend this latter study to include the latest ALP bounds from $\epem$ collisions at BES-III~\cite{BESIII:2024hdv} and Belle~II~\cite{Belle-II:2026znr}, as well as from new UPC results at the LHC~\cite{CMS:2024bnt}.

\subsubsection{Electron-positron collisions}
\label{sec:ee_colls}

The LO cross sections for ALP and GLP production in $\epem$ collisions at \cm\ energy $\sqrts$ through their nonresonant emission from a virtual $\gamma$ or Z boson, leading to the triphoton final state shown in Fig.~\ref{fig:diags} (top right),  neglecting the tiny electron mass $m_\mathrm{e}$, read~\cite{dEnterria:2023}
\begin{eqnarray}
\sigma(\epem\to \mathrm{a}(\gaga) \gamma)=\frac{e^2}{96\pi} \gag^2\left(1-\frac{\ma^2}{s}\right)^3\;\mathcal{B}_{\mathrm{a}\to \gaga},\;\text{for ALPs,}
\label{eq:sigma_ee_ALP}
\end{eqnarray}
\begin{eqnarray}
\sigma(\epem\to \mathrm{G}(\gaga)\gamma)&=&\frac{e^2}{96\pi}\left(\frac{\alphaG}{ \MP}\right)^2\left(1-\frac{\mG^2}{s}\right)^3\;\mathcal{B}_{\mathrm{G}\to \gaga},\;\text{for GLPs,}
\label{eq:belle}
\end{eqnarray}
where $\mathcal{B}_{\mathrm{a,G}\to \gaga}$ are the corresponding decay branching fractions at each given particle mass. By equating both expressions, the coupling conversion factor needed to recast ALP into GLP searches at $\epem$ colliders reads,
\begin{align}\label{eq:ee_conversion}
 \frac{\alphaG}{\MP} \bigg|_{\text{GLP, }\epem} \sim \left(\frac{\mathcal{B}_{\mathrm{a}\to \gaga}}{\mathcal{B}_{\mathrm{G}\to \gaga}}\right)^{1/2} \gag\bigg|_{\text{ALP, }\epem}\,.
\end{align}
The expression above can be used to reinterpret LEP~\cite{L3:1994shn,Baillargeon:1995dg,OPAL:2002vhf} and Belle~II~\cite{Dolan:2017osp,Belle-II:2020jti,Belle-II:2026znr} ALP searches over $\ma\approx 10$~MeV--100~GeV. The numerical conversion factor is $\mathcal{O}(1)$ for $\epem$ limits with $\ma\lesssim 300$~MeV, but is $\mathcal{O}(4)$ above this mass when accounting for the suppressed $\mathcal{B}_{\mathrm{G}\to\gaga}\approx 0.05$ decay of GLPs compared with photophilic ALPs with $\mathcal{B}_{\mathrm{G}\to\gaga}\approx 1$.

More recently, searches in $\epem$ collisions have been performed by BES-III~\cite{BESIII:2022rzz,BESIII:2024hdv}, where an intermediate $\jpsi$ meson is first resonantly produced that decays into the ALP plus a photon leading to the triphoton final state, $\epem\to \jpsi \to \mathrm{a}(\gaga)\gamma$. At leading order, the partial widths of the radiative decay of quarkonium vector mesons ($\rm VM = \jpsi,\Upsilon$) into an ALP or GLP,
read~\cite{dEnterria:2023}
\begin{eqnarray}
\Gamma(\mathrm{VM}\to \mathrm{a}(\gaga)\gamma)&=&\frac{e^2}{324\pi}\gag^2\left(1-\frac{\ma^2}{m_{\mathrm{VM}}^2}\right)^3\langle \mathcal{O}^{\mathrm{VM}}\rangle\;\mathcal{B}_{\mathrm{a}\to \gaga}\,,\;\text{for ALPs,}
\label{eq:GammaJpsi_agamma}
\end{eqnarray}
\begin{eqnarray}
\Gamma(\mathrm{VM}\to \mathrm{G}(\gaga) \gamma)&=&\frac{e^2}{486\pi} \left(\frac{\alphaG}{\MP}\right)^2 F(\mG,m_{\mathrm{VM}}) \; \langle \mathcal{O}^{\mathrm{VM}}\rangle\;\mathcal{B}_{\mathrm{G}\to \gaga},\;\text{for GLPs,}
\label{eq:GammaJpsi_Ggamma}
\end{eqnarray}
where $\langle \mathcal{O}^{\mathrm{VM}}\rangle$ is the long-distance matrix element describing the formation of the quarkonium bound state~\cite{Brambilla:2010cs,Lansberg:2019adr}, and $F(\mG,m_{\mathrm{VM}})=\left(1-\frac{m_\mathrm{G}^2}{m_{\mathrm{VM}}^2}\right)\left(1+3\frac{m_\mathrm{G}^2}{m_{\mathrm{VM}}^2}+6\frac{m_\mathrm{G}^4}{m_{\mathrm{VM}}^4}\right)$. 
By equating these two expressions, one can derive a bound on the GLP universal coupling from any corresponding limit obtained for the ALP-$\gamma$ case via
\begin{align}
\frac{\alphaG}{\MP} \bigg|_{\text{GLP, }\mathrm{VM}} \sim   \left(\frac{1-r^2}{\sqrt{2/3+2r^2+4r^4}}\right) \left(\frac{\mathcal{B}_{\mathrm{a}\to \gaga}}{\mathcal{B}_{\mathrm{G}\to \gaga}}\right)^{1/2}
\gag\bigg|_{\text{ALP, }\mathrm{VM}.}\,,
\label{eq:limitGaFromJpsi}
\end{align}
with $r=\mG/m_{\mathrm{VM}}$. Over the mass values, $\mG\approx0.18$--2.8~GeV, covered by the BES-III analysis for $\rm VM = \jpsi$, the prefactor of this expression amounts to 1.2--0.1, whereas the square-root of branching fractions for photophilic ALPs amounts to 1.5--4 for $\mG\approx0.18$--2.8~GeV. Combining both factors indicates that, relative to ALPs, the constraints on GLPs are approximately a factor of three weaker (stronger) in the low- (high-) mass region explored by BES-III.

%%%%%%%%%%%%%%%%%%%%%%%%%%%%%%%%%%%%%%%%%%
\subsubsection{Photon-photon collisions at hadron colliders}

The cross section for the exclusive production of a generic C-even resonance X of spin $J$ and two-photon decay width $\Gamma_{\mathrm{X}\to\gaga}$, through photon-photon fusion in an UPC of charged hadrons A$_1$ and A$_2$ (Fig.~\ref{fig:diags}, bottom right), reads~\cite{Brodsky:1971ud,Budnev:1975poe}
\begin{equation}
\sigma(\mathrm{A}_1\; \mathrm{A}_2\,\xrightarrow{\gaga} \mathrm{A}_1\; \mathrm{X \; A}_2) = 
4\pi^2 (2 J+1)\frac{\Gamma_{\mathrm{X}\to\gaga}}{m_\mathrm{X}^2} \frac{\mathrm{d}{\Lumi}^{(\mathrm{A}_1\mathrm{A}_2)}_{\gaga}}{\mathrm{d}m_{\gaga}}\bigg|_{m_{\gaga}=m_\mathrm{X}},
\label{eq:sigma_AA_X}
\end{equation}
where $\frac{\mathrm{d}{\Lumi}^{(\mathrm{A}_1\mathrm{A}_2)}_{\gaga}}{\mathrm{d}m_{\gaga}}\bigg|_{m_{\gaga}=m_\mathrm{X}}$ is the value of the effective two-photon luminosity at the X resonance mass $m_\mathrm{a,G}$ given by the convolution of the EPA fluxes of the colliding hadrons in an UPC at nucleon-nucleon \cm\ energy $\sqrtsnn$.
Examples of the application of Eq.~(\ref{eq:sigma_AA_X}) to compute the production cross sections of a large variety of C-even resonances in UPCs at RHIC and at the LHC can be found, \eg\ in Ref.~\cite{dEnterria:2025ecx}, where the effective two-photon luminosities for proton or ion UPCs have been obtained with the \gammaUPC\ code~\cite{Shao:2022cly}, including the survival probabilities of the colliding hadrons derived with a Glauber Monte Carlo model~\cite{dEnterria:2020dwq}. In the case of heavy-ion beams, the action of all the charges in the nucleus adds coherently and the photon flux is enhanced by a $Z^2$ factor compared with the proton case, leading to a $Z_1^2Z_2^2$ increase in the corresponding $\gaga$ effective luminosities and, thereby, cross sections. Such a large enhancement factor provides a very large advantage for ALP searches, appearing as $\gaga\to\mathrm{a}\to\gaga$ resonances~\cite{Knapen:2016moh} over the light-by-light SM continuum~\cite{dEnterria:2013zqi}, in heavy-ion compared with $\epem$ and \pp\ UPCs.

From Eq.~(\ref{eq:sigma_AA_X}), one can first see that the photon-fusion production of GLPs ($J=2$) will be naively enhanced by a factor of $(2 J+1) = 5$ compared with the ALP ($J=0$) case for the same values of masses and diphoton partial decay widths. Such an apparent benefit will be, however, compensated by the comparatively reduced heavy-graviton decay branching fraction into photons. Indeed, the production cross sections of ALPs and GLPs in UPCs decaying back to diphotons, using the diphoton widths of Eqs.~(\ref{eq:Gamma_ALP}) and~(\ref{eq:Gamma_G}), read
\begin{eqnarray}
\sigma(\mathrm{A}_1\; \mathrm{A}_2\,\xrightarrow{\gaga} \mathrm{A}_1\; \mathrm{a}(\gaga) \; \mathrm{A}_2) &=& 
 \frac{\pi}{16} \gag^2\ma \,\mathcal{B}_{\mathrm{a}\to \gaga}\,
  \frac{\mathrm{d}{\Lumi}^{(\mathrm{A}_1\mathrm{A}_2)}_{\gaga}}{\mathrm{d}m_{\gaga}} \bigg|_{m_{\gaga}=\ma}\,,\;\text{for ALPs,}
\label{eq:sigma_AA_ALP}
\end{eqnarray}
\begin{eqnarray}
\sigma(\mathrm{A}_1\; \mathrm{A}_2\,\xrightarrow{\gaga}\mathrm{A}_1\; \mathrm{G}(\gaga) \; \mathrm{A}_2) &=& 
 \frac{\pi}{4} \left(\frac{\alphaG}{\MP}\right)^2\mG \,\mathcal{B}_{\mathrm{G}\to \gaga}\,
   \frac{\mathrm{d}{\Lumi}^{(\mathrm{A}_1\mathrm{A}_2)}_{\gaga}}{\mathrm{d}m_{\gaga}} \bigg|_{m_{\gaga}=\mG}\,,\;\text{for GLPs,}
\label{eq:sigma_AA_GLP}
\end{eqnarray}

The comparison of both expressions allows deriving the ALP-GLP coupling conversion factor for bosons of equal mass, which reads
\begin{align}\label{eq:UPCs}
 \frac{\alphaG}{\MP} \bigg|_{\text{GLP, UPCs}} \sim \frac{1}{2}\cdot\left(\frac{\mathcal{B}_{\mathrm{a}\to \gaga}}{\mathcal{B}_{\mathrm{G}\to \gaga}}\right)^{1/2} \gag\bigg|_{\text{ALP, UPCs}}\,.
\end{align}
This expression neglects small differences in the experimental fiducial acceptances for ALPs and GLPs decaying into pairs of photons. On average, tensor states decay into photons that are softer and more isotropically distributed than those arising from pseudoscalar decays, and are therefore less likely to satisfy the transverse-momentum and pseudorapidity selection criteria applied in experiments. For the general-purpose LHC detectors, such acceptance differences amount to 1\%--50\% for $m_\mathrm{a,G}\approx 1000$--5~GeV masses, as derived in Ref.~\cite{dEnterria:2023} from full simulations performed with the \gammaUPC\ code combined with \madgraph~\cite{Alwall:2014hca,Frederix:2018nkq} where the corresponding Lagrangian densities, Eqs.~(\ref{eq:Lgraviton}) and (\ref{eq:Leff_ALP2}), have been coded as input models in the Universal Feynman Output (\ufo) format~\cite{Degrande:2011ua,Darme:2023jdn}.
Whereas at face value Eq.~(\ref{eq:UPCs}) indicates a factor of two improved sensitivity to GLPs via photon-fusion in UPCs compared with ALPs, in reality the experimental searches often assume photophilic ALPs (\ie\ $\mathcal{B}_{\mathrm{a}\to \gaga}\approx 1$), and the recast limits on the GLP universal coupling turn out to be somewhat less competitive than their ALPs counterparts. For the relevant mass range $m_\mathrm{a,G}>5$~GeV, $\mathcal{B}_{\mathrm{G}\to\gaga}\approx 0.05$ and, therefore, $\alphaUniv \approx 2\,\gag$.

%\clearpage
%%%%%%%%%%%%%%%%%%%%%%%%%%%%%%%%%%%%%%%%%%
\section{Combined results and comparisons with other limits}
\label{sec:results}

Table~\ref{tab:dictionary} summarizes the mapping between the ALP-photon and GLP-universal couplings derived in this study, including the quantitative conversion factors for each search method, the experimental ALP limits recast in this work, and the relative ALP-GLP sensitivity. Out of 17 search methods analyzed, 8 give similar ALP and GLP sensitivity, and 5 (4) of them feature enhanced (reduced) sensitivities to GLPs compared with ALPs. 

\begin{table}[htbp!]
\caption{\small Summary of the quantitative mapping between the ALP-photon coupling ($\gag$) and the GLP universal coupling ($\alphaUniv$) for bosons with mass $\ma=\mG$. For each ALP detection method examined here, the derived coupling conversion factor, recast ALPs searches, and relative GLP vs.\ ALP sensitivity are listed. The last three rows (collider limits) assume photophilic ALPs ($\mathcal{B}_{\mathrm{a}\to\gaga}\approx 1$, see Fig.~\ref{fig:GLP_BRgamgam}). \label{tab:dictionary}}
\resizebox{1.\textwidth}{!}{%
\tabcolsep=1.mm
\begin{tabular}{llll}\hline
ALP detection method & Coupling conversion factor & ALP limits & Relative GLP vs.\ ALP sensitivity \\
\hline
\underline{Dark-matter candidates}: & & & \\
Haloscope & $\frac{\alphaG}{\MP} \leftrightarrow v_\mathrm{DM}^{-1}\,\gag$ & \cite{ADMX:2020ote,Lee:2020cfj,Brubaker:2016ktl,Quiskamp:2022pks,Ahyoune:2024klt,Alesini:2019ajt,Grenet:2021vbb} & Reduced by $v_\text{DM}\approx 10^{-3}$\\
\makecell[l]{Magnetometers:} \\ 
\;\;Solenoidal $\vec B$ & $\frac{\alphaG}{\MP} \leftrightarrow  v_\mathrm{DM}^{-1}\,\gag$  & \cite{Zhang:2021bpa,DMRadio:2022pkf} &
\makecell[l]{Reduced by $v_\text{DM}\approx 10^{-3}$} \\
\;\;Toroidal $\vec B$, circ.\ pickup & $\frac{\alphaG}{\MP} \leftrightarrow v_\mathrm{DM}^{-1}\gag$ & \cite{ Kahn:2016aff,Ouellet:2019tlz,Gramolin:2020ict} &
\makecell[l]{Reduced by  $v_\text{DM}\approx 10^{-3}$} \\
\;\;Toroidal $\vec B$, fig.8 pickup & $\frac{\alphaG}{\MP} \leftrightarrow \left(\omega_\mathrm{G}d_B\right)\gag$ & \cite{DMRadio:2022jfv,Domcke:2022rgu} &
\makecell[l]{Enhanced by $\lambda_\mathrm{G}/d_B \approx (10^{-6}\,\mG/\mathrm{eV})$ \\
(${\sim}1$ m wide magnet)} \\
\makecell[l]{Two-beams interferometry:} \\ 
\;\;Orthogonal Michelson-like & $\frac{\alphaG}{\MP} \leftrightarrow \left(\frac{\omega_\mathrm{G}}{k_\gamma}\right)\gag$  & \cite{Gue:2024onx,Yao25,YaoR25,LIGOScientific:2021ffg,Zhang:2025fck,Manita:2023mnc} &
\makecell[l]{Enhanced by $k_\gamma/\omega_\mathrm{G} \approx (1\,\mathrm{eV}/\mG)$ \\ (optical laser). Circ.\ pol.\ unneeded.} \\
\;\;Longitudinal bow-tie & $\frac{\alphaG}{\MP} \leftrightarrow \left(\frac{\omega_\mathrm{G}}{k_\gamma}\frac{L}{\ell}\right)\gag$ & \cite{Obata:2018vvr} &
\makecell[l]{Enhanced by $k_\gamma \ell /\omega_\mathrm{G} L\approx (10^{-2}\,\mathrm{eV}/\mG)$ \\ (optical laser). Circ.\ pol.\ unneeded.} \\
\makecell[l]{Upconversion:}\\
\;$B_0\parallel E_1$ & $\frac{\alphaG}{\MP} \leftrightarrow \gag$ & \cite{Thomson:2023moc} & Similar\\
\;$B_0\perp E_1$ & $\frac{\alphaG}{\MP} \leftrightarrow \left(\frac{\omega_\mathrm{G}}{\omega_0}\right)\gag$ & \cite{Bourhill:2022alm,Thomson:2023moc,Li:2025pyi} &
\makecell[l]{Enhanced by $\omega_0/\omega_\mathrm{G} \approx (10^{-6} \text{eV}/m_\mathrm{G})$ \\ (microwave background)}\\
Conversion on Earth's $\vec B$ & $\frac{\alphaG}{\MP} \leftrightarrow (\omega_\mathrm{G} R_{\Earth})\,\gag$ & \cite{Sulai:2023zqw,Nishizawa:2025xka,Arza:2021ekq,Friel:2024shg}
& Similar for $\mG \approx 10^{-14}$~eV\\
DM decay (astro)& $\frac{\alphaG}{\MP} \leftrightarrow \frac{\sqrt{5}}{2}\,\gag$ & \cite{Bolliet:2020ofj,Thorpe-Morgan:2020rwc,Foster:2021ngm,Wadekar:2021qae,Roach:2022lgo,Calore:2022pks,Carenza:2023qxh, Wang:2023imi,Todarello:2023hdk,Pinetti:2025owq} & Similar \\ \hline
\underline{Non-dark-matter candidates}: & & & \\
Helioscopes & $\frac{\alphaG}{\MP} \leftrightarrow \left(\frac{1}{2}\right)^{1/2}\,\gag$ & \cite{CAST:2017uph,IAXO:2019mpb} & Similar \\
$\gamma$ spectra modulation (astro) & $\frac{\alphaG}{\MP} \leftrightarrow \left(\frac{3}{14}\right)^{1/2}\, \gag$ & \cite{Wouters:2013hua, Marsh:2017yvc, Reynolds:2019uqt, Reynes:2021bpe, Fermi-LAT:2016nkz, Davies:2022wvj, MAGIC:2024arq, Li:2020pcn, Li:2021gxs, Li:2024zst} & Enhanced by $\sqrt{14/3}\approx 2$.\\
Anomalous fluxes (astro) & $\frac{\alphaG}{\MP} \leftrightarrow \, \gag$ & \cite{Xiao:2020pra, Hoof:2022xbe,Calore:2021hhn,Dessert:2020lil,Ning:2024eky,Meyer:2020vzy,Manzari:2024jns} & Similar \\
Beam dumps & $\frac{\alphaG}{\MP} \leftrightarrow \frac{1}{2} \left(\frac{\mathcal{B}_{\mathrm{a}\to \gaga}}{\mathcal{B}_{\mathrm{G}\to \gaga}}\right)^{1/2} \gag$ & \cite{CHARM:1985anb,Riordan:1987aw,Batell:2014mga,MiniBooNEDM:2018cxm,MiniBooNE:2020pnu,NA64:2020qwq,Dobrich:2019dxc,Capozzi:2023ffu,FASER:2024bbl}
& Similar \\ 
$\epem\to\mathrm{a}(\gaga)\,\gamma$ & $ \frac{\alphaG}{\MP} \leftrightarrow \left(\frac{\mathcal{B}_{\mathrm{a}\to \gaga}}{\mathcal{B}_{\mathrm{G}\to \gaga}}\right)^{1/2}\gag$ & \cite{L3:1994shn,Baillargeon:1995dg,OPAL:2002vhf,BaBar:2010eww,Dolan:2017osp,Belle-II:2020jti}
& \makecell[l]{Similar for $\mG\lesssim 0.3$~GeV\\
Reduced by 1/4 for $\mG\gtrsim 0.3$~GeV}\\
$\epem\to\mathrm{VM}\to \mathrm{a}(\gaga)\,\gamma$ & $\frac{\alphaG}{\MP}  \leftrightarrow f(r)\left(\frac{\mathcal{B}_{\mathrm{a}\to \gaga}}{\mathcal{B}_{\mathrm{G}\to \gaga}}\right)^{1/2} \gag$ & \cite{BESIII:2022rzz,BESIII:2024hdv} & Similar, $f(r)=1/2$--2 for $r=\mG/m_{\mathrm{VM}}$\\
$\gaga\to\mathrm{a}\to\gaga$ & $\frac{\alphaG}{\MP} \leftrightarrow \frac{1}{2} \left(\frac{\mathcal{B}_{\mathrm{a}\to \gaga}}{\mathcal{B}_{\mathrm{G}\to \gaga}}\right)^{1/2} \gag$ & \cite{CMS:2018erd,ATLAS:2020hii,TOTEM:2021zxa,CMS:2022zfd,ATLAS:2023zfc,
CMS:2024bnt} & Reduced by $1/2$ for $\mG\gtrsim 5$~GeV\\ 
\hline
\end{tabular}
}
\end{table}

In the case of searches for DM boson candidates at low masses ($m_\mathrm{a,G}\lesssim 1$~eV), our study indicates a clear pattern. Whenever an experiment uses a static uniform electromagnetic background to convert dark-matter ALPs into photons, the coupling conversion is $\gag \to v_\mathrm{DM}\,(\alphaUniv)$, \ie\ the experiment (haloscopes and solenoidal magnetometers) are $v_\mathrm{DM}^{-1}\approx 10^3$ more sensitive to ALPs than to GLPs of the same mass. By contrast, whenever the conversion is induced by a spacetime-dependent background electromagnetic field, either static (such as toroidal magnetometers) or with frequency $\omega_\gamma$ (such as two-beams interferometers and upconversion devices) the experiment will be more sensitive to GLPs. Specifically, for the former cases, the conversion factor is  $\gag \to 1/(\omega_\mathrm{G}d)\,(\alphaUniv)$, where $d$ is the typical size of the experiment, while for the latter cases, it is $\gag \to (\omega_\gamma/\omega_\mathrm{G})\,(\alphaUniv)$. Since in general, $1/(\omega_\text{G}d),\, \omega_\gamma/\omega_\mathrm{G} \gg 1$, these situations generally enhance the sensitivity to GLPs compared with ALPs. 
In systems where the new boson is produced through interactions of SM particles, \ie\ when the DM assumption is relaxed and the particle is relativistic, the sensitivity of a given probe to massive gravitons and axions becomes essentially the same, up to $\mathcal{O}(1)$ factors. In general, because the GLP couples universally to all SM fields, it can be produced through more processes\footnote{In addition, as pointed out in~\cite{Garcia-Cely:2025ula}, when the energy of the new boson is lower than the temperature of the system producing it, the GLP production through Bremsstrahlung can substantially exceed the ALP production through the Primakoff effect. However, this is not the parameter space probed by the experiments of interest here.}  compared with the photophilic ALP. For higher masses, the ALP searches at accelerators are generically similarly sensitive to GLPs, modulo differences in their (assumed) diphoton decay widths, $\gag \to (\mathcal{B}_{\mathrm{a}\to \gaga}/\mathcal{B}_{\mathrm{G}\to \gaga})^{1/2}\,(\alphaUniv)$.\\

\begin{figure}[htbp!]
    \centering
    \includegraphics[width=0.7\linewidth]{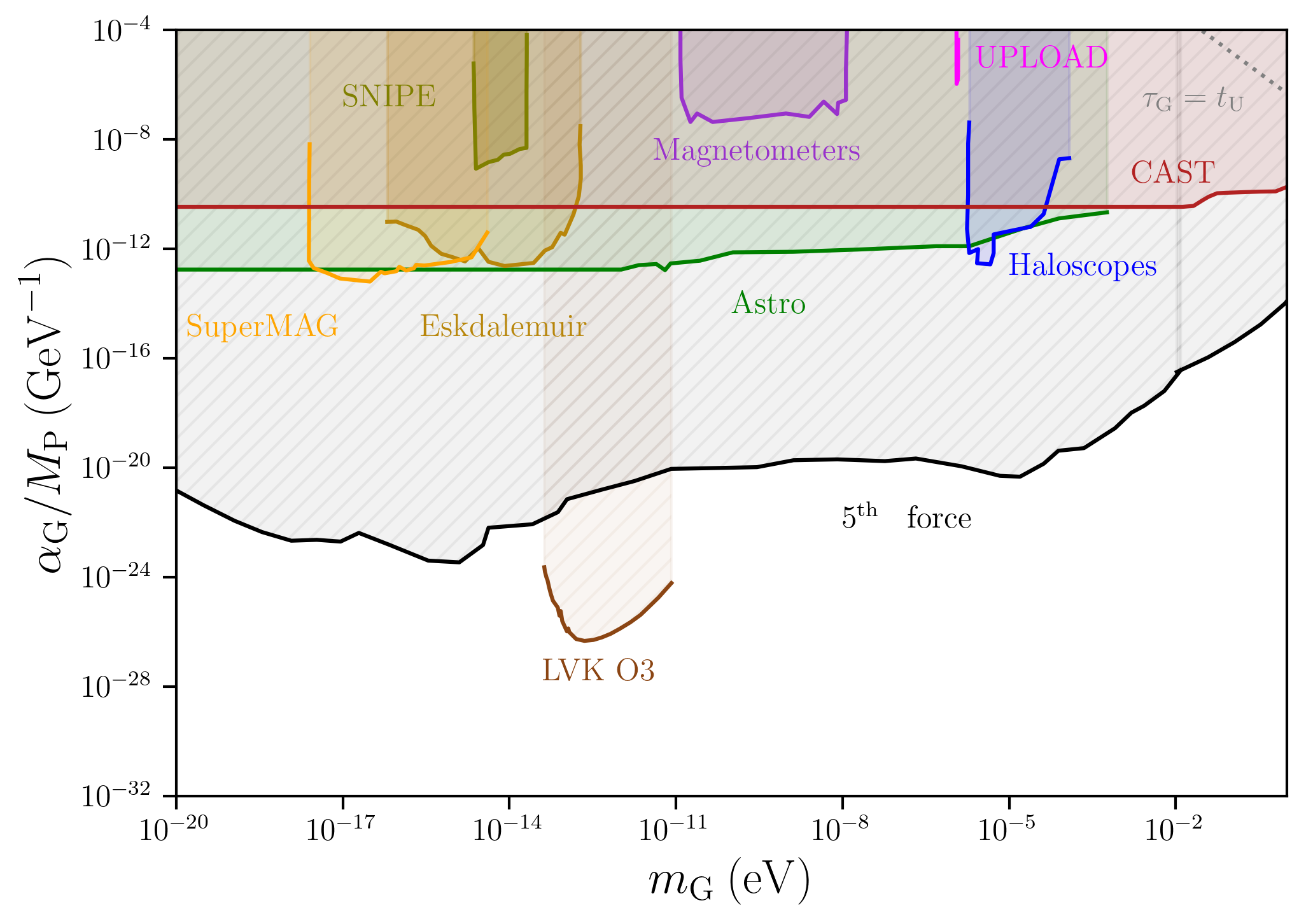}
    \caption{\small Constraints on the GLP mass versus universal coupling plane $(\mG,\alphaUniv)$ derived in this work from existing ALP-searches experiments at low masses ($\ma<1$~eV). The black curve indicates alternative GLP constraints from 5th-force tests~\cite{Cembranos:2017vgi,Chen:2014oda}. The LVK-O3 (LIGO/VIRGO/KAGRA, third observing run) brown curve shows the GLP sensitivity of ground-based GW detectors~\cite{Manita:2023mnc}.}
    \label{fig:bounds}
\end{figure}

Figure~\ref{fig:bounds} collects all the exclusion limits on massive gravitons in the GLP mass versus universal coupling plane $(\mG,\alphaUniv)$ derived from the existing ALPs searches at low masses, $m_\mathrm{a,G}<1$~eV, as obtained with the coupling conversion factors derived in this work. Our curves are compared with alternative bounds derived from fifth-force probes~\cite{Cembranos:2017vgi,Chen:2014oda} (gray dashed area) and GW detectors~\cite{Manita:2023mnc} (brown dashed area). The existence of a massive spin-2 particle naturally induces a universal attractive tensor force, of the Yukawa form $V(r)\propto (\alphaUniv)^2 \exp^{-\mG r}/r$, in addition to gravity. Fifth-force tests search for deviations from the gravitational inverse-square law on solar-system (often on planetary ephemerides~\cite{Fienga:2023ocw}) and laboratory scales, and thereby can provide stringent constraints on the GLPs $(\alphaUniv,\mG)$ parameter space. 

%\vspace{\floatsep}

\begin{figure}[htbp!]
    \centering
    \includegraphics[width=0.7\linewidth]{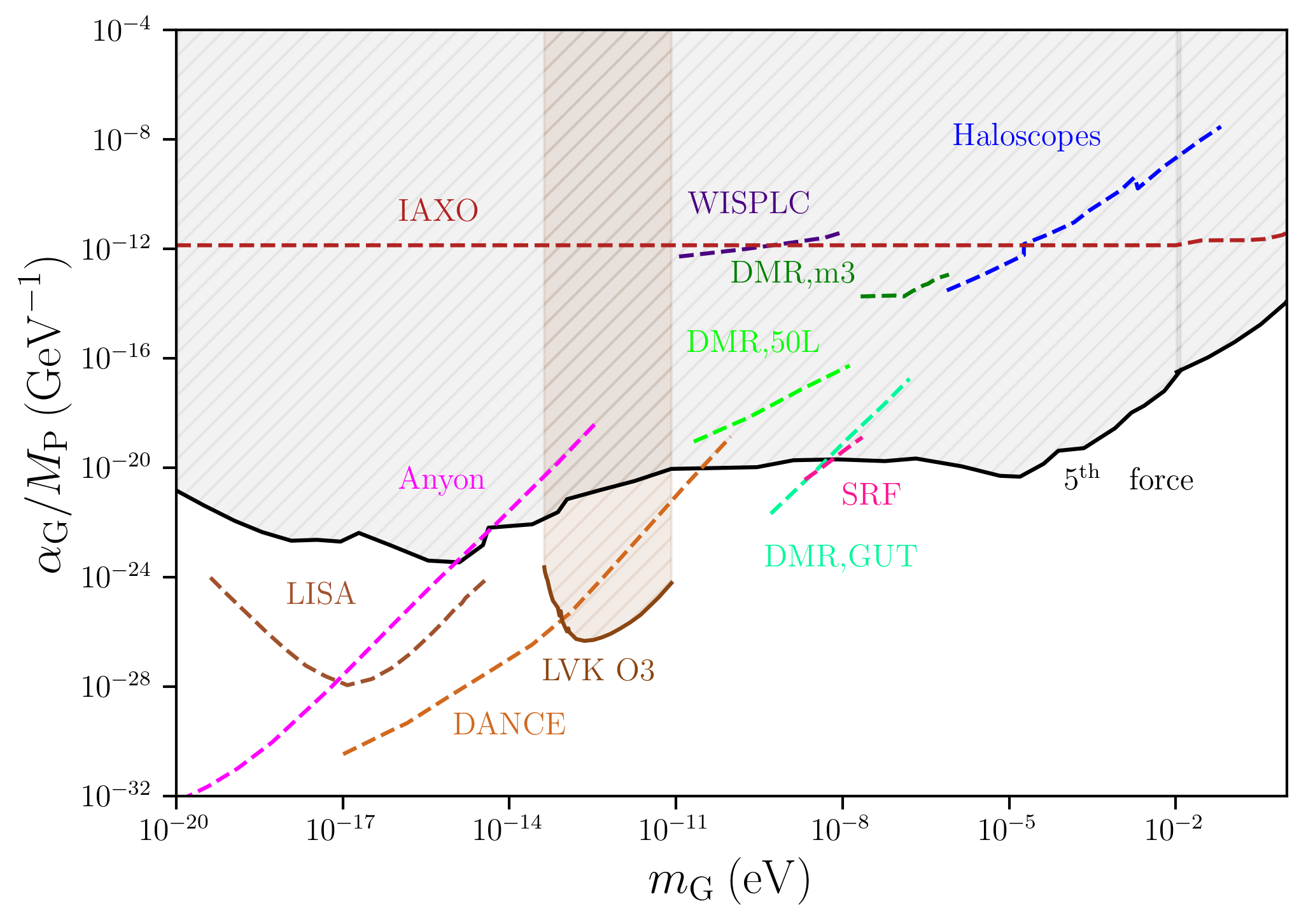}
    \caption{\small Future constraints on the GLP mass versus universal coupling plane $(\mG,\alphaUniv)$ derived here from expected ALP-searches experiments at low masses ($\ma<1$~eV). The acronym `DMR' is used for DMRadio. The LISA (brown) and 5th-force (black) curves show the alternative GLP sensitivity from future space-based GW detectors~\cite{Zhang:2025fck} and from current 5th-force tests~\cite{Cembranos:2017vgi,Chen:2014oda}, respectively.
    \label{fig:bounds_future}}
\end{figure}

The tightest constraints on the interaction strength come from experiments testing large distances and thereby lowest mass GLPs. For the case of interferometric ground-based GW detectors such as the LIGO-VIRGO-KAGRA (LVK) collaboration~\cite{LIGOScientific:2021ffg}, the plotted curve corresponds to the sensitivity to massive spin-2 DM derived\footnote{The bound of~\cite{Manita:2023mnc} is tighter than, \eg\ starting from the $\gag$-sensitivity calculation of LVK from~\cite{Nagano:2019rbw} and then applying Eq.~\eqref{eq:birefringence_corresp_a_g}, because the former accounts for the cross-correlations between all the detectors of the network, which significantly increases the associated sensitivity.} in Ref.~\cite{Manita:2023mnc}.
The 5th-force tests provide the best limits down to $\alphaUniv\approx 10^{-23}$~GeV$^{-1}$ for $\mG<1$~eV (Fig.~\ref{fig:bounds}), except in the $\mG\approx 10^{-13}\mbox{--}10^{-11}$~eV range where ground-based GW detectors can probe couplings as low as $\alphaUniv\approx 10^{-27}$~GeV$^{-1}$.
Whereas our axion-recast limits appear about $10^{10}\mbox{--}10^3$ times less sensitive to GLPs than those derived from 5th-force or GW studies, the situation will change in the future (Fig.~\ref{fig:bounds_future}) thanks to the exploitation of optimized toroidal magnetometers~\cite{DMRadio:2022jfv}, novel upconversion detectors based on anyon resonators~\cite{Bourhill:2022alm} or SRF cavities~\cite{Li:2025pyi}, as well as vacuum birefringence searches such as DANCE in its bow-tie mode~\cite{Obata:2018vvr}. Those methods will provide very strong sensitivity, down to $\alphaUniv\approx 10^{-32}$~GeV$^{-1}$ for GLPs with $\mG\lesssim 10^{-8}$~eV, outperforming the current 5th-force bounds. In this figure, we also show the expected sensitivity of space-based GW detectors such as LISA, as derived in Ref.~\cite{Zhang:2025fck}. Note that such future GLP bounds can be also obtained by using the conversion factor (\ref{eq:birefringence_corresp_a_g}) to transform the $\gag$ sensitivity estimated in Refs.~\cite{Gue25,Yao25, YaoR25}.\\

\begin{figure}[htbp!]
    \centering
    \includegraphics[width=0.75\linewidth]{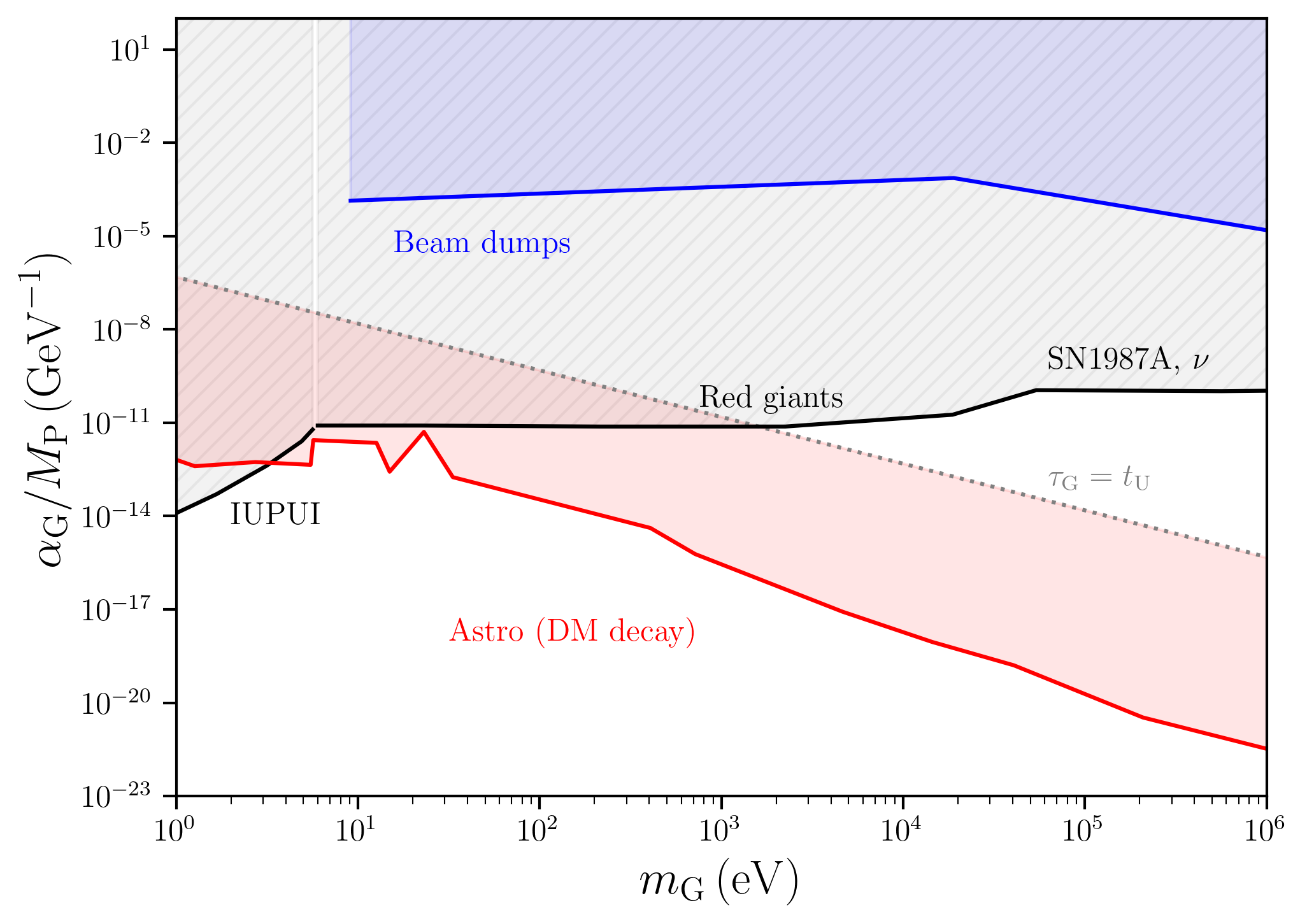}
    \caption{\small Constraints on the GLP mass versus universal coupling plane $(\mG,\alphaUniv)$ derived in this work from existing ALP-searches for masses $\mG = 1$~eV--1~MeV: beam dumps (blue curve) and DM-decay (red curve). The dotted red curve indicates the mass-coupling values for which the GLP lifetime equals the age of the universe $\tau_\mathrm{G}\lesssim t_\mathrm{U}$. The black curve indicates alternative constraints from current 5th-force tests~\cite{Chen:2014oda}, as well as from astrophysical constraints on energy loss in stellar environments~\cite{Cembranos:2017vgi}. 
    \label{fig:himass_bounds_current}}
\end{figure}

Figure~\ref{fig:himass_bounds_current} presents the upper limits on massive gravitons in the $(\mG,\alphaUniv)$ plane, derived from existing ALP searches in the ``intermediate'' mass range $m_\mathrm{a,G}=1$~eV--1~MeV, compared with other existing constraints. The recast GLP limits in this region include the dark-matter decay bounds (red curve) obtained from the non-observation of photon decays from DM GLPs, and the beam-dump constraints (blue curve). The dotted curve indicates the mass-coupling range for which the GLP lifetime equals the age of the universe: no dark-matter bounds can exist above this line, as such particles would have already decayed (see also Fig.~\ref{fig:lifetime}). Both limits are compared with alternative GLP bounds (gray dashed area) derived from fifth-force tests (labeled `IUPUI')~\cite{Chen:2014oda}, and stellar energy losses induced by the emission of massive gravitons (labeled `Red giants' and `SN1987A, neutrinos') calculated in Ref.~\cite{Cembranos:2017vgi}. The processes considered in this latter work are $\epem\to \mathrm{G}$, $\gaga\to \mathrm{G}$, $\mathrm{e}\gamma \to \mathrm{eG}$ (where the photon is either on- or off-shell, and in the latter case, it is provided by a static Coulomb field induced by an external heavy nucleus) and leads to stringent $\alphaUniv\lesssim 10^{-11}\mbox{--}10^{-9}$~GeV$^{-1}$ constraints in the $\mG\approx 1$--$10^{6}$~eV mass range. Our recast DM limit improves upon that previously derived from red giants energy-loss constraints over $\mG\approx 5$~eV--3~keV, although the former assumes that the GLP constitutes dark matter, whereas red-giant-based bounds do not rely on this assumption. In the keV--MeV region, the previous SN1987A constraints are more competitive than the recast beam-dumps results derived here.\\

\begin{figure}[htbp!]
    \centering
    \includegraphics[width=0.7\linewidth]{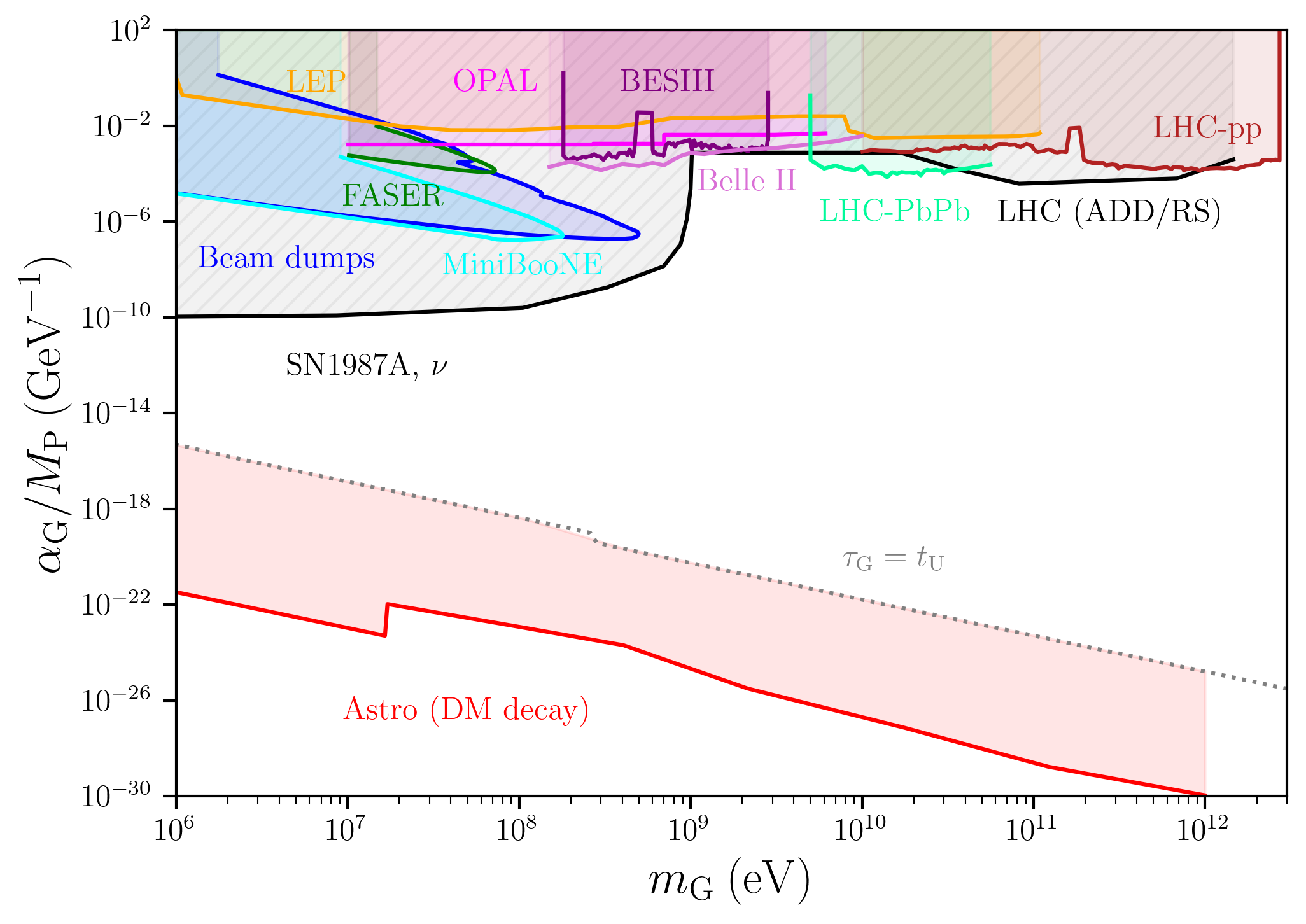}
    \caption{\small Constraints on the GLP mass vs.\ universal coupling plane $(\mG,\alphaUniv)$ derived from existing ALP-searches at masses $\mG=1$~MeV--3~TeV. The red curve follows from the absence of photon decays of DM GLPs; the dotted curve marks the coupling for which the GLP lifetime equals the age of the universe, above which DM would have already decayed. The black curve indicates alternative constraints from the SN1987A neutrino signal~\cite{Cembranos:2017vgi} and from spin-2 (ADD- and RS-type) particle searches at the LHC~\cite{Cembranos:2021vdv}.
    \label{fig:himass_bounds_current_2}}
\end{figure}

Figure~\ref{fig:himass_bounds_current_2} compiles the bounds on heavy gravitons in the $(\mG,\alphaUniv)$ plane over the mass range $\mG\approx 1$~MeV--3~TeV as derived here from existing ALP searches at accelerators (beam dumps, fixed target, colliders), as well as DM decay limits (red area, obtained from the absence of photon decays from DM GLPs, as in Fig.~\ref{fig:himass_bounds_current}). For all masses above 1 MeV, the red line derived under the assumption of GLP dark matter provides the strongest exclusion limits, albeit spanning a band region covering roughly six orders-of-magnitude in the universal couplings, ranging from $\alphaUniv\approx [10^{-22},10^{-15}]$~GeV$^{-1}$ at $\mG=1$~MeV to $\alphaUniv\approx [10^{-25},10^{-30}]$~GeV$^{-1}$ at $\mG=1$~TeV. In the region $\mG\lesssim 1$~GeV, the alternative spin-2 bounds $\alphaUniv\approx 10^{-10}$--$10^{-3}$~GeV$^{-1}$ derived from SN1987A (black curve)~\cite{Cembranos:2017vgi} outperform our newly extracted results from $\epem$ collisions (LEP, OPAL, BES-III areas), beam dumps (blue area, including MiniBooNE), and forward LHC experiments (FASER, in green)\footnote{Note, however, that the SN1987A limits derived in~\cite{Cembranos:2017vgi} most likely lack an upper bound on the coupling, since sufficiently strongly coupled GLPs would become trapped in the supernova core rather than escape freely. This observation further reinforces our conclusion that dedicated searches for GLPs in existing beam-dump experiments, $\epem$ colliders, and LHC experiments are needed to robustly probe this region of parameter space.}. However, in the $\mG\gtrsim 1$--100~GeV range, the Belle~II and LHC PbPb limits for photophilic ALPs, reinterpreted here, appear more competitive than the LHC\,(ADD/RS) black curve from Ref.~\cite{Cembranos:2021vdv}. This latter curve is obtained by recasting searches for ADD/RS massive gravitons in inclusive final states, produced via parton-parton scatterings and followed by $\rm G\to W^+W^-,\,\ell^+\ell^-,\,\gaga$ decays. For masses above $\mG\approx 100$~GeV, the inclusive GLP limits from LHC\,(ADD/RS) are stronger than those from exclusive GLP diphoton searches. This is because the latter suffer from a small branching ratio of about 5\%, whereas inclusive searches ---despite much larger backgrounds~\cite{dEnterria:2023}--- remain sensitive to all decay modes. At the high end of the mass spectrum, $\mG\gtrsim 1$~TeV, exclusive searches in \pp\ collisions become more competitive again, although the LHC~(ADD/RS) curve includes only searches up to a \cm\ energy of 8~TeV.\\

\begin{figure}[htbp!]
    \centering
    \includegraphics[width=0.7\linewidth]{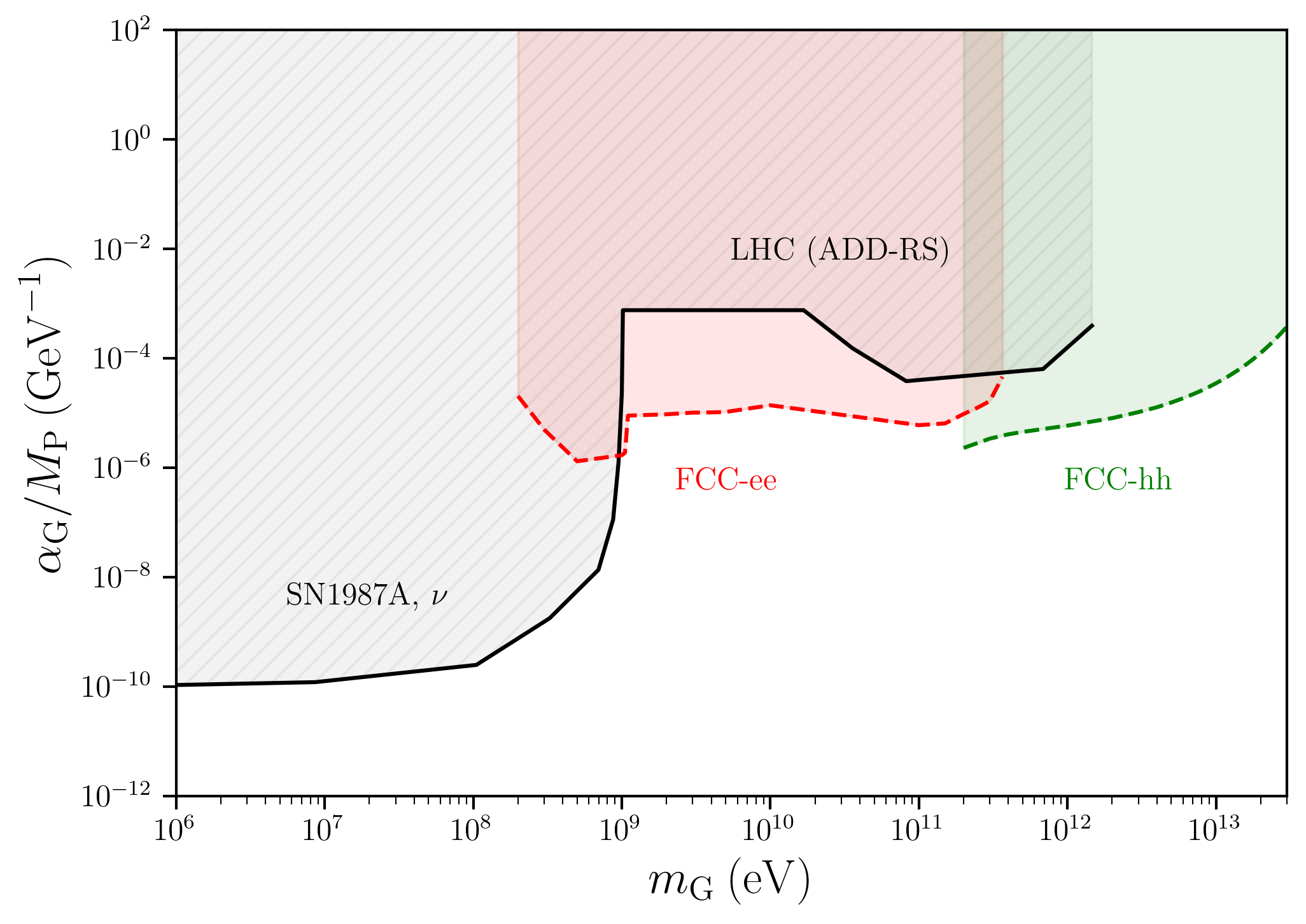}
    \caption{\small Future constraints on the GLP mass vs.\ universal-coupling plane $(\mG,\alphaUniv)$ derived for upcoming collider ALP searches at masses between $1$~MeV and $\sim 10$~TeV at the FCC-ee~\cite{RebelloTeles:2023uig,Polesello:2025gwj} and FCC-hh~\cite{FCC:2025lpp,deBlas:2025gyz,RebelloDdE2026}. The black curve indicates alternative constraints from the SN1987A neutrino signal~\cite{Cembranos:2017vgi} and from spin-2 (ADD- and RS-type) particle searches at the LHC~\cite{Cembranos:2021vdv}.
    \label{fig:himass_bounds_future}}
\end{figure}

Figure~\ref{fig:himass_bounds_future} displays future expected bounds on massive gravitons in the GLP mass–universal coupling plane $(\mG,\alphaUniv)$, derived here from expected ALP searches at the FCC-ee (dashed red curve)~\cite{RebelloTeles:2023uig,Polesello:2025gwj} and FCC-hh (dashed green curve)~\cite{FCC:2025lpp,deBlas:2025gyz,RebelloDdE2026} over the mass range $\mG\approx 1$~MeV--30~TeV, compared with existing spin-2 limits from SN1987A and the LHC\,(ADD/RS) (black curve)~\cite{Cembranos:2021vdv}. Above $\mG\approx 1$~GeV, the FCC sensitivity in exclusive GLP searches via diphoton decays will improve the current limits by about two order-of-magnitude and also probe the unexplored very heavy mass range, $\mG \approx 2$--30~TeV~\cite{RebelloDdE2026}.

\clearpage
%%%%%%%%%%%%%%%%%%%%%%%%%%%%%%%%%%%%%%%%%%
\section{Summary}
\label{sec:summ}

This work has presented a comprehensive reinterpretation of existing limits on spin-0 axion-like particles (ALPs) coupled to photons as constraints on massive spin-2 graviton-like particles (GLPs) with universal coupling ($\alphaUniv$) to Standard Model (SM) fields. The assumption of GLP universal couplings represents the simplest and most theoretically consistent framework, avoiding pathologies such as violations of perturbative unitarity in the $\mG\to0$ limit for scatterings with SM particles. The recasting exploits the close analogy between the production mechanisms of both particles: the Primakoff and Gertsenshtein effects (photon conversion into ALPs and GLPs), respectively, as well as their dominant diphoton decays, with appropriate modifications to account for differences in their spin and interaction structure. The analysis has considered scenarios in which GLPs constitute dark matter as well as cases where this assumption is relaxed. 

Using the correspondence between the two classes of bosons, established on a case-by-case basis, experimental bounds obtained on the ALP mass vs.\ photon-coupling plane $(\ma,\gag)$ have been translated into the corresponding limits in the GLP $(\mG,\alphaUniv)$ parameter space across the full range of masses probed (or to be probed) in laboratory experiments, over $\ma,\mG \approx 10^{-20}$--$10^{14}$ eV. This range of masses encompasses cavity-based detectors (haloscopes and resonant upconversion devices), helioscopes, magnetometers, optical interferometers, beam-dump and fixed-target experiments, collider searches, as well as astrophysical observations and cosmological constraints. 
A detailed dictionary between ALP and GLP limits, as well as a quantitative conversion mapping between the $\gag\Leftrightarrow\alphaUniv$ couplings, have been derived for 17 different present and future ALP searches. Out of the search methods analyzed, eight exhibit comparable sensitivity to ALPs and GLPs, while five (four) of them show relatively enhanced (reduced) sensitivity to GLPs.

\begin{figure}[htbp!]
    \centering
    \includegraphics[width=0.99\linewidth]{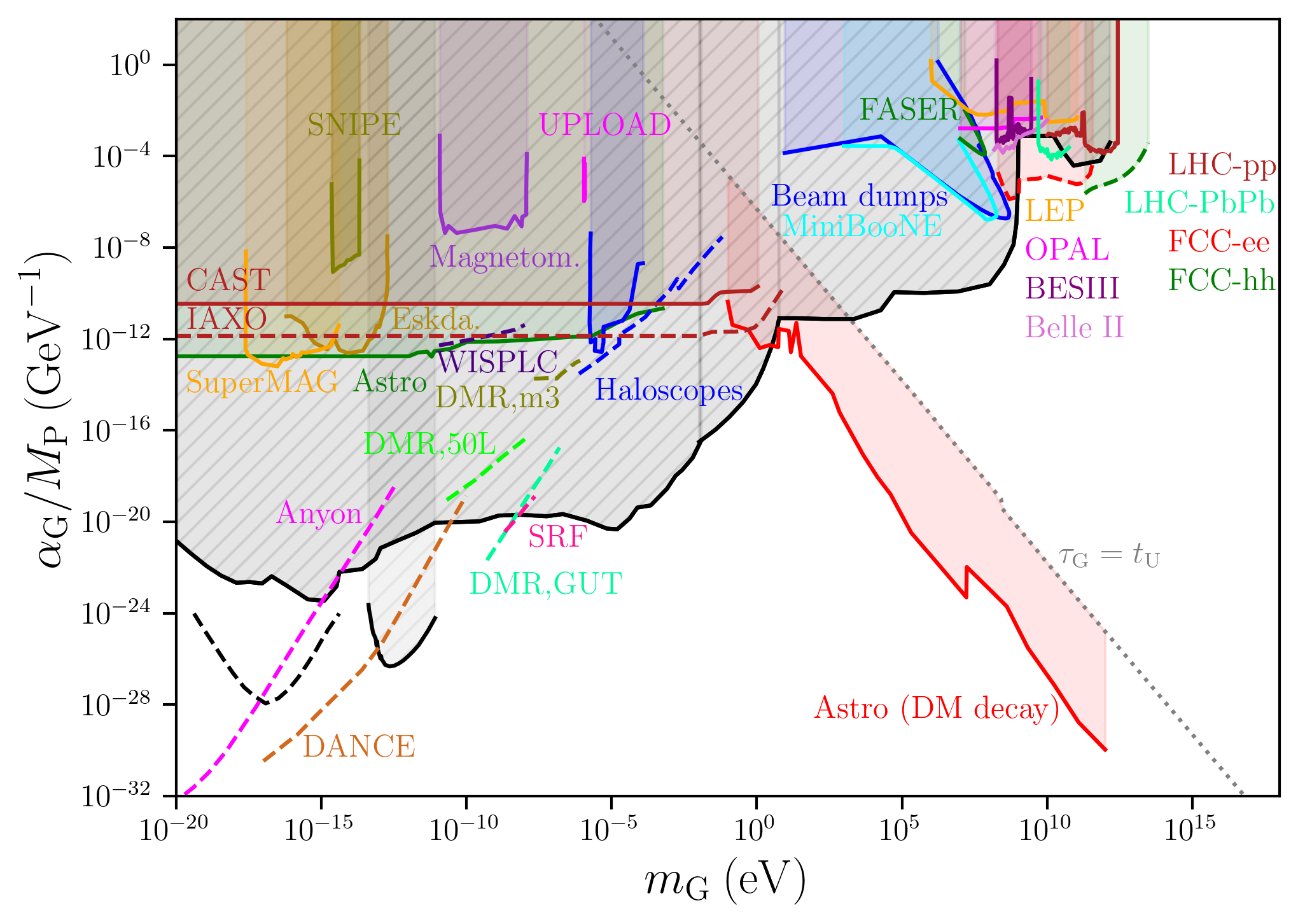}
    \caption{\small Full current (solid lines) and future (dashed lines) constraints in the GLP mass vs.\ universal coupling plane $(\mG,\alphaUniv)$ derived in this work for $\mG=10^{-20}$~eV--30~TeV. The black solid curves indicate alternative spin-2 constraints from 5th-force tests~\cite{Chen:2014oda,Cembranos:2017vgi}, energy loss in stellar environments~\cite{Cembranos:2017vgi}, the neutrino signal of SN1987A~\cite{Cembranos:2017vgi}, ADD/RS particle searches at the LHC~\cite{Cembranos:2021vdv}, as well as LVK data bounds~\cite{Manita:2023mnc}. The dotted black curve indicates the expected LISA sensitivity~\cite{Zhang:2025fck}.
    \label{fig:full_plot}}
\end{figure}

Figure~\ref{fig:full_plot} provides a comprehensive combination of all the results obtained in our study. Our newly extracted GLP bounds in the $(\mG,\alphaUniv)$ plane are compared with previous spin-2 particle limits derived in different mass ranges from a variety of studies: fifth-force probes, GW detectors, SN1987A energy losses, and heavy (ADD/RS) graviton searches. In the low-mass region ($\mG<1$~eV), current ALP searches do not yet surpass the sensitivity of fifth-force experiments in constraining massive spin-2 particles, but future experiments covering the $\mG \lesssim 10^{-8}$~eV range ---such as magnetometers, two-beam interferometers, and upconversion setups--- are expected to probe universal couplings as small as $\alphaUniv \approx 10^{-32}$~GeV$^{-1}$. Notably, these future detectors may exhibit stronger sensitivity to spin-2 particles than to axions, and this fact should be taken into account when analyzing and interpreting the upcoming data. For masses $\mG\approx 1$~eV--1~GeV, the recast ALP limits from beam-dumps, fixed-target and $\epem$ collisions, $\alphaUniv \approx 10^{-1}\mbox{--}10^{-5}$~GeV$^{-1}$, are less competitive than massive graviton bounds derived from the analysis of the SN1987A neutrino signal. Nevertheless, laboratory-based searches remain worthwhile, as they provide a complementary probe that does not rely on the astrophysical assumptions underlying the SN1987A analysis.
In the region $\mG>1$~keV, assuming GLP dark matter, the absence of astrophysical photons from GLP decays provides the most stringent exclusion limits over a decreasing band spanning roughly six orders-of-magnitude in very small universal couplings. The reinterpretation of exclusive searches of ALPs in photon-fusion performed at the LHC provide the most competitive GLP limits, $\alphaUniv \approx 10^{-4}$~GeV$^{-1}$, over $\mG\approx 5$--100~GeV, whereas for heavier masses they provide a complementary strategy to conventional inclusive searches for spin-2 resonances in dijet final states. 

More broadly, and as a final conclusion, we encourage current and future ALP experiments to also consider the potential impact of their apparatus and/or analyses to spin-2 particles and search for, or set limits on, both types of new bosons. We have shown that extending experimental axion-like-particle results to additional beyond-the-SM scenarios can be achieved with relatively modest effort through well-controlled modifications of the interpretative theoretical framework.\\

\paragraph*{Acknowledgments --}

%We thank XXX and YYY for useful discussions on ZZZ.
JG acknowledges support by Grant No.\ CNS2023-143767, funded by MICIU/AEI/10.13039/501100011033 and by the European Union Next-GenerationEU/PRTR. DdE acknowledges collaborative work with Hua-Sheng Shao on the collider studies. 

%%%%%%%%%%%%%%%%%%%%%
\clearpage

\bibliographystyle{myutphys}
\bibliography{main}

\providecommand{\href}[2]{#2}\begingroup\raggedright\begin{thebibliography}{100}

\bibitem{Peccei:1977hh}
R.~D. Peccei and H.~R. Quinn, ``{CP Conservation in the Presence of Instantons},'' \href{http://dx.doi.org/10.1103/PhysRevLett.38.1440}{{\it Phys. Rev. Lett.} {\bfseries 38} (1977) 1440--1443}.

\bibitem{Weinberg:1977ma}
S.~Weinberg, ``{A New Light Boson?},'' \href{http://dx.doi.org/10.1103/PhysRevLett.40.223}{{\it Phys. Rev. Lett.} {\bfseries 40} (1978) 223--226}.

\bibitem{Wilczek:1977pj}
F.~Wilczek, ``{Problem of Strong $P$ and $T$ Invariance in the Presence of Instantons},'' \href{http://dx.doi.org/10.1103/PhysRevLett.40.279}{{\it Phys. Rev. Lett.} {\bfseries 40} (1978) 279--282}.

\bibitem{Duffy:2009ig}
L.~D. Duffy and K.~van Bibber, ``{Axions as Dark Matter Particles},'' \href{http://dx.doi.org/10.1088/1367-2630/11/10/105008}{{\it New J. Phys.} {\bfseries 11} (2009) 105008}, \href{http://arxiv.org/abs/0904.3346}{{\ttfamily arXiv:0904.3346 [hep-ph]}}.

\bibitem{Marsh:2015xka}
D.~J.~E. Marsh, ``{Axion Cosmology},'' \href{http://dx.doi.org/10.1016/j.physrep.2016.06.005}{{\it Phys. Rept.} {\bfseries 643} (2016) 1--79}, \href{http://arxiv.org/abs/1510.07633}{{\ttfamily arXiv:1510.07633 [astro-ph.CO]}}.

\bibitem{Nomura:2008ru}
Y.~Nomura and J.~Thaler, ``{Dark Matter through the Axion Portal},'' \href{http://dx.doi.org/10.1103/PhysRevD.79.075008}{{\it Phys. Rev. D} {\bfseries 79} (2009) 075008}, \href{http://arxiv.org/abs/0810.5397}{{\ttfamily arXiv:0810.5397 [hep-ph]}}.

\bibitem{Dolan:2014ska}
M.~J. Dolan, F.~Kahlhoefer, C.~McCabe, and K.~Schmidt-Hoberg, ``{A taste of dark matter: Flavour constraints on pseudoscalar mediators},'' \href{http://dx.doi.org/10.1007/JHEP03(2015)171}{{\it JHEP} {\bfseries 03} (2015) 171}, \href{http://arxiv.org/abs/1412.5174}{{\ttfamily arXiv:1412.5174 [hep-ph]}}. [Erratum: JHEP 07, 103 (2015)].

\bibitem{Kozaczuk:2015bea}
J.~Kozaczuk and T.~A.~W. Martin, ``{Extending LHC Coverage to Light Pseudoscalar Mediators and Coy Dark Sectors},'' \href{http://dx.doi.org/10.1007/JHEP04(2015)046}{{\it JHEP} {\bfseries 04} (2015) 046}, \href{http://arxiv.org/abs/1501.07275}{{\ttfamily arXiv:1501.07275 [hep-ph]}}.

\bibitem{Cacciapaglia:2019bqz}
G.~Cacciapaglia, G.~Ferretti, T.~Flacke, and H.~Ser{\^o}dio, ``{Light scalars in composite Higgs models},'' \href{http://dx.doi.org/10.3389/fphy.2019.00022}{{\it Front. in Phys.} {\bfseries 7} (2019) 22}, \href{http://arxiv.org/abs/1902.06890}{{\ttfamily arXiv:1902.06890 [hep-ph]}}.

\bibitem{Branco:2011iw}
G.~C. Branco, P.~M. Ferreira, L.~Lavoura, M.~N. Rebelo, M.~Sher, and J.~P. Silva, ``{Theory and phenomenology of two-Higgs-doublet models},'' \href{http://dx.doi.org/10.1016/j.physrep.2012.02.002}{{\it Phys. Rept.} {\bfseries 516} (2012) 1--102}, \href{http://arxiv.org/abs/1106.0034}{{\ttfamily arXiv:1106.0034 [hep-ph]}}.

\bibitem{Ringwald:2012cu}
A.~Ringwald, ``{Searching for axions and ALPs from string theory},'' \href{http://dx.doi.org/10.1088/1742-6596/485/1/012013}{{\it J. Phys. Conf. Ser.} {\bfseries 485} (2014) 012013}, \href{http://arxiv.org/abs/1209.2299}{{\ttfamily arXiv:1209.2299 [hep-ph]}}.

\bibitem{Graham:2015cka}
P.~W. Graham, D.~E. Kaplan, and S.~Rajendran, ``{Cosmological Relaxation of the Electroweak Scale},'' \href{http://dx.doi.org/10.1103/PhysRevLett.115.221801}{{\it Phys. Rev. Lett.} {\bfseries 115} (2015) 221801}, \href{http://arxiv.org/abs/1504.07551}{{\ttfamily arXiv:1504.07551 [hep-ph]}}.

\bibitem{Choi:2020rgn}
K.~Choi, S.~H. Im, and C.~Sub~Shin, ``{Recent Progress in the Physics of Axions and Axion-Like Particles},'' \href{http://dx.doi.org/10.1146/annurev-nucl-120720-031147}{{\it Ann. Rev. Nucl. Part. Sci.} {\bfseries 71} (2021) 225--252}, \href{http://arxiv.org/abs/2012.05029}{{\ttfamily arXiv:2012.05029 [hep-ph]}}.

\bibitem{Hinterbichler:2011tt}
K.~Hinterbichler, ``{Theoretical Aspects of Massive Gravity},'' \href{http://dx.doi.org/10.1103/RevModPhys.84.671}{{\it Rev. Mod. Phys.} {\bfseries 84} (2012) 671--710}, \href{http://arxiv.org/abs/1105.3735}{{\ttfamily arXiv:1105.3735 [hep-th]}}.

\bibitem{Hassan:2011vm}
S.~F. Hassan and R.~A. Rosen, ``{On Non-Linear Actions for Massive Gravity},'' \href{http://dx.doi.org/10.1007/JHEP07(2011)009}{{\it JHEP} {\bfseries 07} (2011) 009}, \href{http://arxiv.org/abs/1103.6055}{{\ttfamily arXiv:1103.6055 [hep-th]}}.

\bibitem{Bellazzini:2023nqj}
B.~Bellazzini, G.~Isabella, S.~Ricossa, and F.~Riva, ``{Massive gravity is not positive},'' \href{http://dx.doi.org/10.1103/PhysRevD.109.024051}{{\it Phys. Rev. D} {\bfseries 109} (2024) 024051}, \href{http://arxiv.org/abs/2304.02550}{{\ttfamily arXiv:2304.02550 [hep-th]}}.

\bibitem{Blas:2024kps}
D.~Blas, J.~Carlton, and C.~McCabe, ``{Massive graviton dark matter searches with long-baseline atom interferometers},'' \href{http://dx.doi.org/10.1103/zxtk-bwnf}{{\it Phys. Rev. D} {\bfseries 111} (2025) 115020}, \href{http://arxiv.org/abs/2412.14282}{{\ttfamily arXiv:2412.14282 [hep-ph]}}.

\bibitem{Babichev:2016bxi}
E.~Babichev, L.~Marzola, M.~Raidal, A.~Schmidt-May, F.~Urban, H.~Veerm{\"a}e, and M.~von Strauss, ``{Heavy spin-2 Dark Matter},'' \href{http://dx.doi.org/10.1088/1475-7516/2016/09/016}{{\it JCAP} {\bfseries 09} (2016) 016}, \href{http://arxiv.org/abs/1607.03497}{{\ttfamily arXiv:1607.03497 [hep-th]}}.

\bibitem{Marzola:2017lbt}
L.~Marzola, M.~Raidal, and F.~R. Urban, ``{Oscillating Spin-2 Dark Matter},'' \href{http://dx.doi.org/10.1103/PhysRevD.97.024010}{{\it Phys. Rev. D} {\bfseries 97} (2018) 024010}, \href{http://arxiv.org/abs/1708.04253}{{\ttfamily arXiv:1708.04253 [hep-ph]}}.

\bibitem{Armaleo:2020}
J.~M. Armaleo, D.~L. Nacir, and F.~R. Urban, ``{Pulsar timing array constraints on spin-2 ULDM},'' \href{http://dx.doi.org/10.1088/1475-7516/2020/09/031}{{\it JCAP} {\bfseries 2020} (2020) 031}.

\bibitem{Cembranos:2017vgi}
J.~A.~R. Cembranos, A.~L. Maroto, and H.~Villarrubia-Rojo, ``{Constraints on hidden gravitons from fifth-force experiments and stellar energy loss},'' \href{http://dx.doi.org/10.1007/JHEP09(2017)104}{{\it JHEP} {\bfseries 09} (2017) 104}, \href{http://arxiv.org/abs/1706.07818}{{\ttfamily arXiv:1706.07818 [hep-ph]}}.

\bibitem{Arkani-Hamed:1998sfv}
N.~Arkani-Hamed, S.~Dimopoulos, and G.~R. Dvali, ``{Phenomenology, astrophysics and cosmology of theories with submillimeter dimensions and TeV scale quantum gravity},'' \href{http://dx.doi.org/10.1103/PhysRevD.59.086004}{{\it Phys. Rev. D} {\bfseries 59} (1999) 086004}, \href{http://arxiv.org/abs/hep-ph/9807344}{{\ttfamily arXiv:hep-ph/9807344}}.

\bibitem{Randall:1999ee}
L.~Randall and R.~Sundrum, ``{A large mass hierarchy from a small extra dimension},'' \href{http://dx.doi.org/10.1103/PhysRevLett.83.3370}{{\it Phys. Rev. Lett.} {\bfseries 83} (1999) 3370--3373}, \href{http://arxiv.org/abs/hep-ph/9905221}{{\ttfamily arXiv:hep-ph/9905221}}.

\bibitem{Aoki:2016zgp}
K.~Aoki and S.~Mukohyama, ``{Massive gravitons as dark matter and gravitational waves},'' \href{http://dx.doi.org/10.1103/PhysRevD.94.024001}{{\it Phys. Rev. D} {\bfseries 94} (2016) 024001}, \href{http://arxiv.org/abs/1604.06704}{{\ttfamily arXiv:1604.06704 [hep-th]}}.

\bibitem{Gonzalo:2022jac}
E.~Gonzalo, M.~Montero, G.~Obied, and C.~Vafa, ``{Dark dimension gravitons as dark matter},'' \href{http://dx.doi.org/10.1007/JHEP11(2023)109}{{\it JHEP} {\bfseries 11} (2023) 109}, \href{http://arxiv.org/abs/2209.09249}{{\ttfamily arXiv:2209.09249 [hep-ph]}}.

\bibitem{deRham:2016nuf}
C.~de~Rham, J.~T. Deskins, A.~J. Tolley, and S.-Y. Zhou, ``{Graviton Mass Bounds},'' \href{http://dx.doi.org/10.1103/RevModPhys.89.025004}{{\it Rev. Mod. Phys.} {\bfseries 89} (2017) 025004}, \href{http://arxiv.org/abs/1606.08462}{{\ttfamily arXiv:1606.08462 [astro-ph.CO]}}.

\bibitem{Graham:2015ouw}
P.~W. Graham, I.~G. Irastorza, S.~K. Lamoreaux, A.~Lindner, and K.~A. van Bibber, ``{Experimental Searches for the Axion and Axion-Like Particles},'' \href{http://dx.doi.org/10.1146/annurev-nucl-102014-022120}{{\it Ann. Rev. Nucl. Part. Sci.} {\bfseries 65} (2015) 485--514}, \href{http://arxiv.org/abs/1602.00039}{{\ttfamily arXiv:1602.00039 [hep-ex]}}.

\bibitem{Jaeckel:2015jla}
J.~Jaeckel and M.~Spannowsky, ``{Probing MeV to 90 GeV axion-like particles with LEP and LHC},'' \href{http://dx.doi.org/10.1016/j.physletb.2015.12.037}{{\it Phys. Lett. B} {\bfseries 753} (2016) 482--487}, \href{http://arxiv.org/abs/1509.00476}{{\ttfamily arXiv:1509.00476 [hep-ph]}}.

\bibitem{Dobrich:2015jyk}
B.~D{\"o}brich, J.~Jaeckel, F.~Kahlhoefer, A.~Ringwald, and K.~Schmidt-Hoberg, ``{ALPtraum: ALP production in proton beam dump experiments},'' \href{http://dx.doi.org/10.1007/JHEP02(2016)018}{{\it JHEP} {\bfseries 02} (2016) 018}, \href{http://arxiv.org/abs/1512.03069}{{\ttfamily arXiv:1512.03069 [hep-ph]}}.

\bibitem{Bauer:2017ris}
M.~Bauer, M.~Neubert, and A.~Thamm, ``{Collider Probes of Axion-Like Particles},'' \href{http://dx.doi.org/10.1007/JHEP12(2017)044}{{\it JHEP} {\bfseries 12} (2017) 044}, \href{http://arxiv.org/abs/1708.00443}{{\ttfamily arXiv:1708.00443 [hep-ph]}}.

\bibitem{Knapen:2016moh}
S.~Knapen, T.~Lin, H.~K. Lou, and T.~Melia, ``{Searching for Axionlike Particles with Ultraperipheral Heavy-Ion Collisions},'' \href{http://dx.doi.org/10.1103/PhysRevLett.118.171801}{{\it Phys. Rev. Lett.} {\bfseries 118} (2017) 171801}, \href{http://arxiv.org/abs/1607.06083}{{\ttfamily arXiv:1607.06083 [hep-ph]}}.

\bibitem{Irastorza:2021tdu}
I.~G. Irastorza, ``{An introduction to axions and their detection},'' \href{http://dx.doi.org/10.21468/SciPostPhysLectNotes.45}{{\it SciPost Phys. Lect. Notes} {\bfseries 45} (2022) 1}, \href{http://arxiv.org/abs/2109.07376}{{\ttfamily arXiv:2109.07376 [hep-ph]}}.

\bibitem{Agrawal:2021dbo}
P.~Agrawal {\it et~al.}, ``{Feebly-interacting particles: FIPs 2020 workshop report},'' \href{http://dx.doi.org/10.1140/epjc/s10052-021-09703-7}{{\it Eur. Phys. J. C} {\bfseries 81} (2021) 1015}, \href{http://arxiv.org/abs/2102.12143}{{\ttfamily arXiv:2102.12143 [hep-ph]}}.

\bibitem{dEnterria:2021ljz}
D.~d'Enterria, ``{Collider constraints on axion-like particles},'' \href{http://arxiv.org/abs/2102.08971}{{\ttfamily arXiv:2102.08971 [hep-ex]}}.

\bibitem{Gertsenshtein:1962kfm}
M.~E. Gertsenshtein and V.~I. Pustovoit, ``{On the detection of low frequency gravitational waves},'' {\it Sov. Phys. JETP} {\bfseries 16} (1962) 433--435. \url{http://www.jetp.ras.ru/cgi-bin/dn/e_016_02_0433.pdf}.

\bibitem{Sikivie:2020zpn}
P.~Sikivie, ``{Invisible Axion Search Methods},'' \href{http://dx.doi.org/10.1103/RevModPhys.93.015004}{{\it Rev. Mod. Phys.} {\bfseries 93} (2021) 015004}, \href{http://arxiv.org/abs/2003.02206}{{\ttfamily arXiv:2003.02206 [hep-ph]}}.

\bibitem{Caputo:2024oqc}
A.~Caputo and G.~Raffelt, ``{Astrophysical Axion Bounds: The 2024 Edition},'' \href{http://dx.doi.org/10.22323/1.454.0041}{{\it PoS} {\bfseries COSMICWISPers} (2024) 041}, \href{http://arxiv.org/abs/2401.13728}{{\ttfamily arXiv:2401.13728 [hep-ph]}}.

\bibitem{Blumlein:1990ay}
J.~Blumlein {\it et~al.}, ``{Limits on neutral light scalar and pseudoscalar particles in a proton beam dump experiment},'' \href{http://dx.doi.org/10.1007/BF01548556}{{\it Z. Phys. C} {\bfseries 51} (1991) 341--350}.

\bibitem{Dobrich:2019dxc}
B.~D{\"o}brich, J.~Jaeckel, and T.~Spadaro, ``{Light in the beam dump - ALP production from decay photons in proton beam-dumps},'' \href{http://dx.doi.org/10.1007/JHEP05(2019)213}{{\it JHEP} {\bfseries 05} (2019) 213}, \href{http://arxiv.org/abs/1904.02091}{{\ttfamily arXiv:1904.02091 [hep-ph]}}. [Erratum: JHEP 10, 046 (2020)].

\bibitem{Jerhot:2022chi}
J.~Jerhot, B.~D{\"o}brich, F.~Ertas, F.~Kahlhoefer, and T.~Spadaro, ``{ALPINIST: Axion-Like Particles In Numerous Interactions Simulated and Tabulated},'' \href{http://dx.doi.org/10.1007/JHEP07(2022)094}{{\it JHEP} {\bfseries 07} (2022) 094}, \href{http://arxiv.org/abs/2201.05170}{{\ttfamily arXiv:2201.05170 [hep-ph]}}.

\bibitem{Harland-Lang:2019zur}
L.~Harland-Lang, J.~Jaeckel, and M.~Spannowsky, ``{A fresh look at ALP searches in fixed target experiments},'' \href{http://dx.doi.org/10.1016/j.physletb.2019.04.045}{{\it Phys. Lett. B} {\bfseries 793} (2019) 281--289}, \href{http://arxiv.org/abs/1902.04878}{{\ttfamily arXiv:1902.04878 [hep-ph]}}.

\bibitem{Mimasu:2014nea}
K.~Mimasu and V.~Sanz, ``{ALPs at Colliders},'' \href{http://dx.doi.org/10.1007/JHEP06(2015)173}{{\it JHEP} {\bfseries 06} (2015) 173}, \href{http://arxiv.org/abs/1409.4792}{{\ttfamily arXiv:1409.4792 [hep-ph]}}.

\bibitem{Brivio:2017ije}
I.~Brivio, M.~B. Gavela, L.~Merlo, K.~Mimasu, J.~M. No, R.~del Rey, and V.~Sanz, ``{ALPs Effective Field Theory and Collider Signatures},'' \href{http://dx.doi.org/10.1140/epjc/s10052-017-5111-3}{{\it Eur. Phys. J. C} {\bfseries 77} (2017) 572}, \href{http://arxiv.org/abs/1701.05379}{{\ttfamily arXiv:1701.05379 [hep-ph]}}.

\bibitem{Biekotter:2025fll}
A.~Biek{\"o}tter and K.~Mimasu, {\it {Axions and Axion-like particles: collider searches}}.
\newblock 8, 2025.
\newblock \href{http://arxiv.org/abs/2508.19358}{{\ttfamily arXiv:2508.19358 [hep-ph]}}.

\bibitem{Redondo:2010dp}
J.~Redondo and A.~Ringwald, ``{Light shining through walls},'' \href{http://dx.doi.org/10.1080/00107514.2011.563516}{{\it Contemp. Phys.} {\bfseries 52} (2011) 211--236}, \href{http://arxiv.org/abs/1011.3741}{{\ttfamily arXiv:1011.3741 [hep-ph]}}.

\bibitem{OSQAR:2015qdv}
{\bfseries OSQAR} Collaboration, R.~Ballou {\it et~al.}, ``{New exclusion limits on scalar and pseudoscalar axionlike particles from light shining through a wall},'' \href{http://dx.doi.org/10.1103/PhysRevD.92.092002}{{\it Phys. Rev. D} {\bfseries 92} (2015) 092002}, \href{http://arxiv.org/abs/1506.08082}{{\ttfamily arXiv:1506.08082 [hep-ex]}}.

\bibitem{ALPSII:2025eri}
{\bfseries ALPS II} Collaboration, D.~C. Brotherton {\it et~al.}, ``{Any Light Particle Searches with ALPS II: first science results},'' \href{http://arxiv.org/abs/2512.14110}{{\ttfamily arXiv:2512.14110 [hep-ex]}}.

\bibitem{Fierz:1939ix}
M.~Fierz and W.~Pauli, ``{On relativistic wave equations for particles of arbitrary spin in an electromagnetic field},'' \href{http://dx.doi.org/10.1098/rspa.1939.0140}{{\it Proc. Roy. Soc. Lond. A} {\bfseries 173} (1939) 211--232}.

\bibitem{Armaleo:2019gil}
J.~M. Armaleo, D.~L{\'o}pez~Nacir, and F.~R. Urban, ``{Binary pulsars as probes for spin-2 ultralight dark matter},'' \href{http://dx.doi.org/10.1088/1475-7516/2020/01/053}{{\it JCAP} {\bfseries 01} (2020) 053}, \href{http://arxiv.org/abs/1909.13814}{{\ttfamily arXiv:1909.13814 [astro-ph.HE]}}.

\bibitem{Das:2016pbk}
G.~Das, C.~Degrande, V.~Hirschi, F.~Maltoni, and H.-S. Shao, ``{NLO predictions for the production of a spin-two particle at the LHC},'' \href{http://dx.doi.org/10.1016/j.physletb.2017.05.007}{{\it Phys. Lett. B} {\bfseries 770} (2017) 507--513}, \href{http://arxiv.org/abs/1605.09359}{{\ttfamily arXiv:1605.09359 [hep-ph]}}.

\bibitem{dEnterria:2023}
D.~d'Enterria, M.~A. Tamlihat, L.~Schoeffel, H.-S. Shao, and Y.~Tayalati, ``{Collider constraints on massive gravitons coupling to photons},'' \href{http://dx.doi.org/10.1016/j.physletb.2023.138237}{{\it Phys. Lett. B} {\bfseries 846} (2023) 138237}, \href{http://arxiv.org/abs/2306.15558}{{\ttfamily arXiv:2306.15558 [hep-ph]}}.

\bibitem{Georgi:1986df}
H.~Georgi, D.~B. Kaplan, and L.~Randall, ``{Manifesting the invisible axion at low energies},'' \href{http://dx.doi.org/10.1016/0370-2693(86)90688-X}{{\it Phys. Lett. B} {\bfseries 169} (1986) 73--78}.

\bibitem{DiLuzio:2020wdo}
L.~Di~Luzio, M.~Giannotti, E.~Nardi, and L.~Visinelli, ``{The landscape of QCD axion models},'' \href{http://dx.doi.org/10.1016/j.physrep.2020.06.002}{{\it Phys. Rept.} {\bfseries 870} (2020) 1--117}, \href{http://arxiv.org/abs/2003.01100}{{\ttfamily arXiv:2003.01100 [hep-ph]}}.

\bibitem{Craig:2018kne}
N.~Craig, A.~Hook, and S.~Kasko, ``{The Photophobic ALP},'' \href{http://dx.doi.org/10.1007/JHEP09(2018)028}{{\it JHEP} {\bfseries 09} (2018) 028}, \href{http://arxiv.org/abs/1805.06538}{{\ttfamily arXiv:1805.06538 [hep-ph]}}.

\bibitem{Blinov:2021say}
N.~Blinov, E.~Kowalczyk, and M.~Wynne, ``{Axion-like particle searches at DarkQuest},'' \href{http://dx.doi.org/10.1007/JHEP02(2022)036}{{\it JHEP} {\bfseries 02} (2022) 036}, \href{http://arxiv.org/abs/2112.09814}{{\ttfamily arXiv:2112.09814 [hep-ph]}}.

\bibitem{Alda:2025nsz}
J.~Alda, M.~Fuentes~Zamoro, L.~Merlo, X.~Ponce~D{\'\i}az, and S.~Rigolin, ``{ALPaca: The ALP Automatic Computing Algorithm},'' \href{http://arxiv.org/abs/2508.08354}{{\ttfamily arXiv:2508.08354 [hep-ph]}}.

\bibitem{Darme:2020sjf}
L.~Darm{\'e}, F.~Giacchino, E.~Nardi, and M.~Raggi, ``{Invisible decays of axion-like particles: constraints and prospects},'' \href{http://dx.doi.org/10.1007/JHEP06(2021)009}{{\it JHEP} {\bfseries 06} (2021) 009}, \href{http://arxiv.org/abs/2012.07894}{{\ttfamily arXiv:2012.07894 [hep-ph]}}.

\bibitem{Spira:1995rr}
M.~Spira, A.~Djouadi, D.~Graudenz, and P.~M. Zerwas, ``{Higgs boson production at the LHC},'' \href{http://dx.doi.org/10.1016/0550-3213(95)00379-7}{{\it Nucl. Phys. B} {\bfseries 453} (1995) 17--82}, \href{http://arxiv.org/abs/hep-ph/9504378}{{\ttfamily arXiv:hep-ph/9504378}}.

\bibitem{Higuchi:1986py}
A.~Higuchi, ``{Forbidden Mass Range for Spin-2 Field Theory in De Sitter Space-time},'' \href{http://dx.doi.org/10.1016/0550-3213(87)90691-2}{{\it Nucl. Phys. B} {\bfseries 282} (1987) 397--436}.

\bibitem{Zimmermann:2024xvd}
T.~Zimmermann, J.~Alvey, D.~J.~E. Marsh, M.~Fairbairn, and J.~I. Read, ``{Dwarf Galaxies Imply Dark Matter is Heavier than $2.2{\times}10^{-21}$~eV},'' \href{http://dx.doi.org/10.1103/PhysRevLett.134.151001}{{\it Phys. Rev. Lett.} {\bfseries 134} (2025) 151001}, \href{http://arxiv.org/abs/2405.20374}{{\ttfamily arXiv:2405.20374 [astro-ph.CO]}}.

\bibitem{Cirelli:2024ssz}
M.~Cirelli, A.~Strumia, and J.~Zupan, ``{Dark Matter},'' \href{http://arxiv.org/abs/2406.01705}{{\ttfamily arXiv:2406.01705 [hep-ph]}}.

\bibitem{Freese:2012xd}
K.~Freese, M.~Lisanti, and C.~Savage, ``{Colloquium: Annual modulation of dark matter},'' \href{http://dx.doi.org/10.1103/RevModPhys.85.1561}{{\it Rev. Mod. Phys.} {\bfseries 85} (2013) 1561--1581}, \href{http://arxiv.org/abs/1209.3339}{{\ttfamily arXiv:1209.3339 [astro-ph.CO]}}.

\bibitem{Kuhlen:2013tra}
M.~Kuhlen, A.~Pillepich, J.~Guedes, and P.~Madau, ``{The Distribution of Dark Matter in the Milky Way's Disk},'' \href{http://dx.doi.org/10.1088/0004-637X/784/2/161}{{\it Astrophys. J.} {\bfseries 784} (2014) 161}, \href{http://arxiv.org/abs/1308.1703}{{\ttfamily arXiv:1308.1703 [astro-ph.GA]}}.

\bibitem{Domcke:2022rgu}
V.~Domcke, C.~Garcia-Cely, and N.~L. Rodd, ``{Novel Search for High-Frequency Gravitational Waves with Low-Mass Axion Haloscopes},'' \href{http://dx.doi.org/10.1103/PhysRevLett.129.041101}{{\it Phys. Rev. Lett.} {\bfseries 129} (2022) 041101}, \href{http://arxiv.org/abs/2202.00695}{{\ttfamily arXiv:2202.00695 [hep-ph]}}.

\bibitem{Berlin:2022}
A.~Berlin, D.~Blas, R.~T. D'Agnolo, S.~A. Ellis, R.~Harnik, Y.~Kahn, and J.~Schütte-Engel, ``{Detecting high-frequency gravitational waves with microwave cavities},'' \href{http://dx.doi.org/10.1103/PhysRevD.105.116011}{{\it Phys. Rev. D} {\bfseries 105} (2022) 116011}.

\bibitem{ADMX:2020ote}
{\bfseries ADMX} Collaboration, R.~Khatiwada {\it et~al.}, ``{Axion Dark Matter Experiment: Detailed design~and operations},'' \href{http://dx.doi.org/10.1063/5.0037857}{{\it Rev. Sci. Instrum.} {\bfseries 92} (2021) 124502}, \href{http://arxiv.org/abs/2010.00169}{{\ttfamily arXiv:2010.00169 [astro-ph.IM]}}.

\bibitem{Lee:2020cfj}
S.~Lee, S.~Ahn, J.~Choi, B.~R. Ko, and Y.~K. Semertzidis, ``{Axion Dark Matter Search around 6.7 $\mu$eV},'' \href{http://dx.doi.org/10.1103/PhysRevLett.124.101802}{{\it Phys. Rev. Lett.} {\bfseries 124} (2020) 101802}, \href{http://arxiv.org/abs/2001.05102}{{\ttfamily arXiv:2001.05102 [hep-ex]}}.

\bibitem{Brubaker:2016ktl}
B.~M. Brubaker {\it et~al.}, ``{First results from a microwave cavity axion search at 24 $\mu$eV},'' \href{http://dx.doi.org/10.1103/PhysRevLett.118.061302}{{\it Phys. Rev. Lett.} {\bfseries 118} (2017) 061302}, \href{http://arxiv.org/abs/1610.02580}{{\ttfamily arXiv:1610.02580 [astro-ph.CO]}}.

\bibitem{Quiskamp:2022pks}
{\bfseries ORGAN} Collaboration, A.~P. Quiskamp, B.~T. McAllister, P.~Altin, E.~N. Ivanov, M.~Goryachev, and M.~E. Tobar, ``{Direct search for dark matter axions excluding ALP cogenesis in the 63- to 67-{\ensuremath{\mu}}eV range with the ORGAN experiment},'' \href{http://dx.doi.org/10.1126/sciadv.abq3765}{{\it Sci. Adv.} {\bfseries 8} (2022) abq3765}, \href{http://arxiv.org/abs/2203.12152}{{\ttfamily arXiv:2203.12152 [hep-ex]}}.

\bibitem{Ahyoune:2024klt}
S.~Ahyoune {\it et~al.}, ``{RADES axion search results with a high-temperature superconducting cavity in an 11.7 T magnet},'' \href{http://dx.doi.org/10.1007/JHEP04(2025)113}{{\it JHEP} {\bfseries 04} (2025) 113}, \href{http://arxiv.org/abs/2403.07790}{{\ttfamily arXiv:2403.07790 [hep-ex]}}.

\bibitem{Alesini:2019ajt}
D.~Alesini {\it et~al.}, ``{Galactic axions search with a superconducting resonant cavity},'' \href{http://dx.doi.org/10.1103/PhysRevD.99.101101}{{\it Phys. Rev. D} {\bfseries 99} (2019) 101101}, \href{http://arxiv.org/abs/1903.06547}{{\ttfamily arXiv:1903.06547 [physics.ins-det]}}.

\bibitem{Grenet:2021vbb}
T.~Grenet, R.~Ballou, Q.~Basto, K.~Martineau, P.~Perrier, P.~Pugnat, J.~Quevillon, N.~Roch, and C.~Smith, ``{The Grenoble Axion Haloscope platform (GrAHal): development plan and first results},'' \href{http://arxiv.org/abs/2110.14406}{{\ttfamily arXiv:2110.14406 [hep-ex]}}.

\bibitem{Gue:2026kga}
J.~Gu{\'e}, T.~Krokotsch, and G.~Moortgat-Pick, ``{Covariant eigenmode overlap formalism for gravitational wave signals in electromagnetic cavities},'' \href{http://arxiv.org/abs/2602.08507}{{\ttfamily arXiv:2602.08507 [gr-qc]}}.

\bibitem{Kahn:2016aff}
Y.~Kahn, B.~R. Safdi, and J.~Thaler, ``{Broadband and Resonant Approaches to Axion Dark Matter Detection},'' \href{http://dx.doi.org/10.1103/PhysRevLett.117.141801}{{\it Phys. Rev. Lett.} {\bfseries 117} (2016) 141801}, \href{http://arxiv.org/abs/1602.01086}{{\ttfamily arXiv:1602.01086 [hep-ph]}}.

\bibitem{Ouellet:2019tlz}
J.~L. Ouellet {\it et~al.}, ``{Design and implementation of the ABRACADABRA-10 cm axion dark matter search},'' \href{http://dx.doi.org/10.1103/PhysRevD.99.052012}{{\it Phys. Rev. D} {\bfseries 99} (2019) 052012}, \href{http://arxiv.org/abs/1901.10652}{{\ttfamily arXiv:1901.10652 [physics.ins-det]}}.

\bibitem{Gramolin:2020ict}
A.~V. Gramolin, D.~Aybas, D.~Johnson, J.~Adam, and A.~O. Sushkov, ``{Search for axion-like dark matter with ferromagnets},'' \href{http://dx.doi.org/10.1038/s41567-020-1006-6}{{\it Nature Phys.} {\bfseries 17} (2021) 79--84}, \href{http://arxiv.org/abs/2003.03348}{{\ttfamily arXiv:2003.03348 [hep-ex]}}.

\bibitem{DMRadio:2022pkf}
{\bfseries DMRadio} Collaboration, L.~Brouwer {\it et~al.}, ``{Projected sensitivity of DMRadio-m3: A search for the QCD axion below 1{\,}{\,}{\ensuremath{\mu}}eV},'' \href{http://dx.doi.org/10.1103/PhysRevD.106.103008}{{\it Phys. Rev. D} {\bfseries 106} (2022) 103008}, \href{http://arxiv.org/abs/2204.13781}{{\ttfamily arXiv:2204.13781 [hep-ex]}}.

\bibitem{DMRadio:2022jfv}
{\bfseries DMRadio} Collaboration, L.~Brouwer {\it et~al.}, ``{Proposal for a definitive search for GUT-scale QCD axions},'' \href{http://dx.doi.org/10.1103/PhysRevD.106.112003}{{\it Phys. Rev. D} {\bfseries 106} (2022) 112003}, \href{http://arxiv.org/abs/2203.11246}{{\ttfamily arXiv:2203.11246 [hep-ex]}}.

\bibitem{Zhang:2021bpa}
Z.~Zhang, D.~Horns, and O.~Ghosh, ``{Search for dark matter with an LC circuit},'' \href{http://dx.doi.org/10.1103/PhysRevD.106.023003}{{\it Phys. Rev. D} {\bfseries 106} (2022) 023003}, \href{http://arxiv.org/abs/2111.04541}{{\ttfamily arXiv:2111.04541 [hep-ex]}}.

\bibitem{Sulai:2023zqw}
I.~A. Sulai {\it et~al.}, ``{Hunt for magnetic signatures of hidden-photon and axion dark matter in the wilderness},'' \href{http://dx.doi.org/10.1103/PhysRevD.108.096026}{{\it Phys. Rev. D} {\bfseries 108} (2023) 096026}, \href{http://arxiv.org/abs/2306.11575}{{\ttfamily arXiv:2306.11575 [hep-ph]}}.

\bibitem{Nishizawa:2025xka}
A.~Nishizawa, A.~Taruya, and Y.~Himemoto, ``{Axion dark matter search from terrestrial magnetic fields at extremely low frequencies},'' \href{http://arxiv.org/abs/2504.07559}{{\ttfamily arXiv:2504.07559 [hep-ph]}}.

\bibitem{Arza:2021ekq}
A.~Arza, M.~A. Fedderke, P.~W. Graham, D.~F.~J. Kimball, and S.~Kalia, ``{Earth as a transducer for axion dark-matter detection},'' \href{http://dx.doi.org/10.1103/PhysRevD.105.095007}{{\it Phys. Rev. D} {\bfseries 105} (2022) 095007}, \href{http://arxiv.org/abs/2112.09620}{{\ttfamily arXiv:2112.09620 [hep-ph]}}.

\bibitem{Friel:2024shg}
M.~Friel, J.~W. Gjerloev, S.~Kalia, and A.~Zamora, ``{Search for ultralight dark matter in the SuperMAG high-fidelity dataset},'' \href{http://dx.doi.org/10.1103/PhysRevD.110.115036}{{\it Phys. Rev. D} {\bfseries 110} (2024) 115036}, \href{http://arxiv.org/abs/2408.16045}{{\ttfamily arXiv:2408.16045 [hep-ph]}}.

\bibitem{Obata:2018vvr}
I.~Obata, T.~Fujita, and Y.~Michimura, ``{Optical Ring Cavity Search for Axion Dark Matter},'' \href{http://dx.doi.org/10.1103/PhysRevLett.121.161301}{{\it Phys. Rev. Lett.} {\bfseries 121} (2018) 161301}, \href{http://arxiv.org/abs/1805.11753}{{\ttfamily arXiv:1805.11753 [astro-ph.CO]}}.

\bibitem{Pandey:2024dcd}
S.~Pandey, E.~D. Hall, and M.~Evans, ``{First Results from the Axion Dark-Matter Birefringent Cavity (ADBC) Experiment},'' \href{http://dx.doi.org/10.1103/PhysRevLett.133.111003}{{\it Phys. Rev. Lett.} {\bfseries 133} (2024) 111003}, \href{http://arxiv.org/abs/2404.12517}{{\ttfamily arXiv:2404.12517 [hep-ex]}}.

\bibitem{Heinze:2023nfb}
J.~Heinze, A.~Gill, A.~Dmitriev, J.~Smetana, T.~Yan, V.~Boyer, D.~Martynov, and M.~Evans, ``{First Results of the Laser-Interferometric Detector for Axions (LIDA)},'' \href{http://dx.doi.org/10.1103/PhysRevLett.132.191002}{{\it Phys. Rev. Lett.} {\bfseries 132} (2024) 191002}, \href{http://arxiv.org/abs/2307.01365}{{\ttfamily arXiv:2307.01365 [astro-ph.CO]}}.

\bibitem{Nagano:2019rbw}
K.~Nagano, T.~Fujita, Y.~Michimura, and I.~Obata, ``{Axion Dark Matter Search with Interferometric Gravitational Wave Detectors},'' \href{http://dx.doi.org/10.1103/PhysRevLett.123.111301}{{\it Phys. Rev. Lett.} {\bfseries 123} (2019) 111301}, \href{http://arxiv.org/abs/1903.02017}{{\ttfamily arXiv:1903.02017 [hep-ph]}}.

\bibitem{Gue25}
J.~Gu{\'e}, A.~Hees, and P.~Wolf, ``{Probing the axion{\textendash}photon coupling with space-based gravitational wave detectors},'' \href{http://dx.doi.org/10.1088/1361-6382/adb23c}{{\it Class. Quant. Grav.} {\bfseries 42} (2025) 055015}, \href{http://arxiv.org/abs/2410.17763}{{\ttfamily arXiv:2410.17763 [hep-ph]}}.

\bibitem{Yao25}
Y.-H. Yao, T.~Jiang, and Y.~Tang, ``{Prospects for axion dark matter searches at LISA-like interferometers},'' \href{http://dx.doi.org/10.1103/PhysRevD.111.055031}{{\it Phys. Rev. D} {\bfseries 111} (2025) 055031}, \href{http://arxiv.org/abs/2410.22072}{{\ttfamily arXiv:2410.22072 [hep-ph]}}.

\bibitem{YaoR25}
R.-M. Yao, X.-J. Bi, P.-F. Yin, and Q.-G. Huang, ``{Axion-like Dark Matter Search with Space-based Gravitational Wave Detectors},'' \href{http://arxiv.org/abs/2504.10083}{{\ttfamily arXiv:2504.10083 [hep-ph]}}.

\bibitem{Heinze:2024bdc}
J.~Heinze {\it et~al.}, ``{DarkGEO: a large-scale laser-interferometric axion detector},'' \href{http://dx.doi.org/10.1088/1367-2630/ad48ac}{{\it New J. Phys.} {\bfseries 26} (2024) 055002}, \href{http://arxiv.org/abs/2401.11907}{{\ttfamily arXiv:2401.11907 [astro-ph.CO]}}.

\bibitem{Oshima:2023csb}
Y.~Oshima, H.~Fujimoto, J.~Kume, S.~Morisaki, K.~Nagano, T.~Fujita, I.~Obata, A.~Nishizawa, Y.~Michimura, and M.~Ando, ``{First results of axion dark matter search with DANCE},'' \href{http://dx.doi.org/10.1103/PhysRevD.108.072005}{{\it Phys. Rev. D} {\bfseries 108} (2023) 072005}, \href{http://arxiv.org/abs/2303.03594}{{\ttfamily arXiv:2303.03594 [hep-ex]}}.

\bibitem{Thomson:2023moc}
C.~A. Thomson, M.~Goryachev, B.~T. McAllister, E.~N. Ivanov, P.~Altin, and M.~E. Tobar, ``{Searching for low-mass axions using resonant upconversion},'' \href{http://dx.doi.org/10.1103/PhysRevD.107.112003}{{\it Phys. Rev. D} {\bfseries 107} (2023) 112003}, \href{http://arxiv.org/abs/2301.06778}{{\ttfamily arXiv:2301.06778 [hep-ex]}}.

\bibitem{Bourhill:2022alm}
J.~F. Bourhill, E.~C.~I. Paterson, M.~Goryachev, and M.~E. Tobar, ``{Searching for ultralight axions with twisted cavity resonators of anyon rotational symmetry with bulk modes of nonzero helicity},'' \href{http://dx.doi.org/10.1103/PhysRevD.108.052014}{{\it Phys. Rev. D} {\bfseries 108} (2023) 052014}, \href{http://arxiv.org/abs/2208.01640}{{\ttfamily arXiv:2208.01640 [hep-ph]}}.

\bibitem{Li:2025pyi}
Z.~Li, K.~Zhou, M.~Oriunno, A.~Berlin, S.~Calatroni, R.~Tito~D'Agnolo, S.~A.~R. Ellis, P.~Schuster, S.~G. Tantawi, and N.~Toro, ``{A Prototype Hybrid Mode Cavity for Heterodyne Axion Detection},'' \href{http://arxiv.org/abs/2507.07173}{{\ttfamily arXiv:2507.07173 [physics.ins-det]}}.

\bibitem{Bolliet:2020ofj}
B.~Bolliet, J.~Chluba, and R.~Battye, ``{Spectral distortion constraints on photon injection from low-mass decaying particles},'' \href{http://dx.doi.org/10.1093/mnras/stab1997}{{\it Mon. Not. Roy. Astron. Soc.} {\bfseries 507} (2021) 3148--3178}, \href{http://arxiv.org/abs/2012.07292}{{\ttfamily arXiv:2012.07292 [astro-ph.CO]}}.

\bibitem{Thorpe-Morgan:2020rwc}
C.~Thorpe-Morgan, D.~Malyshev, A.~Santangelo, J.~Jochum, B.~J{\"a}ger, M.~Sasaki, and S.~Saeedi, ``{THESEUS insights into axionlike particles, dark photon, and sterile neutrino dark matter},'' \href{http://dx.doi.org/10.1103/PhysRevD.102.123003}{{\it Phys. Rev. D} {\bfseries 102} (2020) 123003}, \href{http://arxiv.org/abs/2008.08306}{{\ttfamily arXiv:2008.08306 [astro-ph.HE]}}.

\bibitem{Foster:2021ngm}
J.~W. Foster, M.~Kongsore, C.~Dessert, Y.~Park, N.~L. Rodd, K.~Cranmer, and B.~R. Safdi, ``{Deep Search for Decaying Dark Matter with XMM-Newton Blank-Sky Observations},'' \href{http://dx.doi.org/10.1103/PhysRevLett.127.051101}{{\it Phys. Rev. Lett.} {\bfseries 127} (2021) 051101}, \href{http://arxiv.org/abs/2102.02207}{{\ttfamily arXiv:2102.02207 [astro-ph.CO]}}.

\bibitem{Wadekar:2021qae}
D.~Wadekar and Z.~Wang, ``{Strong constraints on decay and annihilation of dark matter from heating of gas-rich dwarf galaxies},'' \href{http://dx.doi.org/10.1103/PhysRevD.106.075007}{{\it Phys. Rev. D} {\bfseries 106} (2022) 075007}, \href{http://arxiv.org/abs/2111.08025}{{\ttfamily arXiv:2111.08025 [hep-ph]}}.

\bibitem{Roach:2022lgo}
B.~M. Roach, S.~Rossland, K.~C.~Y. Ng, K.~Perez, J.~F. Beacom, B.~W. Grefenstette, S.~Horiuchi, R.~Krivonos, and D.~R. Wik, ``{Long-exposure NuSTAR constraints on decaying dark matter in the Galactic halo},'' \href{http://dx.doi.org/10.1103/PhysRevD.107.023009}{{\it Phys. Rev. D} {\bfseries 107} (2023) 023009}, \href{http://arxiv.org/abs/2207.04572}{{\ttfamily arXiv:2207.04572 [astro-ph.HE]}}.

\bibitem{Calore:2022pks}
F.~Calore, A.~Dekker, P.~D. Serpico, and T.~Siegert, ``{Constraints on light decaying dark matter candidates from 16~yr of INTEGRAL/SPI observations},'' \href{http://dx.doi.org/10.1093/mnras/stad457}{{\it Mon. Not. Roy. Astron. Soc.} {\bfseries 520} (2023) 4167--4172}, \href{http://arxiv.org/abs/2209.06299}{{\ttfamily arXiv:2209.06299 [hep-ph]}}. [Erratum: Mon.Not.Roy.Astron.Soc. 538, 132 (2025)].

\bibitem{Carenza:2023qxh}
P.~Carenza, G.~Lucente, and E.~Vitagliano, ``{Probing the blue axion with cosmic optical background anisotropies},'' \href{http://dx.doi.org/10.1103/PhysRevD.107.083032}{{\it Phys. Rev. D} {\bfseries 107} (2023) 083032}, \href{http://arxiv.org/abs/2301.06560}{{\ttfamily arXiv:2301.06560 [hep-ph]}}.

\bibitem{Wang:2023imi}
H.~Wang {\it et~al.}, ``{Spectroscopic search for optical emission lines from dark matter decay},'' \href{http://dx.doi.org/10.1103/PhysRevD.110.103007}{{\it Phys. Rev. D} {\bfseries 110} (2024) 103007}, \href{http://arxiv.org/abs/2311.05476}{{\ttfamily arXiv:2311.05476 [astro-ph.CO]}}.

\bibitem{Todarello:2023hdk}
E.~Todarello, M.~Regis, J.~Reynoso-Cordova, M.~Taoso, D.~Vaz, J.~Brinchmann, M.~Steinmetz, and S.~L. Zoutendijke, ``{Robust bounds on ALP dark matter from dwarf spheroidal galaxies in the optical MUSE-Faint survey},'' \href{http://dx.doi.org/10.1088/1475-7516/2024/05/043}{{\it JCAP} {\bfseries 05} (2024) 043}, \href{http://arxiv.org/abs/2307.07403}{{\ttfamily arXiv:2307.07403 [astro-ph.CO]}}.

\bibitem{Pinetti:2025owq}
E.~Pinetti, ``{First constraints on QCD axion dark matter using James Webb Space Telescope observations},'' \href{http://arxiv.org/abs/2503.11753}{{\ttfamily arXiv:2503.11753 [hep-ph]}}.

\bibitem{Sun:2023acy}
Y.~Sun, J.~W. Foster, H.~Liu, J.~B. Mu{\~n}oz, and T.~R. Slatyer, ``{Inhomogeneous energy injection in the 21-cm power spectrum: Sensitivity to dark matter decay},'' \href{http://dx.doi.org/10.1103/PhysRevD.111.043015}{{\it Phys. Rev. D} {\bfseries 111} (2025) 043015}, \href{http://arxiv.org/abs/2312.11608}{{\ttfamily arXiv:2312.11608 [hep-ph]}}.

\bibitem{Saha:2025any}
A.~K. Saha, S.~Bouri, A.~Das, A.~Dubey, and R.~Laha, ``{Shedding Infrared Light on QCD Axion and ALP Dark Matter with JWST},'' \href{http://arxiv.org/abs/2503.14582}{{\ttfamily arXiv:2503.14582 [hep-ph]}}.

\bibitem{CAST:2017uph}
{\bfseries CAST} Collaboration, V.~Anastassopoulos {\it et~al.}, ``{New CAST Limit on the Axion-Photon Interaction},'' \href{http://dx.doi.org/10.1038/nphys4109}{{\it Nature Phys.} {\bfseries 13} (2017) 584--590}, \href{http://arxiv.org/abs/1705.02290}{{\ttfamily arXiv:1705.02290 [hep-ex]}}.

\bibitem{IAXO:2019mpb}
{\bfseries IAXO} Collaboration, E.~Armengaud {\it et~al.}, ``{Physics potential of the International Axion Observatory (IAXO)},'' \href{http://dx.doi.org/10.1088/1475-7516/2019/06/047}{{\it JCAP} {\bfseries 06} (2019) 047}, \href{http://arxiv.org/abs/1904.09155}{{\ttfamily arXiv:1904.09155 [hep-ph]}}.

\bibitem{Guarini:2020hps}
E.~Guarini, P.~Carenza, J.~Galan, M.~Giannotti, and A.~Mirizzi, ``{Production of axionlike particles from photon conversions in large-scale solar magnetic fields},'' \href{http://dx.doi.org/10.1103/PhysRevD.102.123024}{{\it Phys. Rev. D} {\bfseries 102} (2020) 123024}, \href{http://arxiv.org/abs/2010.06601}{{\ttfamily arXiv:2010.06601 [hep-ph]}}.

\bibitem{Garcia-Cely:2025ula}
C.~Garc{\'\i}a-Cely and A.~Ringwald, ``{Stellar Bounds on Light Spin-2 Particles in Bimetric Theories},'' \href{http://arxiv.org/abs/2511.03707}{{\ttfamily arXiv:2511.03707 [hep-ph]}}.

\bibitem{Wouters:2013hua}
D.~Wouters and P.~Brun, ``{Constraints on Axion-like Particles from X-Ray Observations of the Hydra Galaxy Cluster},'' \href{http://dx.doi.org/10.1088/0004-637X/772/1/44}{{\it Astrophys. J.} {\bfseries 772} (2013) 44}, \href{http://arxiv.org/abs/1304.0989}{{\ttfamily arXiv:1304.0989 [astro-ph.HE]}}.

\bibitem{Marsh:2017yvc}
M.~C.~D. Marsh, H.~R. Russell, A.~C. Fabian, B.~P. McNamara, P.~Nulsen, and C.~S. Reynolds, ``{A New Bound on Axion-Like Particles},'' \href{http://dx.doi.org/10.1088/1475-7516/2017/12/036}{{\it JCAP} {\bfseries 12} (2017) 036}, \href{http://arxiv.org/abs/1703.07354}{{\ttfamily arXiv:1703.07354 [hep-ph]}}.

\bibitem{Reynolds:2019uqt}
C.~S. Reynolds, M.~C.~D. Marsh, H.~R. Russell, A.~C. Fabian, R.~Smith, F.~Tombesi, and S.~Veilleux, ``{Astrophysical limits on very light axion-like particles from Chandra grating spectroscopy of NGC 1275},'' \href{http://dx.doi.org/10.3847/1538-4357/ab6a0c}{{\it Astrophys. J.} {\bfseries 890} (2020) 59}, \href{http://arxiv.org/abs/1907.05475}{{\ttfamily arXiv:1907.05475 [hep-ph]}}.

\bibitem{Reynes:2021bpe}
J.~S. Reyn{\'e}s, J.~H. Matthews, C.~S. Reynolds, H.~R. Russell, R.~N. Smith, and M.~C.~D. Marsh, ``{New constraints on light axion-like particles using Chandra transmission grating spectroscopy of the powerful cluster-hosted quasar H1821+643},'' \href{http://dx.doi.org/10.1093/mnras/stab3464}{{\it Mon. Not. Roy. Astron. Soc.} {\bfseries 510} (2021) 1264--1277}, \href{http://arxiv.org/abs/2109.03261}{{\ttfamily arXiv:2109.03261 [astro-ph.HE]}}.

\bibitem{Fermi-LAT:2016nkz}
{\bfseries Fermi-LAT} Collaboration, M.~Ajello {\it et~al.}, ``{Search for Spectral Irregularities due to Photon{\textendash}Axionlike-Particle Oscillations with the Fermi Large Area Telescope},'' \href{http://dx.doi.org/10.1103/PhysRevLett.116.161101}{{\it Phys. Rev. Lett.} {\bfseries 116} (2016) 161101}, \href{http://arxiv.org/abs/1603.06978}{{\ttfamily arXiv:1603.06978 [astro-ph.HE]}}.

\bibitem{Davies:2022wvj}
J.~Davies, M.~Meyer, and G.~Cotter, ``{Constraints on axionlike particles from a combined analysis of three flaring Fermi flat-spectrum radio quasars},'' \href{http://dx.doi.org/10.1103/PhysRevD.107.083027}{{\it Phys. Rev. D} {\bfseries 107} (2023) 083027}, \href{http://arxiv.org/abs/2211.03414}{{\ttfamily arXiv:2211.03414 [astro-ph.HE]}}.

\bibitem{MAGIC:2024arq}
{\bfseries MAGIC} Collaboration, H.~Abe {\it et~al.}, ``{Constraints on axion-like particles with the Perseus Galaxy Cluster with MAGIC},'' \href{http://dx.doi.org/10.1016/j.dark.2024.101425}{{\it Phys. Dark Univ.} {\bfseries 44} (2024) 101425}, \href{http://arxiv.org/abs/2401.07798}{{\ttfamily arXiv:2401.07798 [astro-ph.HE]}}.

\bibitem{Li:2020pcn}
H.-J. Li, J.-G. Guo, X.-J. Bi, S.-J. Lin, and P.-F. Yin, ``{Limits on axion-like particles from Mrk 421 with 4.5-year period observations by ARGO-YBJ and Fermi-LAT},'' \href{http://dx.doi.org/10.1103/PhysRevD.103.083003}{{\it Phys. Rev. D} {\bfseries 103} (2021) 083003}, \href{http://arxiv.org/abs/2008.09464}{{\ttfamily arXiv:2008.09464 [astro-ph.HE]}}.

\bibitem{Li:2021gxs}
H.-J. Li, X.-J. Bi, and P.-F. Yin, ``{Searching for axion-like particles with the blazar observations of MAGIC and Fermi-LAT *},'' \href{http://dx.doi.org/10.1088/1674-1137/ac6d4f}{{\it Chin. Phys. C} {\bfseries 46} (2022) 085105}, \href{http://arxiv.org/abs/2110.13636}{{\ttfamily arXiv:2110.13636 [astro-ph.HE]}}.

\bibitem{Li:2024zst}
H.-J. Li, W.~Chao, and Y.-F. Zhou, ``{Upper limit on the axion-photon coupling from Markarian 421},'' \href{http://dx.doi.org/10.1016/j.physletb.2024.139075}{{\it Phys. Lett. B} {\bfseries 858} (2024) 139075}, \href{http://arxiv.org/abs/2406.00387}{{\ttfamily arXiv:2406.00387 [hep-ph]}}.

\bibitem{Raffelt:1987im}
G.~Raffelt and L.~Stodolsky, ``{Mixing of the Photon with Low Mass Particles},'' \href{http://dx.doi.org/10.1103/PhysRevD.37.1237}{{\it Phys. Rev. D} {\bfseries 37} (1988) 1237}.

\bibitem{Biggio:2006im}
C.~Biggio, E.~Masso, and J.~Redondo, ``{Mixing of photons with massive spin-two particles in a magnetic field},'' \href{http://dx.doi.org/10.1103/PhysRevD.79.015012}{{\it Phys. Rev. D} {\bfseries 79} (2009) 015012}, \href{http://arxiv.org/abs/hep-ph/0604062}{{\ttfamily arXiv:hep-ph/0604062}}.

\bibitem{Dessert:2022yqq}
C.~Dessert, D.~Dunsky, and B.~R. Safdi, ``{Upper limit on the axion-photon coupling from magnetic white dwarf polarization},'' \href{http://dx.doi.org/10.1103/PhysRevD.105.103034}{{\it Phys. Rev. D} {\bfseries 105} (2022) 103034}, \href{http://arxiv.org/abs/2203.04319}{{\ttfamily arXiv:2203.04319 [hep-ph]}}.

\bibitem{Xiao:2020pra}
M.~Xiao, K.~M. Perez, M.~Giannotti, O.~Straniero, A.~Mirizzi, B.~W. Grefenstette, B.~M. Roach, and M.~Nynka, ``{Constraints on Axionlike Particles from a Hard X-Ray Observation of Betelgeuse},'' \href{http://dx.doi.org/10.1103/PhysRevLett.126.031101}{{\it Phys. Rev. Lett.} {\bfseries 126} (2021) 031101}, \href{http://arxiv.org/abs/2009.09059}{{\ttfamily arXiv:2009.09059 [astro-ph.HE]}}.

\bibitem{Hoof:2022xbe}
S.~Hoof and L.~Schulz, ``{Updated constraints on axion-like particles from temporal information in supernova SN1987A gamma-ray data},'' \href{http://dx.doi.org/10.1088/1475-7516/2023/03/054}{{\it JCAP} {\bfseries 03} (2023) 054}, \href{http://arxiv.org/abs/2212.09764}{{\ttfamily arXiv:2212.09764 [hep-ph]}}.

\bibitem{Calore:2021hhn}
F.~Calore, P.~Carenza, C.~Eckner, T.~Fischer, M.~Giannotti, J.~Jaeckel, K.~Kotake, T.~Kuroda, A.~Mirizzi, and F.~Sivo, ``{3D template-based Fermi-LAT constraints on the diffuse supernova axion-like particle background},'' \href{http://dx.doi.org/10.1103/PhysRevD.105.063028}{{\it Phys. Rev. D} {\bfseries 105} (2022) 063028}, \href{http://arxiv.org/abs/2110.03679}{{\ttfamily arXiv:2110.03679 [astro-ph.HE]}}.

\bibitem{Dessert:2020lil}
C.~Dessert, J.~W. Foster, and B.~R. Safdi, ``{X-ray Searches for Axions from Super Star Clusters},'' \href{http://dx.doi.org/10.1103/PhysRevLett.125.261102}{{\it Phys. Rev. Lett.} {\bfseries 125} (2020) 261102}, \href{http://arxiv.org/abs/2008.03305}{{\ttfamily arXiv:2008.03305 [hep-ph]}}.

\bibitem{Ning:2024eky}
O.~Ning and B.~R. Safdi, ``{Leading Axion-Photon Sensitivity with NuSTAR Observations of M82 and M87},'' \href{http://dx.doi.org/10.1103/PhysRevLett.134.171003}{{\it Phys. Rev. Lett.} {\bfseries 134} (2025) 171003}, \href{http://arxiv.org/abs/2404.14476}{{\ttfamily arXiv:2404.14476 [hep-ph]}}.

\bibitem{Meyer:2020vzy}
M.~Meyer and T.~Petrushevska, ``{Search for Axionlike-Particle-Induced Prompt $\gamma$-Ray Emission from Extragalactic Core-Collapse Supernovae with the $Fermi$ Large Area Telescope},'' \href{http://dx.doi.org/10.1103/PhysRevLett.124.231101}{{\it Phys. Rev. Lett.} {\bfseries 124} (2020) 231101}, \href{http://arxiv.org/abs/2006.06722}{{\ttfamily arXiv:2006.06722 [astro-ph.HE]}}. [Erratum: Phys.Rev.Lett. 125, 119901 (2020)].

\bibitem{Manzari:2024jns}
C.~A. Manzari, Y.~Park, B.~R. Safdi, and I.~Savoray, ``{Supernova Axions Convert to Gamma Rays in Magnetic Fields of Progenitor Stars},'' \href{http://dx.doi.org/10.1103/PhysRevLett.133.211002}{{\it Phys. Rev. Lett.} {\bfseries 133} (2024) 211002}, \href{http://arxiv.org/abs/2405.19393}{{\ttfamily arXiv:2405.19393 [hep-ph]}}.

\bibitem{Fiorillo:2025gnd}
D.~F.~G. Fiorillo, {\'A}.~Gil~Muyor, H.-T. Janka, G.~G. Raffelt, and E.~Vitagliano, ``{Axion-photon conversion in transient compact stars: Systematics, constraints, and opportunities},'' \href{http://dx.doi.org/10.1088/1475-7516/2026/03/053}{{\it JCAP} {\bfseries 03} (2026) 053}, \href{http://arxiv.org/abs/2509.13322}{{\ttfamily arXiv:2509.13322 [hep-ph]}}.

\bibitem{Dolan:2022kul}
M.~J. Dolan, F.~J. Hiskens, and R.~R. Volkas, ``{Advancing globular cluster constraints on the axion-photon coupling},'' \href{http://dx.doi.org/10.1088/1475-7516/2022/10/096}{{\it JCAP} {\bfseries 10} (2022) 096}, \href{http://arxiv.org/abs/2207.03102}{{\ttfamily arXiv:2207.03102 [hep-ph]}}.

\bibitem{Fiorillo:2025yzf}
D.~F.~G. Fiorillo, T.~Pitik, and E.~Vitagliano, ``{Energy Transfer by Feebly Interacting Particles in Supernovae: The Trapping Regime},'' \href{http://dx.doi.org/10.1103/cz94-dqxt}{{\it Phys. Rev. Lett.} {\bfseries 135} (2025) 071005}, \href{http://arxiv.org/abs/2503.13653}{{\ttfamily arXiv:2503.13653 [hep-ph]}}.

\bibitem{Caputo:2022mah}
A.~Caputo, H.-T. Janka, G.~Raffelt, and E.~Vitagliano, ``{Low-Energy Supernovae Severely Constrain Radiative Particle Decays},'' \href{http://dx.doi.org/10.1103/PhysRevLett.128.221103}{{\it Phys. Rev. Lett.} {\bfseries 128} (2022) 221103}, \href{http://arxiv.org/abs/2201.09890}{{\ttfamily arXiv:2201.09890 [astro-ph.HE]}}.

\bibitem{Muller:2023vjm}
E.~M{\"u}ller, F.~Calore, P.~Carenza, C.~Eckner, and M.~C.~D. Marsh, ``{Investigating the gamma-ray burst from decaying MeV-scale axion-like particles produced in supernova explosions},'' \href{http://dx.doi.org/10.1088/1475-7516/2023/07/056}{{\it JCAP} {\bfseries 07} (2023) 056}, \href{http://arxiv.org/abs/2304.01060}{{\ttfamily arXiv:2304.01060 [astro-ph.HE]}}.

\bibitem{Lucente:2020whw}
G.~Lucente, P.~Carenza, T.~Fischer, M.~Giannotti, and A.~Mirizzi, ``{Heavy axion-like particles and core-collapse supernovae: constraints and impact on the explosion mechanism},'' \href{http://dx.doi.org/10.1088/1475-7516/2020/12/008}{{\it JCAP} {\bfseries 12} (2020) 008}, \href{http://arxiv.org/abs/2008.04918}{{\ttfamily arXiv:2008.04918 [hep-ph]}}.

\bibitem{Beaufort:2023zuj}
C.~Beaufort, M.~Bastero-Gil, T.~Luce, and D.~Santos, ``{New solar x-ray constraints on keV axionlike particles},'' \href{http://dx.doi.org/10.1103/PhysRevD.108.L081302}{{\it Phys. Rev. D} {\bfseries 108} (2023) L081302}, \href{http://arxiv.org/abs/2303.06968}{{\ttfamily arXiv:2303.06968 [hep-ph]}}.

\bibitem{Candon:2025ypl}
F.~R. Cand{\'o}n, D.~F.~G. Fiorillo, H.-T. Janka, B.~F.~A. van Baal, and E.~Vitagliano, ``{Small Progenitors, Large Couplings: Type Ic Supernova Constraints on Radiatively Decaying Particles},'' \href{http://arxiv.org/abs/2509.18253}{{\ttfamily arXiv:2509.18253 [hep-ph]}}.

\bibitem{Diamond:2023cto}
M.~Diamond, D.~F.~G. Fiorillo, G.~Marques-Tavares, I.~Tamborra, and E.~Vitagliano, ``{Multimessenger Constraints on Radiatively Decaying Axions from GW170817},'' \href{http://dx.doi.org/10.1103/PhysRevLett.132.101004}{{\it Phys. Rev. Lett.} {\bfseries 132} (2024) 101004}, \href{http://arxiv.org/abs/2305.10327}{{\ttfamily arXiv:2305.10327 [hep-ph]}}.

\bibitem{Vinyoles:2015aba}
N.~Vinyoles, A.~Serenelli, F.~L. Villante, S.~Basu, J.~Redondo, and J.~Isern, ``{New axion and hidden photon constraints from a solar data global fit},'' \href{http://dx.doi.org/10.1088/1475-7516/2015/10/015}{{\it JCAP} {\bfseries 10} (2015) 015}, \href{http://arxiv.org/abs/1501.01639}{{\ttfamily arXiv:1501.01639 [astro-ph.SR]}}.

\bibitem{Langhoff:2022bij}
K.~Langhoff, N.~J. Outmezguine, and N.~L. Rodd, ``{Irreducible Axion Background},'' \href{http://dx.doi.org/10.1103/PhysRevLett.129.241101}{{\it Phys. Rev. Lett.} {\bfseries 129} (2022) 241101}, \href{http://arxiv.org/abs/2209.06216}{{\ttfamily arXiv:2209.06216 [hep-ph]}}.

\bibitem{Landsberg:2015pka}
G.~Landsberg, ``{Searches for Extra Spatial Dimensions with the CMS Detector at the LHC},'' \href{http://dx.doi.org/10.1142/S0217732315400179}{{\it Mod. Phys. Lett. A} {\bfseries 30} (2015) 1540017}, \href{http://arxiv.org/abs/1506.00024}{{\ttfamily arXiv:1506.00024 [hep-ex]}}.

\bibitem{ATLAS:2024fdw}
{\bfseries ATLAS} Collaboration, G.~Aad {\it et~al.}, ``{Exploration at the high-energy frontier: ATLAS Run~2 searches investigating the exotic jungle beyond the Standard Model},'' \href{http://dx.doi.org/10.1016/j.physrep.2024.10.001}{{\it Phys. Rept.} {\bfseries 1116} (2025) 301--385}, \href{http://arxiv.org/abs/2403.09292}{{\ttfamily arXiv:2403.09292 [hep-ex]}}.

\bibitem{CMS:2024nht}
{\bfseries CMS} Collaboration, A.~Hayrapetyan {\it et~al.}, ``{Search for new physics in high-mass diphoton events from proton-proton collisions at $ \sqrt{\textrm{s}} $ = 13 TeV},'' \href{http://dx.doi.org/10.1007/JHEP08(2024)215}{{\it JHEP} {\bfseries 08} (2024) 215}, \href{http://arxiv.org/abs/2405.09320}{{\ttfamily arXiv:2405.09320 [hep-ex]}}.

\bibitem{Atwood:1999zg}
D.~Atwood, S.~Bar-Shalom, and A.~Soni, ``{Graviton production by two photon processes in Kaluza-Klein theories with large extra dimensions},'' \href{http://arxiv.org/abs/hep-ph/9903538}{{\ttfamily arXiv:hep-ph/9903538}}.

\bibitem{Ahern:2000jn}
S.~C. Ahern, J.~W. Norbury, and W.~J. Poyser, ``{Graviton production in relativistic heavy ion collisions},'' \href{http://dx.doi.org/10.1103/PhysRevD.62.116001}{{\it Phys. Rev. D} {\bfseries 62} (2000) 116001}, \href{http://arxiv.org/abs/gr-qc/0009059}{{\ttfamily arXiv:gr-qc/0009059}}.

\bibitem{Fichet:2014uka}
S.~Fichet, G.~von Gersdorff, B.~Lenzi, C.~Royon, and M.~Saimpert, ``{Light-by-light scattering with intact protons at the LHC: from Standard Model to New Physics},'' \href{http://dx.doi.org/10.1007/JHEP02(2015)165}{{\it JHEP} {\bfseries 02} (2015) 165}, \href{http://arxiv.org/abs/1411.6629}{{\ttfamily arXiv:1411.6629 [hep-ph]}}.

\bibitem{Inan:2018jza}
S.~C.~t. Inan and A.~V. Kisselev, ``{Search for the RS model with a small curvature through photon-induced process at the LHC},'' \href{http://dx.doi.org/10.1140/epjc/s10052-018-6210-5}{{\it Eur. Phys. J. C} {\bfseries 78} (2018) 729}, \href{http://arxiv.org/abs/1805.01441}{{\ttfamily arXiv:1805.01441 [hep-ph]}}.

\bibitem{Cembranos:2021vdv}
J.~A.~R. Cembranos, R.~L. Delgado, and H.~Villarrubia-Rojo, ``{LHC constraints on hidden gravitons},'' \href{http://dx.doi.org/10.1007/JHEP01(2022)129}{{\it JHEP} {\bfseries 01} (2022) 129}, \href{http://arxiv.org/abs/2108.00930}{{\ttfamily arXiv:2108.00930 [hep-ph]}}.

\bibitem{Shao:2022cly}
H.-S. Shao and D.~d'Enterria, ``{gamma-UPC: automated generation of exclusive photon-photon processes in ultraperipheral proton and nuclear collisions with varying form factors},'' \href{http://dx.doi.org/10.1007/JHEP09(2022)248}{{\it JHEP} {\bfseries 09} (2022) 248}, \href{http://arxiv.org/abs/2207.03012}{{\ttfamily arXiv:2207.03012 [hep-ph]}}.

\bibitem{Voronchikhin:2022rwc}
I.~V. Voronchikhin and D.~V. Kirpichnikov, ``{Probing hidden spin-2 mediator of dark matter with NA64e, LDMX, NA64\ensuremath{\mu}, and M3},'' \href{http://dx.doi.org/10.1103/PhysRevD.106.115041}{{\it Phys. Rev. D} {\bfseries 106} (2022) 115041}, \href{http://arxiv.org/abs/2210.00751}{{\ttfamily arXiv:2210.00751 [hep-ph]}}.

\bibitem{Jodlowski:2023yne}
K.~Jod\l{}owski, ``{Looking forward to photon-coupled long-lived particles I: massive spin-2 portal},'' \href{http://arxiv.org/abs/2305.05710}{{\ttfamily arXiv:2305.05710 [hep-ph]}}.

\bibitem{Baltz:2007kq}
A.~J. Baltz, ``{The Physics of Ultraperipheral Collisions at the LHC},'' \href{http://dx.doi.org/10.1016/j.physrep.2007.12.001}{{\it Phys. Rept.} {\bfseries 458} (2008) 1--171}, \href{http://arxiv.org/abs/0706.3356}{{\ttfamily arXiv:0706.3356 [nucl-ex]}}.

\bibitem{dEnterria:2013zqi}
D.~d'Enterria and G.~G. da~Silveira, ``{Observing light-by-light scattering at the Large Hadron Collider},'' \href{http://dx.doi.org/10.1103/PhysRevLett.111.080405}{{\it Phys. Rev. Lett.} {\bfseries 111} (2013) 080405}, \href{http://arxiv.org/abs/1305.7142}{{\ttfamily arXiv:1305.7142 [hep-ph]}}. [Erratum: Phys.Rev.Lett. 116, 129901 (2016)].

\bibitem{Bruce:2018yzs}
R.~Bruce {\it et~al.}, ``{New physics searches with heavy-ion collisions at the CERN Large Hadron Collider},'' \href{http://dx.doi.org/10.1088/1361-6471/ab7ff7}{{\it J. Phys. G} {\bfseries 47} (2020) 060501}, \href{http://arxiv.org/abs/1812.07688}{{\ttfamily arXiv:1812.07688 [hep-ph]}}.

\bibitem{dEnterria:2022sut}
D.~d'Enterria {\it et~al.}, ``{Opportunities for new physics searches with heavy ions at colliders},'' \href{http://dx.doi.org/10.1088/1361-6471/acc197}{{\it J. Phys. G} {\bfseries 50} (2023) 050501}, \href{http://arxiv.org/abs/2203.05939}{{\ttfamily arXiv:2203.05939 [hep-ph]}}.

\bibitem{Riordan:1987aw}
E.~M. Riordan {\it et~al.}, ``{A Search for Short Lived Axions in an Electron Beam Dump Experiment},'' \href{http://dx.doi.org/10.1103/PhysRevLett.59.755}{{\it Phys. Rev. Lett.} {\bfseries 59} (1987) 755}.

\bibitem{Batell:2014mga}
B.~Batell, R.~Essig, and Z.~Surujon, ``{Strong Constraints on Sub-GeV Dark Sectors from SLAC Beam Dump E137},'' \href{http://dx.doi.org/10.1103/PhysRevLett.113.171802}{{\it Phys. Rev. Lett.} {\bfseries 113} (2014) 171802}, \href{http://arxiv.org/abs/1406.2698}{{\ttfamily arXiv:1406.2698 [hep-ph]}}.

\bibitem{MiniBooNEDM:2018cxm}
{\bfseries MiniBooNE DM} Collaboration, A.~A. Aguilar-Arevalo {\it et~al.}, ``{Dark Matter Search in Nucleon, Pion, and Electron Channels from a Proton Beam Dump with MiniBooNE},'' \href{http://dx.doi.org/10.1103/PhysRevD.98.112004}{{\it Phys. Rev. D} {\bfseries 98} (2018) 112004}, \href{http://arxiv.org/abs/1807.06137}{{\ttfamily arXiv:1807.06137 [hep-ex]}}.

\bibitem{MiniBooNE:2020pnu}
{\bfseries MiniBooNE} Collaboration, A.~A. Aguilar-Arevalo {\it et~al.}, ``{Updated MiniBooNE neutrino oscillation results with increased data and new background studies},'' \href{http://dx.doi.org/10.1103/PhysRevD.103.052002}{{\it Phys. Rev. D} {\bfseries 103} (2021) 052002}, \href{http://arxiv.org/abs/2006.16883}{{\ttfamily arXiv:2006.16883 [hep-ex]}}.

\bibitem{Capozzi:2023ffu}
F.~Capozzi, B.~Dutta, G.~Gurung, W.~Jang, I.~M. Shoemaker, A.~Thompson, and J.~Yu, ``{New constraints on ALP couplings to electrons and photons from ArgoNeuT and the MiniBooNE beam dump},'' \href{http://dx.doi.org/10.1103/PhysRevD.108.075019}{{\it Phys. Rev. D} {\bfseries 108} (2023) 075019}, \href{http://arxiv.org/abs/2307.03878}{{\ttfamily arXiv:2307.03878 [hep-ph]}}.

\bibitem{CHARM:1985anb}
{\bfseries CHARM} Collaboration, F.~Bergsma {\it et~al.}, ``{Search for Axion Like Particle Production in 400-{GeV} Proton - Copper Interactions},'' \href{http://dx.doi.org/10.1016/0370-2693(85)90400-9}{{\it Phys. Lett. B} {\bfseries 157} (1985) 458--462}.

\bibitem{NA64:2020qwq}
{\bfseries NA64} Collaboration, D.~Banerjee {\it et~al.}, ``{Search for Axionlike and Scalar Particles with the NA64 Experiment},'' \href{http://dx.doi.org/10.1103/PhysRevLett.125.081801}{{\it Phys. Rev. Lett.} {\bfseries 125} (2020) 081801}, \href{http://arxiv.org/abs/2005.02710}{{\ttfamily arXiv:2005.02710 [hep-ex]}}.

\bibitem{FASER:2024bbl}
{\bfseries FASER} Collaboration, R.~Mammen~Abraham {\it et~al.}, ``{Shining light on the dark sector: search for axion-like particles and other new physics in photonic final states with FASER},'' \href{http://dx.doi.org/10.1007/JHEP01(2025)199}{{\it JHEP} {\bfseries 01} (2025) 199}, \href{http://arxiv.org/abs/2410.10363}{{\ttfamily arXiv:2410.10363 [hep-ex]}}.

\bibitem{SHiP:2021nfo}
{\bfseries SHiP} Collaboration, C.~Ahdida {\it et~al.}, ``{The SHiP experiment at the proposed CERN SPS Beam Dump Facility},'' \href{http://dx.doi.org/10.1140/epjc/s10052-022-10346-5}{{\it Eur. Phys. J. C} {\bfseries 82} (2022) 486}, \href{http://arxiv.org/abs/2112.01487}{{\ttfamily arXiv:2112.01487 [physics.ins-det]}}.

\bibitem{Brodsky:1971ud}
S.~J. Brodsky, T.~Kinoshita, and H.~Terazawa, ``{Two Photon Mechanism of Particle Production by High-Energy Colliding Beams},'' \href{http://dx.doi.org/10.1103/PhysRevD.4.1532}{{\it Phys. Rev. D} {\bfseries 4} (1971) 1532--1557}.

\bibitem{Budnev:1975poe}
V.~M. Budnev, I.~F. Ginzburg, G.~V. Meledin, and V.~G. Serbo, ``{The Two photon particle production mechanism. Physical problems. Applications. Equivalent photon approximation},'' \href{http://dx.doi.org/10.1016/0370-1573(75)90009-5}{{\it Phys. Rept.} {\bfseries 15} (1975) 181--281}.

\bibitem{Bjorken:2009mm}
J.~D. Bjorken, R.~Essig, P.~Schuster, and N.~Toro, ``{New Fixed-Target Experiments to Search for Dark Gauge Forces},'' \href{http://dx.doi.org/10.1103/PhysRevD.80.075018}{{\it Phys. Rev. D} {\bfseries 80} (2009) 075018}, \href{http://arxiv.org/abs/0906.0580}{{\ttfamily arXiv:0906.0580 [hep-ph]}}.

\bibitem{L3:1994shn}
{\bfseries L3} Collaboration, M.~Acciarri {\it et~al.}, ``{Search for anomalous Z --{\ensuremath{>}} gamma gamma gamma events at LEP},'' \href{http://dx.doi.org/10.1016/0370-2693(95)01612-T}{{\it Phys. Lett. B} {\bfseries 345} (1995) 609--616}.

\bibitem{Baillargeon:1995dg}
M.~Baillargeon, F.~Boudjema, E.~Chopin, and V.~Lafage, ``{New physics with three photon events at LEP},'' \href{http://dx.doi.org/10.1007/s002880050188}{{\it Z. Phys. C} {\bfseries 71} (1996) 431--442}, \href{http://arxiv.org/abs/hep-ph/9506396}{{\ttfamily arXiv:hep-ph/9506396}}.

\bibitem{OPAL:2002vhf}
{\bfseries OPAL} Collaboration, G.~Abbiendi {\it et~al.}, ``{Multiphoton production in e+ e- collisions at $\sqrts = 181$~GeV to 209~GeV},'' \href{http://dx.doi.org/10.1140/epjc/s2002-01074-5}{{\it Eur. Phys. J. C} {\bfseries 26} (2003) 331--344}, \href{http://arxiv.org/abs/hep-ex/0210016}{{\ttfamily arXiv:hep-ex/0210016}}.

\bibitem{Dolan:2017osp}
M.~J. Dolan, T.~Ferber, C.~Hearty, F.~Kahlhoefer, and K.~Schmidt-Hoberg, ``{Revised constraints and Belle II sensitivity for visible and invisible axion-like particles},'' \href{http://dx.doi.org/10.1007/JHEP12(2017)094}{{\it JHEP} {\bfseries 12} (2017) 094}, \href{http://arxiv.org/abs/1709.00009}{{\ttfamily arXiv:1709.00009 [hep-ph]}}. [Erratum: JHEP 03, 190 (2021)].

\bibitem{Belle-II:2020jti}
{\bfseries Belle~II} Collaboration, F.~Abudin\'en {\it et~al.}, ``{Search for Axion-Like Particles produced in $e^+e^-$ collisions at Belle~II},'' \href{http://dx.doi.org/10.1103/PhysRevLett.125.161806}{{\it Phys. Rev. Lett.} {\bfseries 125} (2020) 161806}, \href{http://arxiv.org/abs/2007.13071}{{\ttfamily arXiv:2007.13071 [hep-ex]}}.

\bibitem{Belle-II:2026znr}
{\bfseries Belle-II} Collaboration, M.~Abumusabh {\it et~al.}, ``{Search for axion-like particles decaying to two photons at Belle II},'' \href{http://arxiv.org/abs/2607.07800}{{\ttfamily arXiv:2607.07800 [hep-ex]}}.

\bibitem{BESIII:2022rzz}
{\bfseries BESIII} Collaboration, M.~Ablikim {\it et~al.}, ``{Search for an axion-like particle in radiative J/\ensuremath{\psi} decays},'' \href{http://dx.doi.org/10.1016/j.physletb.2023.137698}{{\it Phys. Lett. B} {\bfseries 838} (2023) 137698}, \href{http://arxiv.org/abs/2211.12699}{{\ttfamily arXiv:2211.12699 [hep-ex]}}.

\bibitem{BESIII:2024hdv}
{\bfseries BESIII} Collaboration, M.~Ablikim {\it et~al.}, ``{Search for diphoton decays of an axionlike particle in radiative J/{\ensuremath{\psi}} decays},'' \href{http://dx.doi.org/10.1103/PhysRevD.110.L031101}{{\it Phys. Rev. D} {\bfseries 110} (2024) L031101}, \href{http://arxiv.org/abs/2404.04640}{{\ttfamily arXiv:2404.04640 [hep-ex]}}.

\bibitem{CLEO:1994hzy}
{\bfseries CLEO} Collaboration, R.~Balest {\it et~al.}, ``{Upsilon (1s) ---{\ensuremath{>}} gamma + noninteracting particles},'' \href{http://dx.doi.org/10.1103/PhysRevD.51.2053}{{\it Phys. Rev. D} {\bfseries 51} (1995) 2053--2060}.

\bibitem{RebelloTeles:2023uig}
P.~Rebello~Teles, D.~d'Enterria, V.~P. Gon{\c{c}}alves, and D.~E. Martins, ``{Searches for axionlike particles via {\ensuremath{\gamma}}{\ensuremath{\gamma}} fusion at future e+e- colliders},'' \href{http://dx.doi.org/10.1103/PhysRevD.109.055003}{{\it Phys. Rev. D} {\bfseries 109} (2024) 055003}, \href{http://arxiv.org/abs/2310.17270}{{\ttfamily arXiv:2310.17270 [hep-ex]}}.

\bibitem{Polesello:2025gwj}
G.~Polesello, ``{Sensitivity of the FCC-ee to decay of an axion-like particle into two photons},'' \href{http://dx.doi.org/10.1007/JHEP06(2025)239}{{\it JHEP} {\bfseries 06} (2025) 239}, \href{http://arxiv.org/abs/2502.08411}{{\ttfamily arXiv:2502.08411 [hep-ph]}}.

\bibitem{FCC:2025lpp}
{\bfseries FCC} Collaboration, M.~Benedikt {\it et~al.}, ``{Future Circular Collider Feasibility Study Report: Volume 1, Physics, Experiments, Detectors},'' \href{http://dx.doi.org/10.1140/epjc/s10052-025-15077-x}{{\it Eur. Phys. J. C} {\bfseries 85} (2025) 1468}, \href{http://arxiv.org/abs/2505.00272}{{\ttfamily arXiv:2505.00272 [hep-ex]}}.

\bibitem{deBlas:2025gyz}
J.~de~Blas {\it et~al.}, ``{Physics Briefing Book: Input for the 2026 update of the European Strategy for Particle Physics},'' \href{http://arxiv.org/abs/2511.03883}{{\ttfamily arXiv:2511.03883 [hep-ex]}}.

\bibitem{RebelloDdE2026}
P.~Rebello-Teles and D.~d'Enterria, ``\textit{Searches for axion- and graviton-like particles via $\gaga$ fusion at FCC-hh}.'' In preparation, (2026).

\bibitem{CMS:2018erd}
{\bfseries CMS} Collaboration, A.~M. Sirunyan {\it et~al.}, ``{Evidence for light-by-light scattering and searches for axion-like particles in ultraperipheral PbPb collisions at $\sqrt{s_\mathrm{NN}} = 5.02$~TeV},'' \href{http://dx.doi.org/10.1016/j.physletb.2019.134826}{{\it Phys. Lett. B} {\bfseries 797} (2019) 134826}, \href{http://arxiv.org/abs/1810.04602}{{\ttfamily arXiv:1810.04602 [hep-ex]}}.

\bibitem{CMS:2024bnt}
{\bfseries CMS} Collaboration, A.~Hayrapetyan {\it et~al.}, ``{Measurement of light-by-light scattering and the Breit-Wheeler process, and search for axion-like particles in ultraperipheral PbPb collisions at $\sqrt{{s}_{\text{NN}}}$ = 5.02 TeV},'' \href{http://dx.doi.org/10.1007/JHEP08(2025)006}{{\it JHEP} {\bfseries 08} (2025) 006}, \href{http://arxiv.org/abs/2412.15413}{{\ttfamily arXiv:2412.15413 [nucl-ex]}}.

\bibitem{ATLAS:2020hii}
{\bfseries ATLAS} Collaboration, G.~Aad {\it et~al.}, ``{Measurement of light-by-light scattering and search for axion-like particles with 2.2 nb$^{-1}$ of Pb+Pb data with the ATLAS detector},'' \href{http://dx.doi.org/10.1007/JHEP11(2021)050}{{\it JHEP} {\bfseries 03} (2021) 243}, \href{http://arxiv.org/abs/2008.05355}{{\ttfamily arXiv:2008.05355 [hep-ex]}}. [Erratum: JHEP 11, 050 (2021)].

\bibitem{TOTEM:2021zxa}
{\bfseries TOTEM and CMS} Collaboration, A.~Tumasyan {\it et~al.}, ``{First Search for Exclusive Diphoton Production at High Mass with Tagged Protons in Proton-Proton Collisions at $\sqrt{s}= 13$~TeV},'' \href{http://dx.doi.org/10.1103/PhysRevLett.129.011801}{{\it Phys. Rev. Lett.} {\bfseries 129} (2022) 011801}, \href{http://arxiv.org/abs/2110.05916}{{\ttfamily arXiv:2110.05916 [hep-ex]}}.

\bibitem{CMS:2022zfd}
{\bfseries TOTEM, CMS} Collaboration, A.~Tumasyan {\it et~al.}, ``{Search for high-mass exclusive diphoton production with tagged protons in proton-proton collisions at $\sqrts=13$~TeV},'' \href{http://dx.doi.org/10.1103/PhysRevD.110.012010}{{\it Phys. Rev. D} {\bfseries 110} (2024) 012010}, \href{http://arxiv.org/abs/2311.02725}{{\ttfamily arXiv:2311.02725 [hep-ex]}}.

\bibitem{ATLAS:2023zfc}
{\bfseries ATLAS} Collaboration, G.~Aad {\it et~al.}, ``{Search for an axion-like particle with forward proton scattering in association with photon pairs at ATLAS},'' \href{http://dx.doi.org/10.1007/JHEP07(2023)234}{{\it JHEP} {\bfseries 07} (2023) 234}, \href{http://arxiv.org/abs/2304.10953}{{\ttfamily arXiv:2304.10953 [hep-ex]}}.

\bibitem{Bauer:2018uxu}
M.~Bauer, M.~Heiles, M.~Neubert, and A.~Thamm, ``{Axion-Like Particles at Future Colliders},'' \href{http://dx.doi.org/10.1140/epjc/s10052-019-6587-9}{{\it Eur. Phys. J. C} {\bfseries 79} (2019) 74}, \href{http://arxiv.org/abs/1808.10323}{{\ttfamily arXiv:1808.10323 [hep-ph]}}.

\bibitem{Brambilla:2010cs}
N.~Brambilla {\it et~al.}, ``{Heavy Quarkonium: Progress, Puzzles, and Opportunities},'' \href{http://dx.doi.org/10.1140/epjc/s10052-010-1534-9}{{\it Eur. Phys. J. C} {\bfseries 71} (2011) 1534}, \href{http://arxiv.org/abs/1010.5827}{{\ttfamily arXiv:1010.5827 [hep-ph]}}.

\bibitem{Lansberg:2019adr}
J.-P. Lansberg, ``{New Observables in Inclusive Production of Quarkonia},'' \href{http://dx.doi.org/10.1016/j.physrep.2020.08.007}{{\it Phys. Rept.} {\bfseries 889} (2020) 1--106}, \href{http://arxiv.org/abs/1903.09185}{{\ttfamily arXiv:1903.09185 [hep-ph]}}.

\bibitem{dEnterria:2025ecx}
D.~d'Enterria and K.~Kang, ``{Exclusive photon-fusion production of even-spin resonances and exotic QED atoms in high-energy hadron collisions},'' \href{http://dx.doi.org/10.1103/rnxl-v6gd}{{\it Phys. Rev. D} {\bfseries 112} (2025) 116022}, \href{http://arxiv.org/abs/2503.10952}{{\ttfamily arXiv:2503.10952 [hep-ph]}}.

\bibitem{dEnterria:2020dwq}
D.~d'Enterria and C.~Loizides, ``{Progress in the Glauber Model at Collider Energies},'' \href{http://dx.doi.org/10.1146/annurev-nucl-102419-060007}{{\it Ann. Rev. Nucl. Part. Sci.} {\bfseries 71} (2021) 315--344}, \href{http://arxiv.org/abs/2011.14909}{{\ttfamily arXiv:2011.14909 [hep-ph]}}.

\bibitem{Alwall:2014hca}
J.~Alwall, R.~Frederix, S.~Frixione, V.~Hirschi, F.~Maltoni, O.~Mattelaer, H.~S. Shao, T.~Stelzer, P.~Torrielli, and M.~Zaro, ``{The automated computation of tree-level and next-to-leading order differential cross sections, and their matching to parton shower simulations},'' \href{http://dx.doi.org/10.1007/JHEP07(2014)079}{{\it JHEP} {\bfseries 07} (2014) 079}, \href{http://arxiv.org/abs/1405.0301}{{\ttfamily arXiv:1405.0301 [hep-ph]}}.

\bibitem{Frederix:2018nkq}
R.~Frederix, S.~Frixione, V.~Hirschi, D.~Pagani, H.~S. Shao, and M.~Zaro, ``{The automation of next-to-leading order electroweak calculations},'' \href{http://dx.doi.org/10.1007/JHEP11(2021)085}{{\it JHEP} {\bfseries 07} (2018) 185}, \href{http://arxiv.org/abs/1804.10017}{{\ttfamily arXiv:1804.10017 [hep-ph]}}. [Erratum: JHEP 11, 085 (2021)].

\bibitem{Degrande:2011ua}
C.~Degrande, C.~Duhr, B.~Fuks, D.~Grellscheid, O.~Mattelaer, and T.~Reiter, ``{UFO - The Universal FeynRules Output},'' \href{http://dx.doi.org/10.1016/j.cpc.2012.01.022}{{\it Comput. Phys. Commun.} {\bfseries 183} (2012) 1201--1214}, \href{http://arxiv.org/abs/1108.2040}{{\ttfamily arXiv:1108.2040 [hep-ph]}}.

\bibitem{Darme:2023jdn}
L.~Darm{\'e} {\it et~al.}, ``{UFO 2.0: the {\textquoteleft}Universal Feynman Output{\textquoteright} format},'' \href{http://dx.doi.org/10.1140/epjc/s10052-023-11780-9}{{\it Eur. Phys. J. C} {\bfseries 83} (2023) 631}, \href{http://arxiv.org/abs/2304.09883}{{\ttfamily arXiv:2304.09883 [hep-ph]}}.

\bibitem{Gue:2024onx}
J.~Gu\'e, A.~Hees, and P.~Wolf, ``{Violation of the equivalence principle induced by oscillating rest mass and transition frequency, and its detection in atom interferometers},'' \href{http://dx.doi.org/10.1103/PhysRevD.110.035005}{{\it Phys. Rev. D} {\bfseries 110} (2024) 035005}, \href{http://arxiv.org/abs/2401.14742}{{\ttfamily arXiv:2401.14742 [hep-ph]}}.

\bibitem{LIGOScientific:2021ffg}
{\bfseries LIGO Scientific, KAGRA, Virgo} Collaboration, R.~Abbott {\it et~al.}, ``{Constraints on dark photon dark matter using data from LIGO{\textquoteright}s and Virgo{\textquoteright}s third observing run},'' \href{http://dx.doi.org/10.1103/PhysRevD.105.063030}{{\it Phys. Rev. D} {\bfseries 105} (2022) 063030}, \href{http://arxiv.org/abs/2105.13085}{{\ttfamily arXiv:2105.13085 [astro-ph.CO]}}. [Erratum: Phys.Rev.D 109, 089902 (2024)].

\bibitem{Zhang:2025fck}
J.-R. Zhang, J.~Chen, H.-S. Jiao, R.-G. Cai, and Y.-L. Zhang, ``{Probing spin-2 ultralight dark matter with space-based gravitational wave detectors in the mHz regime},'' \href{http://dx.doi.org/10.1103/cz2w-5cfj}{{\it Phys. Rev. D} {\bfseries 112} (2025) 064030}, \href{http://arxiv.org/abs/2501.11071}{{\ttfamily arXiv:2501.11071 [gr-qc]}}.

\bibitem{Manita:2023mnc}
Y.~Manita, H.~Takeda, K.~Aoki, T.~Fujita, and S.~Mukohyama, ``{Exploring the spin of ultralight dark matter with gravitational wave detectors},'' \href{http://dx.doi.org/10.1103/PhysRevD.109.095012}{{\it Phys. Rev. D} {\bfseries 109} (2024) 095012}, \href{http://arxiv.org/abs/2310.10646}{{\ttfamily arXiv:2310.10646 [hep-ph]}}.

\bibitem{BaBar:2010eww}
{\bfseries BaBar} Collaboration, P.~del Amo~Sanchez {\it et~al.}, ``{Search for Production of Invisible Final States in Single-Photon Decays of $\Upsilon(1S)$},'' \href{http://dx.doi.org/10.1103/PhysRevLett.107.021804}{{\it Phys. Rev. Lett.} {\bfseries 107} (2011) 021804}, \href{http://arxiv.org/abs/1007.4646}{{\ttfamily arXiv:1007.4646 [hep-ex]}}.

\bibitem{Chen:2014oda}
Y.~J. Chen, W.~K. Tham, D.~E. Krause, D.~Lopez, E.~Fischbach, and R.~S. Decca, ``{Stronger Limits on Hypothetical Yukawa Interactions in the 30{\textendash}8000 nm Range},'' \href{http://dx.doi.org/10.1103/PhysRevLett.116.221102}{{\it Phys. Rev. Lett.} {\bfseries 116} (2016) 221102}, \href{http://arxiv.org/abs/1410.7267}{{\ttfamily arXiv:1410.7267 [hep-ex]}}.

\bibitem{Fienga:2023ocw}
A.~Fienga and O.~Minazzoli, ``{Testing theories of gravity with planetary ephemerides},'' \href{http://dx.doi.org/10.1007/s41114-023-00047-0}{{\it Living Rev. Rel.} {\bfseries 27} (2024) 1}, \href{http://arxiv.org/abs/2303.01821}{{\ttfamily arXiv:2303.01821 [gr-qc]}}.

\end{thebibliography}\endgroup

\end{document}